\documentclass[longauth]{aa}
\usepackage{longtable}
\usepackage{multirow}
\usepackage{lmodern}
\usepackage[varg]{txfonts}
\bibpunct{(}{)}{;}{a}{}{,}
\usepackage{amssymb}
\usepackage{graphicx}
\usepackage{url}
\usepackage{mathrsfs}
\usepackage{color}
\usepackage{lscape}
\newcommand{\myemail}{lgalbany@das.uchile.cl}
\usepackage{amsmath}
\usepackage[citecolor=blue]{hyperref}
\hypersetup{colorlinks=true}
 \newcommand{\avg}[1]{\left< #1 \right>}

\def\ratioR23{([\ion{O}{ii}] $\lambda$3727 +[\ion{O}{iii}] $\lambda\lambda$4959,5007)/H$\beta$}

\def\ratioS23{([\ion{S}{ii}] $\lambda\lambda$6717,6731 +[\ion{S}{iii}] $\lambda\lambda$9069,9532)/H$\beta$}

\def\OIIuu{[\ion{O}{ii}] $\lambda\lambda$3726,3729}
\def\OII{[{\ion{O}{ii}}]}

\def\OIIIt{[\ion{O}{iii}] $\lambda$4363}

\def\OIIIud{[\ion{O}{iii}] $\lambda\lambda$4959,5007}

\def\SII{[{\ion{S}{ii}}]}

\newcommand{\Had}{H$\alpha$}
\newcommand{\Hbd}{H$\beta$}

\def\R23{$R_{23}$}

\def\Msun{$M_\odot$}

\makeatletter
  \newcommand\tinyv{\@setfontsize\tinyv{6pt}{6}}
  \newcommand\tinyvv{\@setfontsize\tinyv{5pt}{5}}
\makeatother

\begin{document}

\title{Nearby supernova host galaxies from the CALIFA Survey:}
\subtitle{II. SN environmental metallicity}
\author{L. Galbany\inst{1,2}, 
V. Stanishev\inst{3}, 
A. M. Mour\~ao\inst{3}, 
M. Rodrigues\inst{4,5}, 
H. Flores\inst{5}, 
C. J. Walcher\inst{6},
S. F. S\'anchez\inst{7},
R. Garc\'ia-Benito\inst{8},
D. Mast\inst{9},
C. Badenes\inst{10},
R. M. Gonz\'alez Delgado\inst{8},
C. Kehrig\inst{8},
M. Lyubenova\inst{11},
R. A. Marino\inst{12,13}
M. Moll\'a\inst{14},
S. Meidt\inst{15},
E. P\'erez\inst{8},
G. van de Ven\inst{15},
J. M. V\'ilchez\inst{8}}

\offprints{\myemail}

\institute{Millennium Institute of Astrophysics, Santiago, Chile.
 \and Departamento de Astronom\'ia, Universidad de Chile, Casilla 36-D, Santiago, Chile.
 \and CENTRA - Centro Multidisciplinar de Astrof\'isica and Departamento de F\'isica, Instituto Superior T\'ecnico, ULisboa, Av. Rovisco Pais 1, 1049-001 Lisbon, Portugal.
 \and European Southern Observatory, Alonso de Cordova 3107 Casilla 19001 - Vitacura -Santiago, Chile.
 \and GEPI, Observatoire de Paris, UMR 8111, CNRS, Universit\'e Paris Diderot, 5 place Jules Janssen, 92190 Meudon, France.
 \and Leibniz-Institut f\"ur Astrophysik Potsdam (AIP), An der Sternwarte 16, D-14482 Potsdam, Germany
 \and Instituto de Astronom\'ia, Universidad Nacional Aut\'onoma de M\'exico, A.P. 70-264, 04510, M\'exico, D.F.
 \and Instituto de Astrof\'sica de Andaluc\'ia (CSIC), Glorieta de la Astronom\'ia s/n, Aptdo. 3004, E18080-Granada, Spain.
 \and Instituto de Cosmologia, Relatividade e Astrof\'isica - ICRA, Centro Brasileiro de Pesquisas F\'isicas, Rua Dr. Xavier Sigaud 150, CEP 22290-180, Rio de Janeiro, RJ, Brazil.
 \and Department of Physics and Astronomy, University of Pittsburgh, Allen Hall, 3941 O'Hara St, Pittsburgh PA 15260, USA.
\and Kapteyn Astronomical Institute, University of Groningen, PO Box 800, NL-9700 AV Groningen, the Netherlands.
\and Department of Physics, Institute for Astronomy, ETH Z\"urich, CH-8093 Z\"urich, Switzerland
\and Departamento de Astrof\'isica y CC. de la Atm\'osfera, Facultad de CC. F\'isicas, Universidad Complutense de Madrid, Avda. Complutense s/n, 28040 Madrid, Spain.
\and Departamento de Investigaci\'on B\'asica, CIEMAT, Avda. Complutense 40, E-28040 Madrid, Spain.
\and Max-Planck-Institut f\"ur Astronomie / K\"onigstuhl 17 D-69117 Heidelberg, Germany.
 }
 
\date{Received December 26th  / Accepted -------------}

\abstract{ 
The metallicity of a supernova progenitor, together with its mass, is one of the main parameters that can rule their outcome.
We present the second study of nearby supernova (SN) host galaxies ($0.005<z<0.03$) using Integral Field Spectroscopy (IFS) from the CALIFA survey. 
We analyze the metallicity of 115 galaxies, which hosted 132 SNe within and 10 SNe outside the field-of-view (FoV) of the instrument.
 Further 18 galaxies, which hosted only SNe outside the FoV were also studied. 
Using the O3N2 calibrator from Marino et al. (2013) we found no statistically significant differences between the gas-phase metallicities at the locations of the three main SN types -- Ia, Ib/c and II, all having $12+\log(O/H)\simeq8.50$ within 0.02~dex.
The total galaxy metallicities are also very similar and we argue that this is because our sample consists only of SNe discovered in massive galaxies ($\log(M/M_\sun)>10$ dex) by targeted searches.
We also found no evidence that the metallicity at the SN location differs from the average metallicity at the galactocentric distance of the SNe. 
By extending our SN sample with published metallicities at the SN location, we are able to study the metallicity distributions for all SN subtypes split into SN discovered in targeted and untargeted searches. We confirm a bias toward higher host masses and metallicities in the targeted searches. 
Combining data from targeted and untargeted searches we found a sequence from higher to lower local metallicity: SN Ia, Ic, and II show the highest metallicity, which is significantly higher than SN Ib, IIb, and Ic-BL. Our results support the picture of SN~Ib resulting from binary progenitors and, at least part of, SN Ic being the result of single massive stars stripped of their outer layers by metallicity driven winds.
We studied several proxies of the local metallicity frequently used in the literature and found that the total host metallicity allows for the estimation of the metallicity at the SN location with an accuracy better than 0.08 dex and very small bias.
In addition,  weak AGNs that cannot be seen in the total spectrum may weakly bias (by 0.04~dex) the metallicity estimate from the galaxy integrated spectrum.
}


\authorrunning{L.Galbany}
\titlerunning{IFS of SN host galaxies - Paper II}
\maketitle 

\section{Introduction} \label{sec:intro}

Heavy elements up to the iron-groups are synthesized by fusion of lighter nuclei in the cores of stars. When the nuclear energy source in the core gets exhausted the star enters the final stage of its life. Within 4--40\,Myr after their birth stars heavier than $\sim8M_\sun$ form a heavy iron core, which  gravitationally collapses into a neutron star or a black hole \citep{1979NuPhA.324..487B, 1989ARA&A..27..629A} triggering explosive ejection star's outer envelope, an event called core-collapse supernova (CC~SN). Stars with masses between 1.5 and 8$M_\sun$ form a degenerate carbon-oxygen (C/O) white dwarf (WD) \citep{1980ApJ...237..111B} with mass in the range of $0.5-1.1M_\sun$ \citep{1999ApJ...524..226D}. If a C/O WD is in a binary system, it can  accrete mass from the companion stat (another WD, a main sequence or red giant star). Under certain conditions the WD can increase its mass to $\sim1.4M_\sun$ and thermonuclear reactions can ignite in its center to completely disrupt the star in a very bright thermonuclear explosion \citep{1960ApJ...132..565H} called Type Ia SN (SN~Ia). This general picture is supported by the fact that CC~SNe are only found in active star-forming galaxies, spiral arms and \ion{H}{ii} regions \citep{1992AJ....103.1788V,2012MNRAS.424.1372A,2014A&A...572A..38G}, which confirms that their progenitors should be massive short-lived stars, while SNe~Ia are found in all types of galaxies, including early type galaxies without ongoing star formation.

The exact understanding of the progenitor systems and explosion mechanism of SNe~Ia remains elusive, and no direct progenitor detection has been reported yet (but see \citealt{2014Natur.512...54M}). However, there is growing evidence that the progenitor stars of SN~Ia have wide range of ages following a delay-time distribution (DTD) with a form close to $t^{-1}$ from hundreds of Myr to 11 Gyr \citep{2010ApJ...722.1879M}. It has been suggested that some small heterogeneities in the observed properties of SNe~Ia can be attributed to differences in the mechanism that drives the explosion, such as the nature of the companion star (single or double degenerate scenario) or if the explosion initiates as a detonation or a deflagration.

Type II is the most common class of SN and several progenitor stars have been detected in pre-explosion images, allowing to constrain their initial mass lie between 8.5 to 16.5 \Msun~\citep{2003PASP..115....1V,2015PASA...32...16S}. Although for the peculiar type II SN 1987A a blue supergiant (BSG) was suggested as a progenitor \citep{1987Natur.327..597H}, the most viable candidates are red supergiants (RSG, \citealp{2011ApJ...742....6E, 2011ApJ...739L..37M}); the more massive ones probably resulting in type IIP (plateau or slow-decliners) and those that have lost large amount of their H envelope and with yellow colors in type IIL (linear, or fast-decliners). 
Type IIn SNe show narrow lines in their spectra, which result from interaction between the ejecta and circumstellar matter (CSM). For this reason, it has been proposed that their progenitors could be Luminous Blue Variable (LBV) stars \citep{2008ApJ...686..467S}. Although only a few progenitor detections have been reported \citep{2009Natur.458..865G}, there is evidence that their progenitors could be less massive than the normal SN~II \citep{2012MNRAS.424.1372A, 2014MNRAS.441.2230H}. 

Type Ibc SNe, also called stripped-envelope SNe, are less frequent and less detections are available (See \citealt{2015arXiv151008049L}). At the moment of explosion SNe~Ic have lost their both H and He layers and thus show no H and He lines in their spectra. SNe~Ib have lost only the outer H layer  and show He lines. 
There are two possible channels through which these explosions can occur. In the single-star scenario, the best candidates are massive ($>$25-30\Msun) Wolf-Rayet (WR) stars that have been stripped of their envelopes by strong line-driven winds, which are dependent on metallicity \citep{1986ApJ...306L..77G, 2007ARA&A..45..177C, 2009A&A...502..611G}. The other possibility is lower-mass stars that lose their outer envelopes during the evolution in a binary system \citep{1992ApJ...391..246P, 1996IAUS..165..119N, 2011MNRAS.414.2985D}. Although there should be a combination of both scenarios, the binary scenario presently enjoys significant support \citep{2007PASP..119.1211F, 2011MNRAS.412.1522S}. A direct progenitor detection in pre-explosion images has been reported, but the exact nature of progenitor system could not be well constrained (iPTF13bvn, \citealt{2014AJ....148...68B, 2015MNRAS.446.2689E}).
Finally, SNe~IIb are a chameleon class between II and Ib. Initially they show only hydrogen lines in their spectra, but at later stages helium lines also appear. Thus suggests that SN~IIb progenitors have retained only a thin layer of hydrogen on the surface. Possible detections of SN~IIb progenitor and companion have been reported (1993J, \citealt{1994AJ....107..662A,2004Natur.427..129M}; 2011dh, \citealt{2014ApJ...793L..22F,2014AJ....147...37V}), pointing to red (for 1993J) and yellow (for 2011dh) supergiants as the best candidates. 

Given the difficulty of direct progenitor detection even in high-quality pre-explosion images, the characterization of the SN environment has proved to provide constraints on the possible scenarios that lead to different SN type explosions (See  \citealt{2015PASA...32...19A} for a recent review). Among all the parameters that can be measured in the environment, the metallicity is of particular interest, because it is expected to affect many aspects of the SN explosions, including the luminosity of SNe~Ia and the type of CC~SN produced by the massive stars \citep{2010ApJ...711L..66B,2016ApJ...818L..19M,2011ApJ...731L...4M}.
Most often the observed SN properties have been correlated to global metallicity of their host galaxies either directly measured or inferred from other proxies. \cite{2008ApJ...673..999P,2012ApJ...759..107K,2003A&A...406..259P,2010ApJ...721..777A} all found that the number ratio of SN~Ibc to SN~II increases with the metallicity. Regarding SNe~Ia, \cite{2013ApJ...770..107C} and \cite{2014MNRAS.438.1391P}  found that the Hubble residuals correlated with the host gas-phase metallicity

 \begin{table*}\tiny
\caption{Properties of the 37 SNe added to the sample presented in G14. This corresponds to 34 SNe in 33 galaxies observed from June 2014, and 3 SNe that recently exploded in NGC 5404, NGC 6166 and UGC 04132. 
}
\label{tab:sam}
\begin{center}
\vspace{-0.6cm}
\begin{tabular}{llccrcclccc}
\hline\hline
Galaxy&Morphology&\multicolumn{1}{c}{z}&\multicolumn{1}{c}{E(B-V)}&\multicolumn{1}{c}{PA}&\multicolumn{1}{c}{b/a}& SN   &Type&\multicolumn{1}{c}{RA offset} &\multicolumn{1}{c}{DEC offset} &\multicolumn{1}{c}{Separation}\\
&         &  &    &\multicolumn{1}{c}{[deg]} &              &    &    &\multicolumn{1}{c}{[arcsec]}&\multicolumn{1}{c}{[arcsec]}&\multicolumn{1}{c}{[arcsec]}\\ 
\hline
UGC 04132                                          & Sbc       &   0.017409 & 0.067 & 116.7  & 0.42 &2014ee & IIn &-17.1 & -14.0 &24.9\\
NGC 5406                                            & SAB(rs)bc & 0.017352 & 0.011 &17.9 & 0.94 & PSN J14002117+3854517 & II & +5.0 & +3.0 & 4.9\\
NGC 6166                                            & cD2 pec  & 0.030354 &0.012 &123.8 & 0.95 & PS15aot & Ia-91bg &+1.5 & -8.6 & 8.5\\           
\hline
NGC 0309                                            & SAB(r)c & 0.018886 & 0.035 &26.8 & 0.93& 1999ge & II  & +16.0 & +6.7 & 17.6 \\
NGC 0938                                            & E            & 0.013736  &0.100 & 91.9 & 0.77 & 2015ab	& Ia & -9.3 & +6.3 & 11.7\\
NGC 0991                                            & SAB(rs)c& 0.005110  & 0.024 &36.7 & 0.94& 1984L & Ib  & -32.0 & -22.0 &35.0\\
UGC 02134                                          & 	Sb      & 0.015297  & 0.151 & 8.2 & 0.51& 2011jf  & Ibc & +3.0 & -1.0 & 37.2 \\
NGC 1070                                            & Sb         & 0.013636  & 0.050 & 89.8& 0.75& 2008ie & IIb  & -22.0 & +13.0 &25.4\\
MCG -01-09-006\tablefootmark{$\dagger$}& SB(rs)cd?&0.028977&0.066& 107.5 & 0.32&2005eq & Ia & +15.9 & +26.3 & 30.3 \\
IC 0307\tablefootmark{$\dagger$}       & (R)SB(r)a pec?&0.025981&0.086&149.1&0.61&2005em & IIb & +38.0 & -7.9 & 38.3 \\ 
MCG -01-10-019\tablefootmark{$\dagger$}& SAB(r)cd &0.017505 & 0.050 & 105.7 & 0.72 &2001H  & II  & +5.8 & -2.9 & 4.6\\
NGC 1667                                            & SAB(r)c  &0.015167  & 0.067 & 77.8& 0.65& 1986N & Ia   & -12.0 & -9.0 & 19.6 \\
UGC 04195                                          & SB(r)b  &0.016305  & 0.050 & 111.0& 0.71& 2000ce & Ia   & +15.1 & +17.3 & 22.6 \\
NGC 2554                                            & S0/a      & 0.013870  & 0.050 & 70.0 & 0.77& 2013gq & Ia  & -0.6 & -9.1 & 9.2 \\
NGC 2565\tablefootmark{$\dagger$}  & (R')SBbc?& 0.011948 & 0.042& 81.1& 0.50 & 1992I & II & -28.5 & +7.1 & 28.9 \\
                                                             &                 &                 &          &        &          & 1960M & I & -13.0 & +34.0 & 37.6 \\
NGC 2577\tablefootmark{$\dagger$}   & S0           & 0.00678 & 0.048  &16.8&0.51&  2007ax & Ia &  -2.6 & +5.5 &  6.4 \\
NGC 2595                                            &SAB(rs)c & 0.014443&0.035& 125.9     & 0.95    &1999aa&Ia    & +1.0   & +28.0  &  30.0 \\                 
NGC 2596                                            &	Sb  & 0.019807&0.037& 158.1 & 0.40  &2003bp&Ib    & +17.6   & +11.6  & 22.0  \\                 
NGC 2604                                            &SB(rs)cd  & 0.006930&0.041& 52.4 & 0.96 &2002ce&II    & -16.0   & -12.0  & 23.1  \\      
UGC 04468                                            &S0  & 0.025227&0.030& 74.0 & 0.34 &2006bb&Ia    & +3.6   & -27.5  & 27.7  \\                 
NGC 2916                                            & SA(rs)b? &  0.012442 & 0.023 &105.8&0.63& 1998ar & II & +16.9 &  +38.5 & 42.3 \\
UGC 05520                                          & Scd         & 0.011058  &  0.034 &15.0 & 0.63 & 2000L & II & +6.9 & -19.3 & 22.0\\	
NGC 5425\tablefootmark{$\dagger$}   &Sd           &0.006918&0.018& 38.5 & 0.35 &2011ck&IIP   & -14.8  & +7.3  &  16.1 \\
NGC 5525                                            &S0            &   0.018523 & 0.022 & 109.1 & 0.36 & 2009gf & Ia& -31.0 & -8.5 & 32.0\\
NGC 5557\tablefootmark{$\dagger$}   &E1           &0.010737&0.006& 120.3 & 0.94 &1996aa&Ia   &  -5.0  & +3.0  &  4.2 \\
NGC 5559\tablefootmark{$\dagger$}   &SBd        &0.017232&0.020& 152.1 & 0.33 &2001co&Ibc-pec  &  -13.5  & -12.6  &  18.4 \\
NGC 5587\tablefootmark{$\dagger$}   &S0/a       &0.007682&0.023& 73.1 & 0.32 &2006dy&Ia  &  +10.0  &  +9.0 &  14.1 \\
NGC 5732                                             &Sbc       & 0.012502 & 0.015&  130.4 & 0.56& ASASSN-14jf &II& +8.2 & -17.6 &20.0 \\
NGC 6063                                            & Scd?      &0.00950 & 0.041 & 62.5& 0.56 & 1999ac & Ia-pec   & +23.9 & -29.8 & 38.3 \\
UGC 10123                                          & Sab       & 0.012575 & 0.013 &147.5 &0.36 & 2014cv & IIP & +8.0 & +2.0 & 6.0 \\
NGC 7619                                            & E            & 0.012549& 0.072  &116.5 &0.85& 1970J & Ia   & -27.0 & -30.0 & 37.0 \\
NGC 7691                                            &SAB(rs)bc&0.013479&0.062& 79.1 & 0.98 &2014az&IIP   &  -22.0  &  +18.0 &  28.4 \\ 
\hline
UGC 06517                                           &Sbc        &0.008309&0.029& 124.2 & 0.72  &2006lv&Ibc  &   +10.0 & +12.0  & 15.3  \\ 
IC 0758                                                 & SB(rs)cd? & 0.004253 &0.018& 109.8 & 0.93 & 1999bg & IIP     &  -33.0 & -20.0 &  36.6 \\
UGC 09356\tablefootmark{$\dagger$}  &S?          &0.007419&0.027& 15.2 & 0.46 &2011cj&IIP  &  +4.4  &  +7.5 &  8.2 \\
2MFGC 13321                                      &Sb           &0.026145&0.007& 168.6 & 0.39 &2002aw&Ia   &  -1.6  & +1.5  &  3.4 \\ 
\hline
\end{tabular}
\tablefoot{
The morphological galaxy type, redshift, Milky Way dust reddening, and SN angular separation are from the NED database. SN type and offset (positive in the N and E direction) obtained from the Asiago SN catalogue. The position angle (PA, W to N) and the axis ratio (b/a) are calculated in this work. 
%
\tablefoottext{$\dagger$}{Only observed with the V500 grating.}
%
%
}
\end{center}
\end{table*}

There are also several examples in the literature that went further and tried to characterize the local properties of the SN environment. This is particularly relevant for the CC~SNe, which because of their short-lived progenitors explode closer to their birth place and the metallicity measured at the SN location should be close to that of the progenitor. On contrary, the progenitor stars of SNe~Ia most likely lived long enough to migrate around the galaxy, and the host galaxy properties at the SN explosion may not reflect the properties of the progenitor. \cite{2008AJ....135.1136M} found that SNe Ic with broad lines, which are associated with Gamma Ray Bursts, come from significantly lower-metallicity host environments than their counterparts SNe Ic. Using the radial position as a proxy for local metallicity \cite{2009MNRAS.399..559A} suggested a similar sequence in the progenitor metallicity of CC SNe. \cite{2010MNRAS.407.2660A} and \cite{2011A&A...530A..95L} found no statistically significant difference between the local metallicity of SNe Ib and Ic, obtaining slit or fiber spectra at SNe positions. However, \cite{2011ApJ...731L...4M} used local spectra and central spectra plus a metallicity gradient to claim that larger differences exist between SNe Ic and Ib. \cite{2013A&A...558A.143T} studied the locations of SN1987A-like events whose progenitors are blue supergiants (BSG), and found lower metallicities compared to other CC SNe subtypes. Also for SNe Ia, several studies tried to measure the local properties of the SN environment either using the offset as a proxy for these local properties \citep{2012ApJ...755..125G}, or by indirect approximations of the central metallicity and applying decreasing gradients \citep{2009A&A...503..137B}.

In this series of papers (\citealt{2012A&A...545A..58S} and \citealt{2014A&A...572A..38G}, hereafter Paper {\sc I}) we take a different approach. We used wide-field integral field spectroscopy (IFS) at intermediate spectral resolution provided by the CALIFA survey \citep{2012A&A...538A...8S} combined with other previous observations to measure the properties of the gas and the stellar populations at the location of the SN explosion (in addition to other global host properties). Our goal is to search for differences in environmental parameters among SN types, which would also help to constrain the nature of their progenitors, but we also tested the accuracy of the various proxies to the local metallicity used in the literature. \cite{2013AJ....146...30K, 2013AJ....146...31K, 2015PKAS...30..139K} also used IFS observations, but with a smaller field of view that covered only small part of the galaxies.

In this second paper of the series we focus on the local SN metallicity. The paper is structured a follows. The galaxy sample of SN hosts used in this work is presented in Section \ref{sec:sample}. The methods used to extract the needed information for this study from the observed IFS data-cubes is outlined in Section \ref{sec:analysis}. We present our results in Section \ref{sec:results}. Section \ref{sec:disc} contains the discussion of these results, and finally in Section \ref{sec:conc} we summarize the conclusions.


\section{Galaxy and supernova samples} \label{sec:sample}

The galaxy and SN selection is already described in Paper~I and more details, including on the data reduction, can be found there. 
In this work,  the sample presented in Paper~I was expanded with 33 galaxies (the details are given in Table~\ref{tab:sam}), which hosted 34 discovered SNe. In addition 3 new SNe exploded in two galaxies already presented in Paper~I.
Thus the total sample used in this work consists of 115 galaxies, with mean redshift of 0.015, that hosted 132 SNe (47 type II, 27 type Ib/c+IIb, 58 Ia) which were in the field-of-view (FoV) of PPAK.
81 galaxies were observed by the CALIFA Survey and 34 from other CALIFA-related studies, using the same instrumental configuration.
12 of the 79 galaxies from CALIFA were delivered to the community in the CALIFA Data Release 1 (DR1, \citealt{2013A&A...549A..87H}) and 12 more in the CALIFA DR2 \citep{2015A&A...576A.135G}.
Table \ref{tab:redshift} gives the mean and the standard deviation of the redshift distributions of the whole sample and split by SN type.

Seven galaxies from the sample (NGC~214, NGC~309, NGC~628, NGC~1058, NGC~3184, NGC~3913 and NGC~5557) also hosted 10 SNe, which were outside of the FoV of PPAK.
Further 18 galaxies from the CALIFA sample have SNe associated with them (one SN~In each galaxy), but these SNe are outside of the FoV\footnote{We detected up to twelve supernovae in our galaxy sample that lack from spectroscopic classification. We have not considered them anywhere in this analysis.}.
These SNe were only used to check the results of the analysis of the total galaxy properties of the main sample. We always found that adding these extra SNe to the main sample does not change the results. The parameters of these 25 galaxies with SNe outside the FoV are given in Table \ref{tab:white}.
In Table \ref{tab:fov} the number of SNe in the galaxies in our sample by SN sub-type is summarized. 

Almost all SNe in our sample were discovered by targeted searches. It is known that such searches are biased toward more massive galaxies compared to the untargeted ones. As we will see in Sec.~\ref{sec:totmass}, most galaxies in our sample indeed have masses larger than $\log(M/M_\sun)\simeq10$. Thus, the results presented in this work are representative for SNe discovered in targeted searches.

The 3D datacubes used in this work, including those already presented in Paper~I, were processed with version 1.5 of the CALIFA reduction pipeline \citep{2015A&A...576A.135G}. Nine of the galaxies from Paper~I that at the time were observed only with the red V500 grating (spectral resolution $\sim$6\AA and coverage within 3750- 7300\AA), have now also been observed with the blue V1200 grating (spectral resolution $\sim$2.7\AA and coverage within 3400- 4750\AA). For these objects the combined V1200 + V500 datacubes were used.

\begin{table}
\caption{Statistics of the redshift distributions.}
\label{tab:redshift}                                                                                         
\begin{tabular}{lcccc}
\hline\hline
 & Ia & Ibc/IIb & II & All\\
\hline
$\langle z\rangle$ & 0.0179 & 0.0125 & 0.0128 & 0.0150\\
$\sigma_z$         & 0.0090 & 0.0063 & 0.0074 & 0.0083\\
\hline
 & & & &\\
\end{tabular}
\begin{tabular}{lcccc}
\hline\hline
           & II -- Ibc/IIb & II -- Ia & Ibc/IIb -- Ia & CC -- Ia\\
\hline
KS test   & 0.991 & 0.051 & 0.063 & 0.019\\
($z<0.02$) & 0.951 & 0.120 & 0.189 &0.087\\
\hline
\end{tabular}
\end{table}

\begin{table}
\caption{Statistics of the SNe used in this work. SNe~Ia in star-forming (SF) and passive (P) hosts are given separately.}
\label{tab:fov}                                                                                         
\begin{tabular}{lccccc}
\hline\hline   
                 & II & Ibc/IIb & \multicolumn{2}{c}{Ia} & All\\
                 \cline{4-5}
                 &    &            &     SF &  P  &   \\
\hline                                   
SNe in CALIFA hosts  & 54 & 32 & 59 &  15 &  160 \\         
Inside PPAK FoV      & 47 & 27 & 46 &  12 &  132 \\          
\hline
\end{tabular}
\end{table}


\section{Data analysis} \label{sec:analysis}

The data analysis has been fully described in \cite{2012A&A...545A..58S} and Paper~I. Here we just summarize the main steps and add those not included in the previous papers.
We note that the measurements presented here and in Paper I are independent of the data analysis method used, and a comparison of several methods within the CALIFA collaboration will be matter of a future work (Rosales-Ortega et al. in prep.)

Each galaxy data-cube consists of approximately 4000 spectra spread in square spaxels of 1"$\times$1" in a hexagonal field of view (FoV) of 1.3~arcmin$^2$.
We have applied our analysis procedures to all the individual spectra. 
For most galaxies the spectra in the outer parts have signal-to-noise (S/N) ratio which is insufficient to extract useful information. To analyze the 2D maps of these galaxies spatial binning was applied using adaptive Voronoi tessellations \citep{2003MNRAS.342..345C, 2006MNRAS.368..497D}. The new combined spaxels were required to have S/N around 20 in the continuum band at 4610$\pm$30~\AA. 

The main goal of this paper is to study the metallicity at the SN locations, both stellar and gas-phase. However, in many cases the SN location is on a low S/N spaxel. Because the Vironoi binning is an automatic procedure the new bins are in general not centered at the SN locations. To measure the metallicity at the SN location we followed the same approach as in Paper~I. Series of spectra were extracted in apertures centered at the SN positions and with radii up to 6\arcsec. For some SNe even this was not enough to measure the stellar metallicity and a few cases also the gas metallicity\footnote{Even though metallicity is derived from the spectra at the SN spaxel with 1x1" size, the resolution of the cubes is $\sim$2.57". The effect of this will be further discussed in Sec \ref{sec:locmet}.}. The aperture of the spectrum of each SN used for the analysis in this work is given in Tables \ref{tab:res1} to \ref{tab:res3}.

We also analyzed the total galaxy spectra formed by simply summing the individual spaxel spectra with S/N$\geq1$. This removed the outer low S/N part of the FoV, which contains little light from the galaxy and mostly add noise to the summed spectrum. In several cases, foreground stars in the FoV were also masked out. This analysis was performed in order to compare the properties of the host as derived from integrated spectroscopy, to those derived from spatially resolved spectroscopy.

In order to measure the emission lines flux accurately, the stellar continuum was subtracted using {\tt STARLIGHT} \citep{2005MNRAS.358..363C}.
We performed series of tests comparing the base of 66 single stellar population (SSP) models with different ages and metallicities from \cite{2012A&A...545A..58S} and Paper~I, extracted from the Charlot \& Bruzual 2007 models \citep{2007ASPC..374..303B}, with another basis consisting of 260 SSPs from the basis described in \cite{2014ApJ...791L..16G}. The latter covers ages from 1~Myr to 18~Gyr, four metallicities -- 0.2, 0.4, 1.0 and 1.66~$Z_\sun$, where $Z_\sun=0.019$, and uses \cite{1955ApJ...121..161S} initial mass function (IMF). We found that which basis is used has little effect on the measured emission line fluxes. Thus, to speed up the analysis the 66 SSPs basis was used for the continuum subtraction and measurement of the gas-phase properties.
The most prominent emission lines were fitted using a weighted non-linear least-square fit with a single Gaussian plus a linear term, and corrected for dust attenuation using the ratio of \Had\ and \Hbd\ emission line fluxes. Finally, spatially resolved 2D maps of the flux and error for each line were produced.

For the analysis of the properties of the stellar populations, our test indicated that the properties recovered with the larger basis were more stable and in particular in a given galaxy the stellar metallicity showed smaller scatter. Thus, the estimations of the stellar mass from the total spectra, and the stellar age and metallicity for each particular measurement, were obtained from the {\tt STARLIGHT} fits with the 260 SSPs basis. 
In this work we focus only on the mass-weighted metallicity estimate. For calculating the galaxy masses, the distances inferred from Hubble law were used. Only for the very nearby galaxies distances from Cepheid, Tully-Fisher or other direct methods were used when available.

\subsection{Oxygen abundance} \label{sec:oh}

Nebular emission lines are a good tracer of the young and massive stellar populations that ionize the interstellar medium (ISM), and are the main tool at our disposal for the direct measurement of the metal abundance.
Since oxygen is the most abundant metal in the gas phase and exhibits very strong nebular lines in optical wavelengths, it is usually chosen as a metallicity indicator in ISM studies.
The most accurate method to measure ISM abundances (the so-called direct method) involves determining the ionized gas electron temperature, T$_e$, which is usually estimated from the flux ratios of auroral to nebular emission lines (e.g. \OIIIud/\OIIIt, \citealt{2006A&A...448..955I, 2006A&A...454L.127S}). However, the temperature-sensitive lines such as \OIIIt~are very weak and difficult to measure, especially in metal-rich environments. 
Using V1200 CALIFA data for 150 galaxies, \cite{2013A&A...559A.114M} (M13 hereafter) identified only 16 \ion{H}{ii} regions from which this line can be measured and used them to reliably derive T$_e$. 
Due to the difficulty to find the needed emission lines to estimate the metallicity using the direct method, other alternatives have been developed over the years.
The \mbox{so-called} empirical methods consist of a combination of easily measurable gas emission parameters that are calibrated against the metallicities previously determined by the direct method from \ion{H}{ii} regions and galaxies.
Instead, the theoretical methods are calibrated by comparing the measured line fluxes with those predicted by theoretical photoionization models.
While these two methods have been probed to be sensitive to relative variations of the metallicity, they also showed systematic differences on the absolute metallicity scale (see \citealt{2008ApJ...681.1183K} for a review).
Indeed, the theoretical methods have been showed to overestimate the metallicity by a few tenths of dex, and the empirical methods may underestimate it \citep{2010ApJS..190..233M}. 
Since our aim is to determine the gas oxygen abundances at SN sites within the galaxies, and the direct method cannot be used everywhere reliably with our data, we used several alternative methods described above.

As a primary method, we use the empirical calibration based on the O3N2 index, firstly introduced by \cite{1979A&A....78..200A}, which gives an estimation of the oxygen abundance from the difference of the O3 and N2 line ratios. 
This calibration has the advantage (over other methods) of being insensitive to extinction due to the small separation in wavelength of the emission lines used for the ratio diagnostics and not suffer from differential atmospheric refraction (DAR). 
Anyway, the extinction is not a problem in our case since our emission lines measurements are already extinction corrected.
Recently, M13 presented new calibrations for this O3N2 and for the N2 indices including new direct abundance measurements at the high metallicity regime, which is an advantage compared to the mostly used \citealt{2004MNRAS.348L..59P} (hereafter PP04) O3N2 calibration that used instead photoionization models for that regime.
Although the relative differences between the two calibrations are kept, these new calibrations give more physical results, not allowing for example very high values of the oxygen abundance (e.g. $>$9.0~dex).
For this reason, the main analysis will be done using the empirical calibration introduced in M13. 

\begin{figure*}
\centering
\includegraphics[trim=0.1cm 0cm 0.5cm 0.1cm, clip=true,width=0.30\hsize]{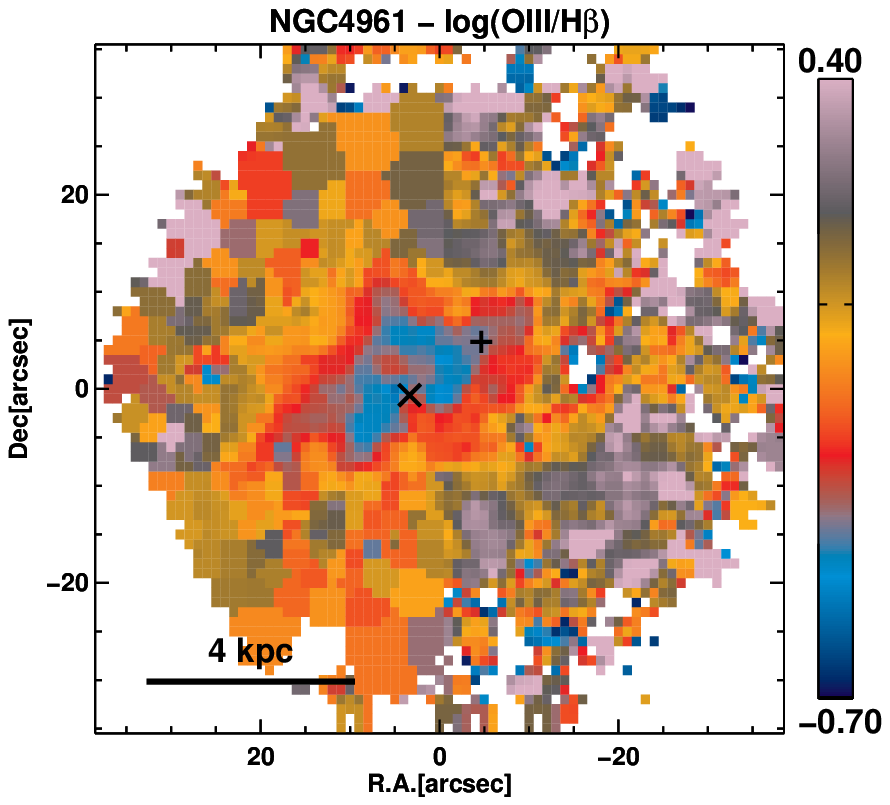}
\includegraphics[trim=0.1cm 0cm 0.5cm 0.1cm, clip=true,width=0.30\hsize]{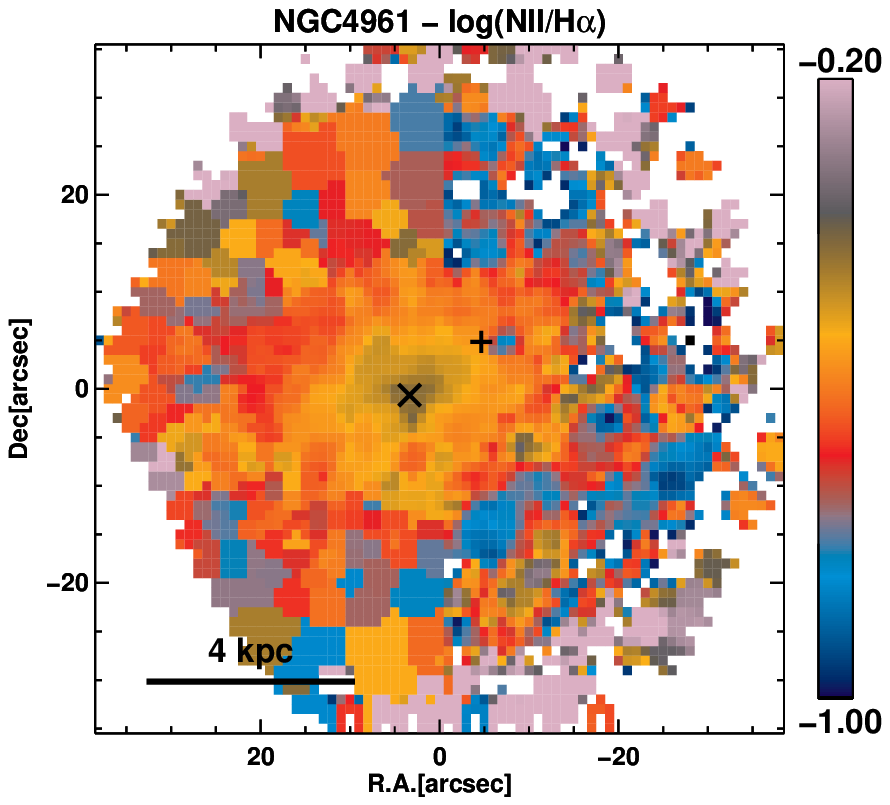}
\includegraphics[trim=0.0cm 0.0cm 0.7cm 0.1cm, clip=true,width=0.30\hsize]{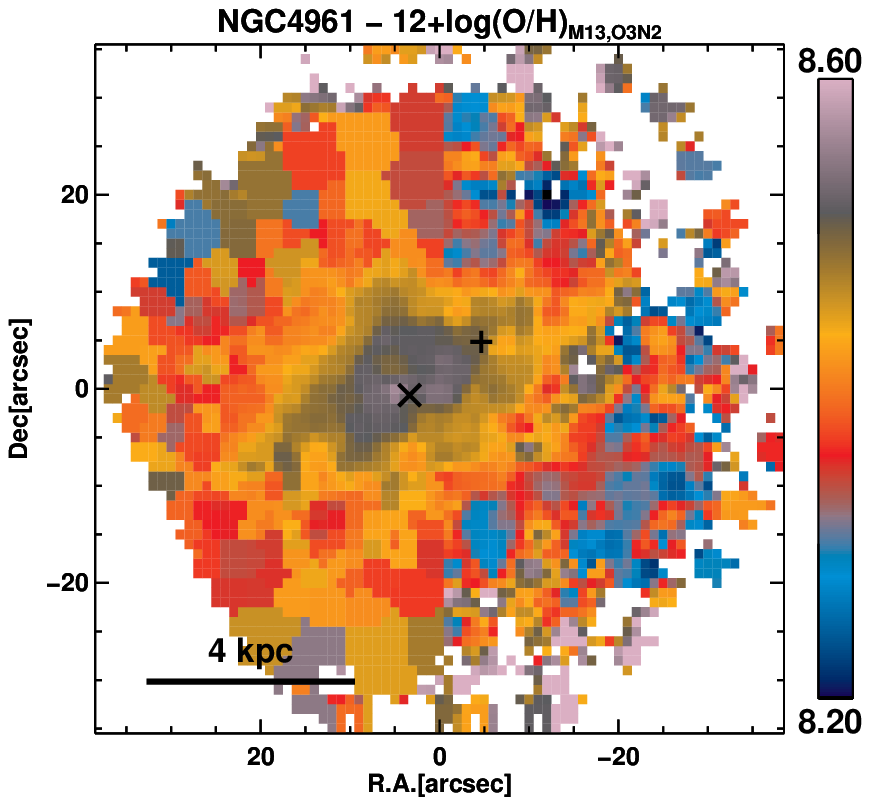}
\includegraphics[trim=0.1cm 0cm 0.5cm 0.1cm, clip=true,width=0.6\hsize]{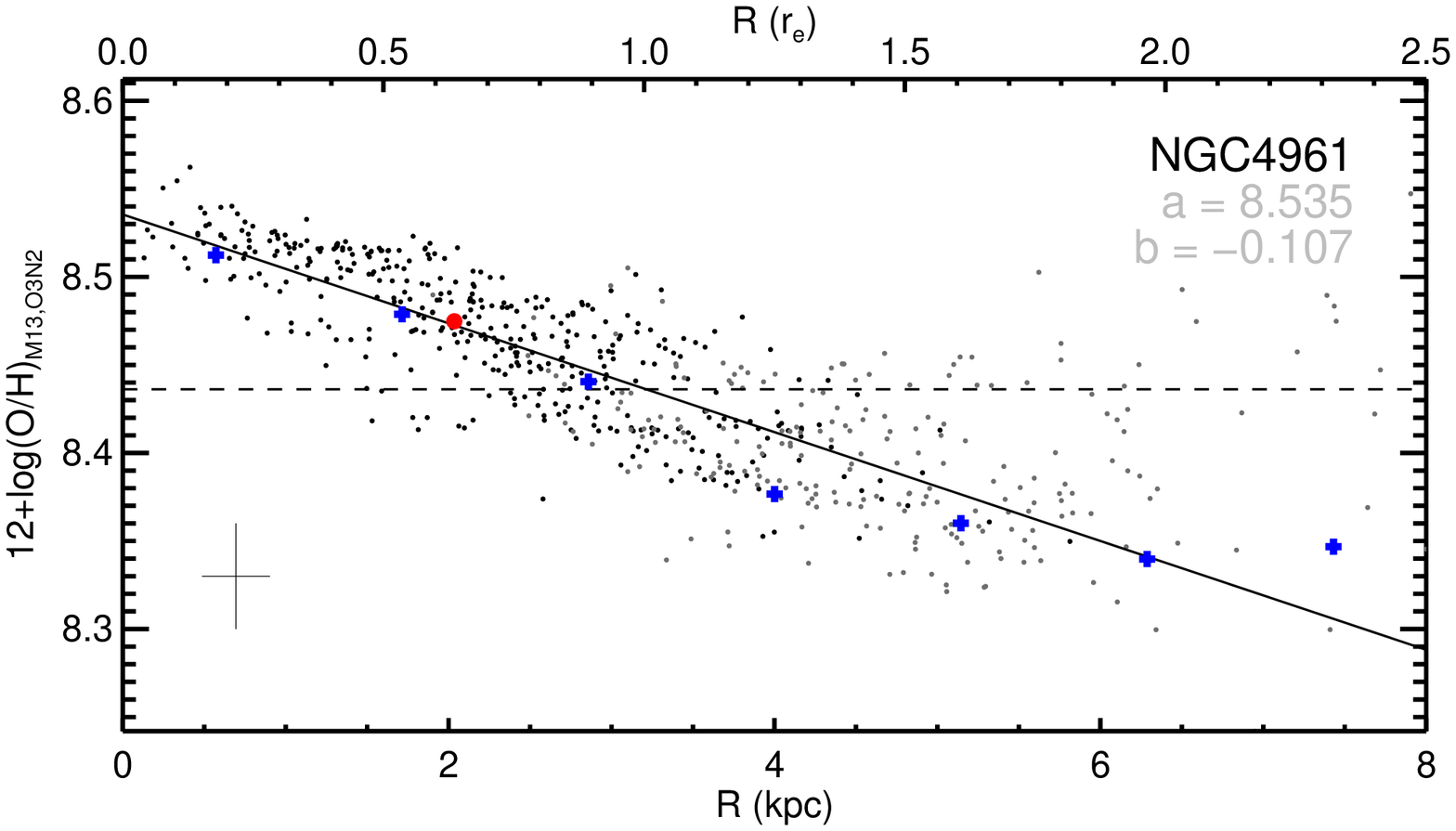}\\
\includegraphics[trim=0.1cm 0cm 0.5cm 0.1cm, clip=true,width=0.3\hsize]{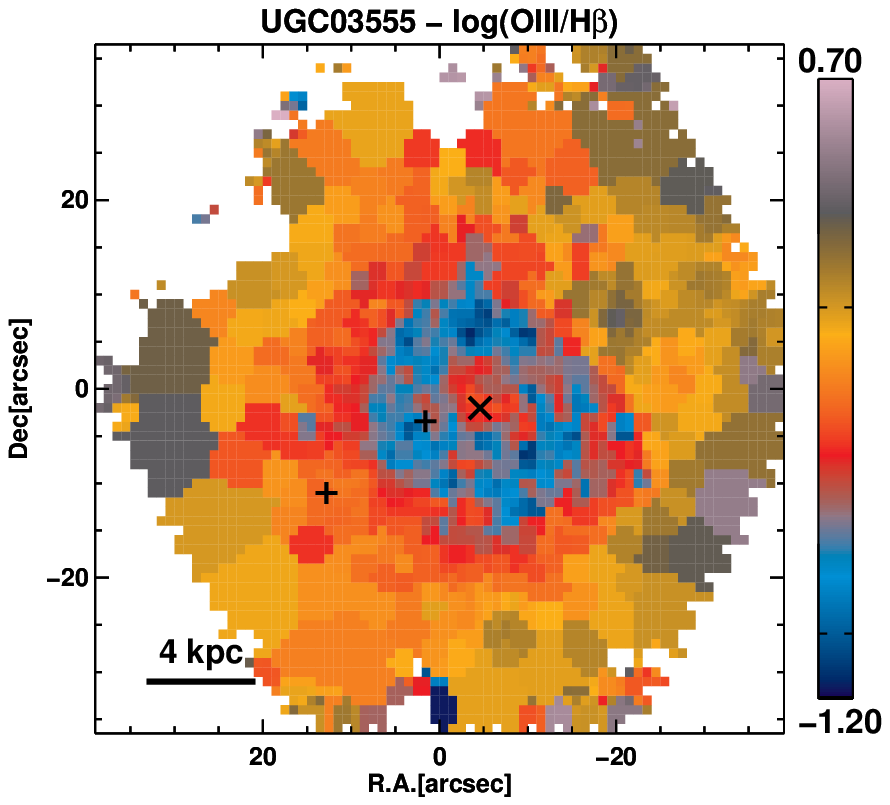}
\includegraphics[trim=0.1cm 0cm 0.5cm 0.1cm, clip=true,width=0.30\hsize]{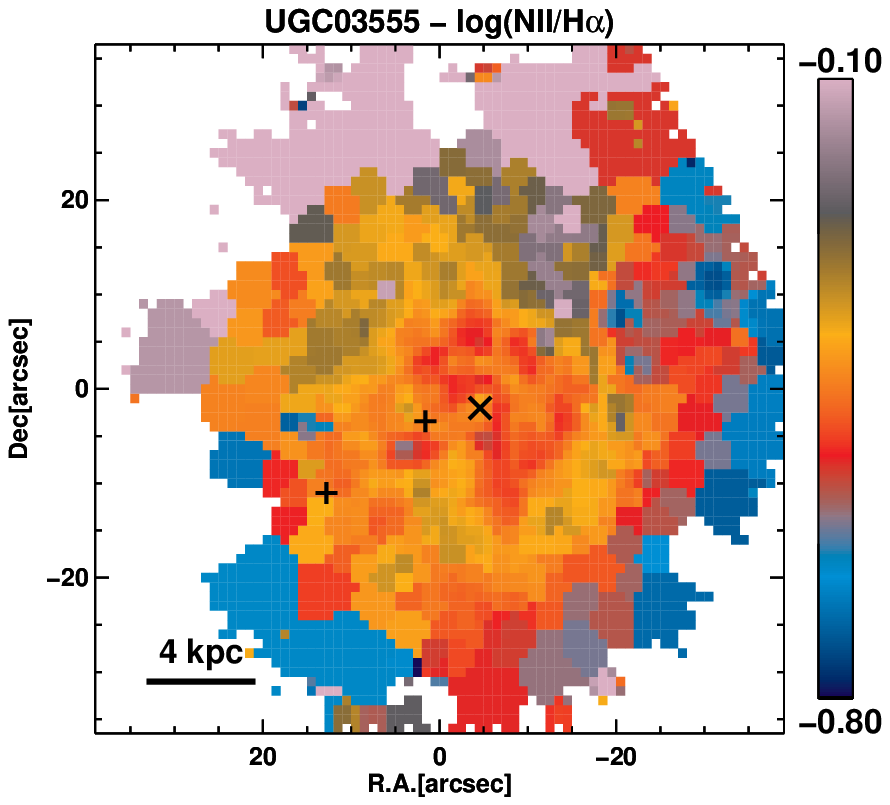}
\includegraphics[trim=0.0cm 0.0cm 0.7cm 0.1cm, clip=true,width=0.30\hsize]{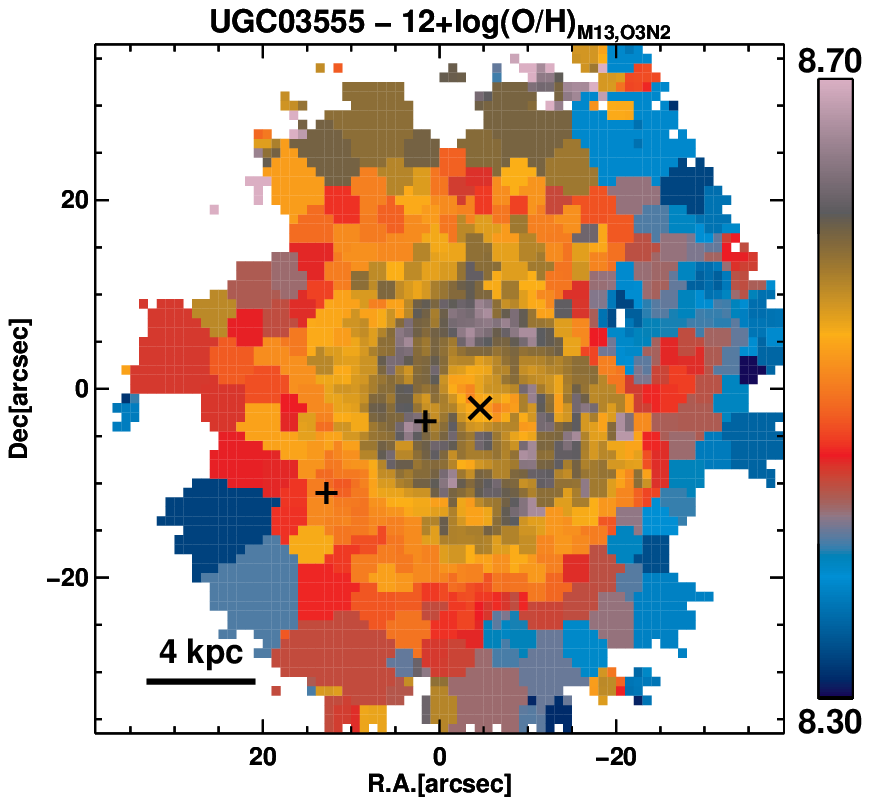}
\includegraphics[trim=0.1cm 0cm 0.5cm 0.1cm, clip=true,width=0.6\hsize]{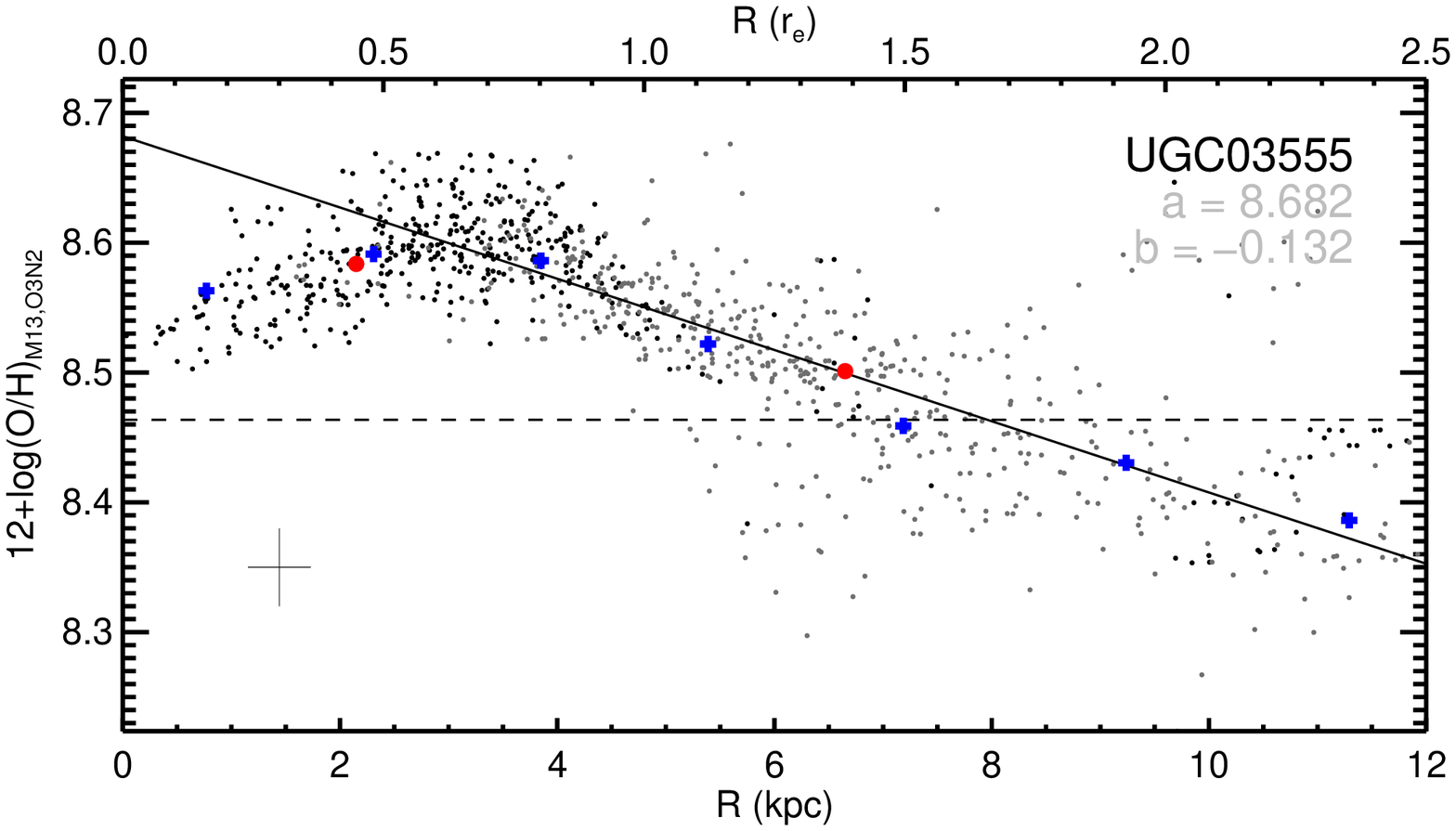}
\caption{NGC 4961 and UGC 03555 OIII/H$\beta$, NII/H$\alpha$, and oxygen abundance maps (top) and gradient fits (bottom). Black dots are measurements in individual spaxels, grey dots are measurements in Voronoi spaxels, blue dots are measurements from azimuthal averaged spectra, and red dots are measurements from the spaxels where SNe exploded (2001ee-II in the top figure, and 2004ge-Ic/1999ed-II in the bottom figure). Mean representative X- and Y- errors are plotted in the lower-left corner for reference.}
\label{fig:metal5}
\end{figure*} 

We used three alternative methods to check the O3N2 results: 
(a) the M13 calibration based only on the N2 index; 
(b) the method described in \citealt{2010ApJ...720.1738P} (hereafter P10) which makes use of the \OIIuu~doublet; and (c) the method described in \citealt{2011MNRAS.412.1145P} (hereafter P11) that uses the [\ion{S}{ii}] doublet.

We note that those emission lines measurements that fall in the AGN region in the BPT diagram \citep{1981PASP...93....5B}  according to the \cite{2001ApJ...556..121K} criterion, were excluded from the analysis everywhere in the paper. This happened basically in the central spaxels of a significant fraction ($\sim$50\%) of our galaxies.

\subsection{Metallicity gradients} \label{sec:grad}

The radial variation of the metal abundance in galaxies has been studied in several works using a wide range of approaches and instruments \citep{1989epg..conf..377D,1994ApJ...420...87Z, 1997ApJ...489...63G, 1999PASP..111..919H, 2000A&A...363..537R, 2009A&A...503..137B, 2012A&A...546A...2S,2016A&A...585A..47M}. 
The existence of radial metallicity decrease towards the galaxy outskirts is now commonly accepted and it can be explained by the combined effects of a radially varying star formation rate and gas in-fall.
It has also been suggested in the past that the magnitude of the radial decrease has a characteristic value \citep{1989epg..conf..377D, 1992MNRAS.259..121V, 2009ApJ...700..309B, 2010ApJ...716L...4Y, 2011MNRAS.415.2439R}. IFS is the technique best suited to study in detail the metal abundance distribution across the galaxy surface and gain insight on its radial dependence. Using {\sc HII} regions of about 300 galaxies observed by CALIFA  \cite{2014A&A...563A..49S} have demonstrated the universality of the gradient, which has been confirmed by \cite{LSM} performing a spaxel by spaxel analysis.

We determined metallicity gradients in our galaxy sample using our methodology to estimate the disk effective radius (see Appendix \ref{sec:re}), and the galaxy position angle and inclination.
For each galaxy the metallicity obtained at each spaxel was plotted as a function of the deprojected galactocentric distance and fitted with a first-order polynomial to estimate the galaxy metallicity gradient.
As stated in Section \ref{sec:oh}, the central spaxels showing emission coming from AGN were removed from the fit because the gas emission lines observed in such regions are not due to ionization coming from young massive stars. 
All the 11 passive/elliptical galaxies in our sample show weak emission in their central regions, which in all cases was classified as coming from an AGN. Five of these galaxies also showed traces of weak emission lines in the central parts outside the AGN-affected region, but those were not strong enough for measuring gradients.

Generally, in most galaxies the metallicity decreases with the radial distance. 
A clear example of the general case is given in Figure \ref{fig:metal5} (top panel) for NGC~4961.
However, in some galaxies a metallicity decrease or flattening close to the center was found (see UGC~03555 in Fig.~\ref{fig:metal5} bottom panel).
This central metallicity decrease was first noted by \cite{2011MNRAS.415.2439R} and \cite{2012A&A...546A...2S}, and was studied in more detail with the whole CALIFA sample by \cite{2014A&A...563A..49S}, who found that about 25\% of the galaxies showed it.
This feature  appears to be associated with a circumnuclear star-formation ring, where gas tends to accumulate due to non-circular motions \citep{2007A&A...466..905F}, and does not seem to be associated with the morphological type of galaxies or with the presence of a bar.
In our sample, 18 out of 58 (31\%) of the SN~Ia, 13 out of 47 (28\%) of the SN~II, and 7 out of 27 (26\%) of the SN~Ibc host galaxies showed the central decrease.
If we include those galaxies that show a flattening rather than a decrease toward the center, all percentages would increase to around 40\%.
It is also interesting to note that $\sim50$\% of the galaxies hosting an AGN (27 out of 58) show the central metallicity decrease and  excluding the 11 elliptical galaxies the percent increases to $\sim$60\% (27 out of 47). Besides, 30\% (8 out of 27) of the galaxies that fall in the composite region of the BPT diagram also show the decrease, but only one located in the star-forming region (UGC~03555) shows this effect. Exploring the possible connection between the central metallicity decrease and the presence of AGN is certainly interesting, but it is beyond the scope of this paper.


\section{Results} \label{sec:results}

In this section we explore the 2D maps of SN host galaxies in search for correlations between the SN type and the properties of their host galaxies regarding the gas elemental abundance and the stellar metallicity.
There are 11 galaxies (host of 12 SNe Ia) in our sample which after subtracting the stellar continuum contribution, do not show emission lines at any position or only show weak emission in their central regions: 
NGC 0495, NGC 1060, NGC 2577, UGC 04468, NGC 4874, NGC 5557, NGC 5611, UGC 10097, NGC 6166, NGC 6173, and NGC 7619.
As in Paper~I, we use these 12 SNe Ia only when the parameter needed for the analysis is measurable.
In the following sections most of the results are presented in the form of cumulative distributions (CDFs). We have performed tests to check whether the differences between the mean values and the standard deviations of the distributions are statistically significant. Two-sample Kolmgorov-Smirnov (KS) tests were also performed to check whether the data was drawn from the same underlaying population.
The measured quantities for individual galaxies/SNe are reported in Tables \ref{tab:res1}, \ref{tab:res2}, and \ref{tab:res3}. In Tables \ref{tab:metal}-\ref{tab:all}  we give the mean and median statistics of the distributions and their asymmetric errors for the three SN subtypes. 

\subsection{Global and local metallicities} \label{local}

\begin{figure*}
\includegraphics*[trim=0.22cm 0.12cm 0.1cm 0.04cm, clip=true,height=5cm]{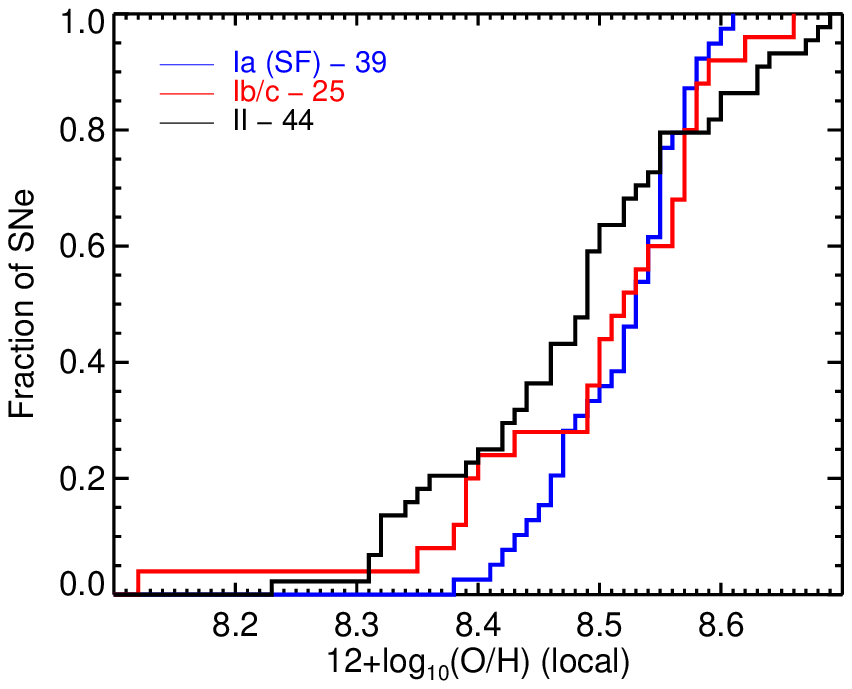}
\includegraphics*[trim=0.6cm 0.12cm 0.08cm 0.04cm, clip=true,height=5cm]{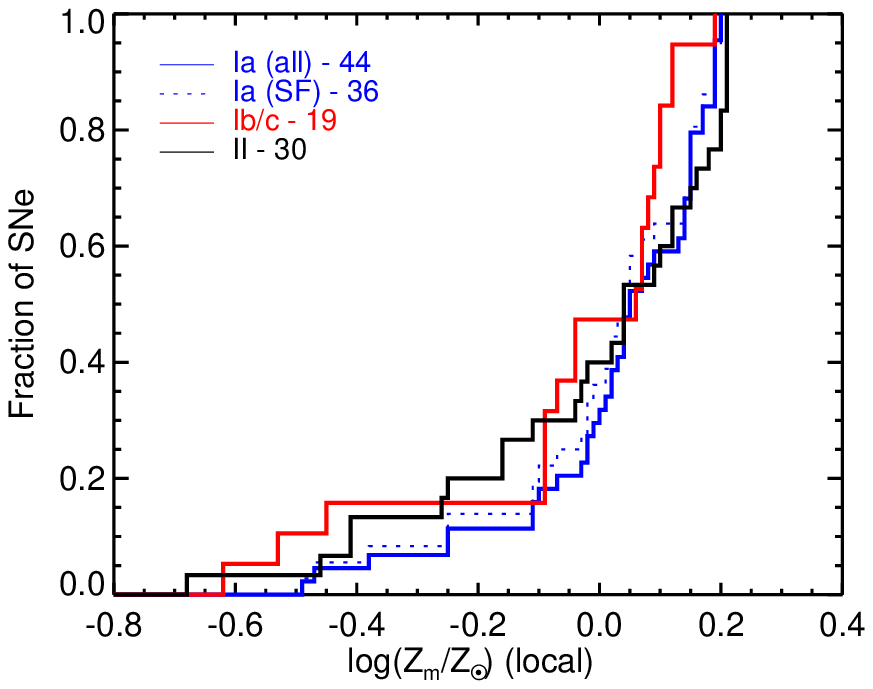}
\end{figure*}
\begin{figure*}
\sidecaption
\includegraphics*[trim=0.22cm 0.12cm 0.1cm 0.04cm, clip=true,height=5cm]{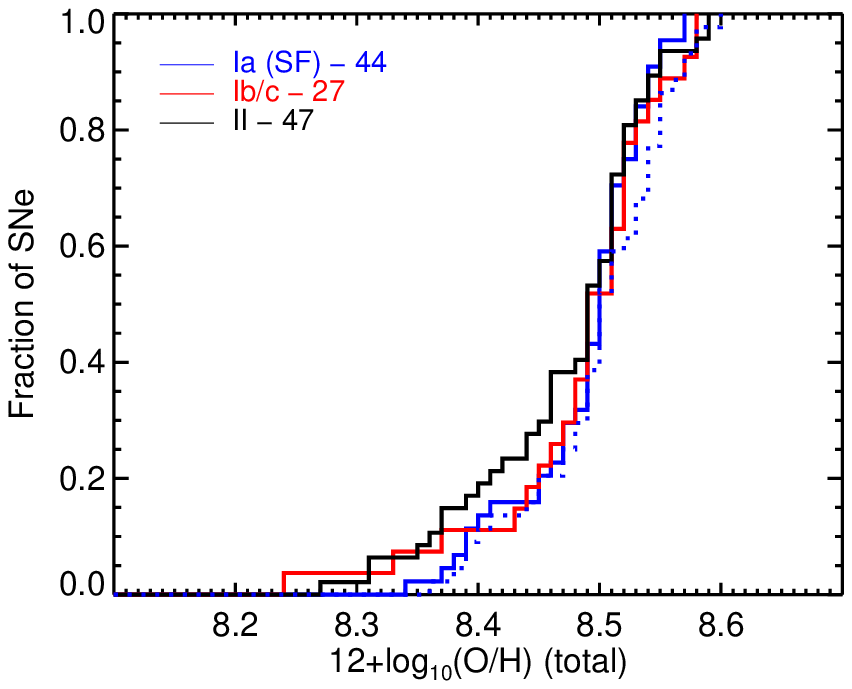}
\includegraphics*[trim=0.6cm 0.12cm 0.08cm 0.04cm, clip=true,height=5cm]{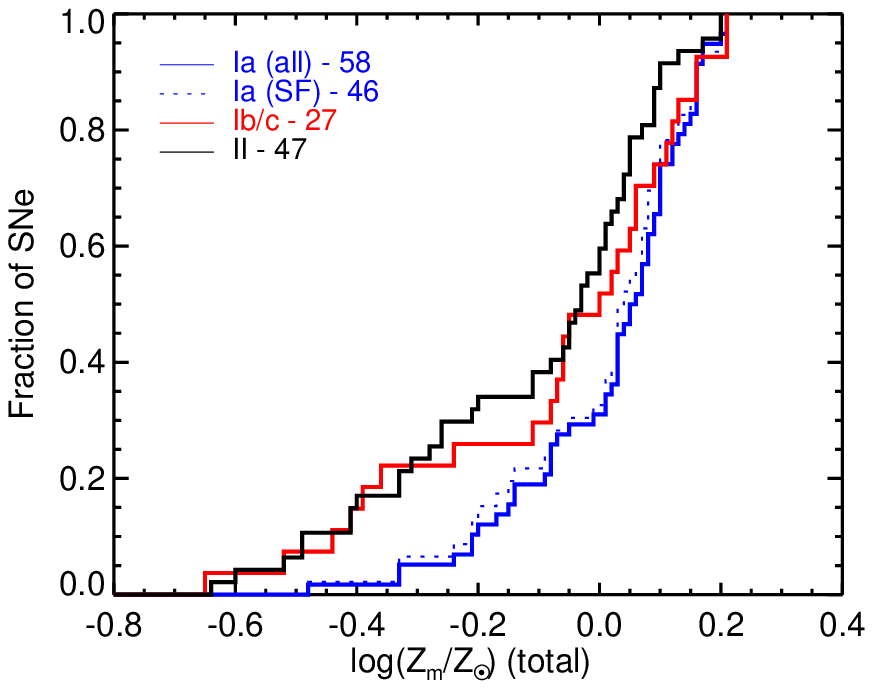}
\caption{SN environment measurements of the oxygen abundance (left) and the mass-weighted stellar metallicity (right) for galaxies hosting different SN types. In the top row we show the distributions of the local metallicity while in the bottom row total metallicity distributions are plotted. The ordinate indicates the fraction of the SN population with metallicities less than the abscissa value.}
\label{fig:metal-loc}
\end{figure*}

The cumulative distributions of the local gas and stellar metallicities for each SN type are shown in the upper row of Fig.~\ref{fig:metal-loc}, and the properties of the distributions are given in Table~\ref{tab:metal}.
There are differences between the distributions of the local gas-phase metallicity among the SN types, but the KS test indicates that only the distribution of SN~Ia and SN~II could be different at 95\% significance. The differences between the mean total metallicities are also small ($\sim$0.04~dex) and the statistical tests again indicate that the differences could be significant only between  SN~Ia and SN~II at 95\% significance. It is also interesting that the standard deviation of the distribution for SNe~Ia is twice smaller than those of the core-collapse SNe, and this difference appears to be statistically significant.
The differences between the mean stellar metallicitiy at the SN location are larger than for the gas-phase metallicity, but the standard deviations of the distributions are also significantly larger. The KS test indicates no statistically significant differences between the distributions. 
Even though the differences between the mean metallicities at the SN locations are small, it is worthy noting that SN~Ia have the highest metallicity of the three SN types. 

The lower row of Fig.~\ref{fig:metal-loc} shows the cumulative distributions of the total galaxy metallicity. The gas-phase metallicity of the three SN types hosts is on average very similar for the three SN types ($\sim$8.48~dex) and the distributions are statistically undistinguishable.
However, similarly  to the metallicity at the SN locations the standard deviation of the distribution of SNe~Ia hosts is smaller than the core-collapse SNe. 
The total stellar metallicity shows noticeably larger differences between the three SN types, with the difference between SN~Ia and SN~II being statistically significant. Again, even tough the differences are small, SN~Ia have the highest metallicity of the three SN types. 

\begin{figure*}
\sidecaption
\includegraphics*[width=6cm]{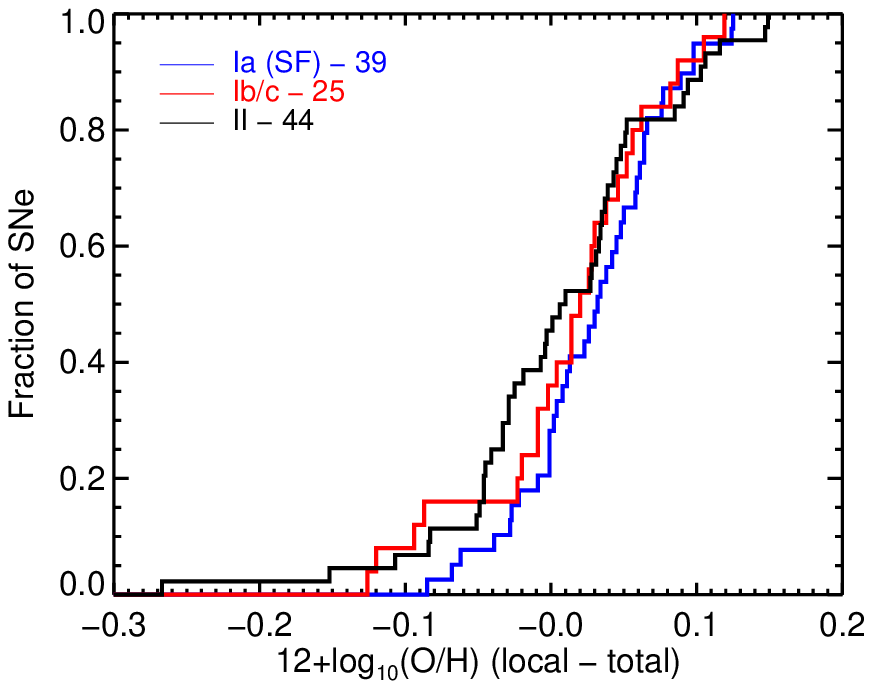}
\includegraphics*[width=6cm]{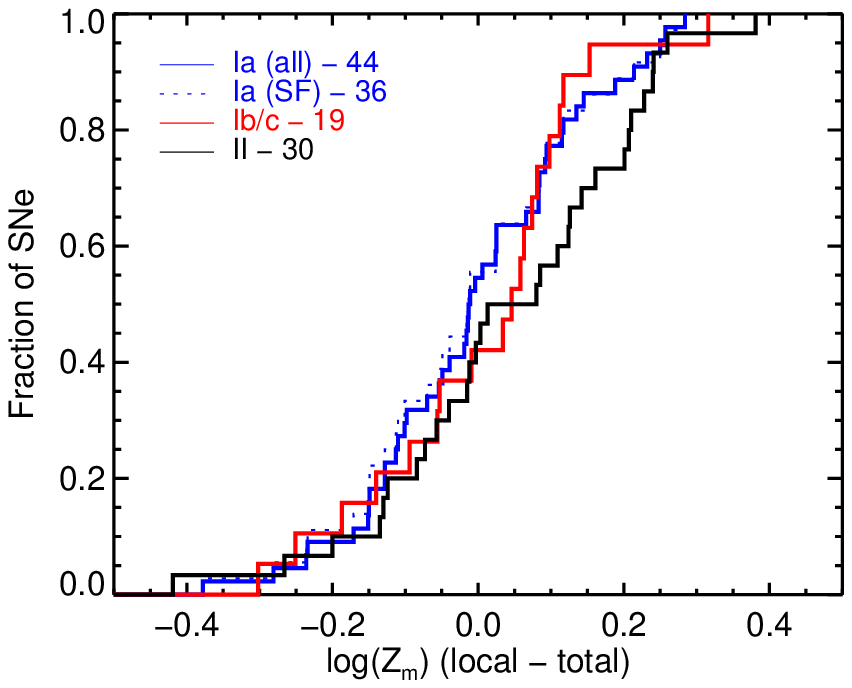} 
\caption{Differences between the local and the total oxygen abundance (left) and mass-weighted stellar metallicity (right) for different SN types.}
\label{fig:metaldiff-loc}
\end{figure*}

Figure~\ref{fig:metaldiff-loc} shows the cumulative distributions of the difference between the local and the global metallicity. The properties of the distributions are given in Table~\ref{tab:proxies}.  All distributions are centered close to the zero, but there are small differences between the SN types. The local gas-phase metallicity of SN~Ia is on average 0.03~dex higher than the total galaxy metallicity and this is the only difference that is statistically significant. 
The differences between the local and the global  stellar metallicities  is not significant for any SN type.

\begin{table}
\centering
\caption{Properties of the distributions of the gas-phase and stellar metallicities at the SN location and the total galaxy. Shown are the mean, the standard deviation $\sigma$, the standard deviation of the mean $\sigma_m$, the median and the number of SNe.}
\label{tab:metal} 
\begin{tabular}{lccccc}
\hline\hline
SN type & mean & $\sigma$ & $\sigma_m$ & median & $N$ \\
\hline
\multicolumn{6}{c}{12+log(O/H)$_\mathrm{local}$} \\
\hline
Ia   &   8.522 &  0.056 &  0.009 &  8.537 & 39 \\
Ib/c &   8.500 &  0.114 &  0.023 &  8.520 & 25 \\
II   &   8.484 &  0.112 &  0.017 &  8.494 & 44 \\
\hline
\multicolumn{6}{c}{12+log(O/H)$_\mathrm{total}$} \\
\hline
Ia   &   8.492 &  0.054 &  0.008 &  8.505 & 44 \\
Ib/c &   8.489 &  0.076 &  0.015 &  8.500 & 27 \\
II   &   8.477 &  0.074 &  0.011 &  8.500 & 47 \\
\hline
\multicolumn{6}{c}{log($Z_m/Z_\sun$)$_\mathrm{local}$} \\
\hline
Ia   &   0.032 &  0.173 &  0.026 &  0.052 & 44 \\
Ib/c &$-$0.049 &  0.232 &  0.053 &  0.067 & 19 \\
II   &$-$0.012 &  0.238 &  0.043 &  0.043 & 30 \\
\hline
\multicolumn{6}{c}{log($Z_m/Z_\sun$)$_\mathrm{total}$} \\
\hline
Ia   &   0.021 &  0.149 &  0.020 &  0.060 & 58 \\
Ib/c &$-$0.069 &  0.239 &  0.046 &  0.007 & 27 \\
II   &$-$0.105 &  0.223 &  0.033 &$-$0.022 & 47\\
\hline
\end{tabular}
\end{table}

\subsection{Other proxies of the local metallicity}

In the past several indirect measurements have been used to estimate the metallicity at the SN location. The spatially resolved spectroscopy nature of our data allowed us to estimate the metallicity at the SN position and at the same time to compare it to the metallicity at the SN location estimated by other approximations. 
Therefore, we are in a position to test whether such methods are good proxies for the local SN metallicity.

In the previous Section we have shown that the metallicity measured from the total galaxy spectrum is a fair approximation of the local metallicity. In this Section we focus on several other proxies. The results of these proxies are given in Table~\ref{tab:proxies} for the three SN types and explained below.

\subsubsection{Central metallicity}\label{sec:galnuc}

The Sloan Digital Sky Survey and other galaxy surveys use fiber spectrographs to provide thousands of spectra of the central regions of galaxies in the nearby Universe. Such surveys can be very useful for SN studies \citep[see e.g., ][]{2008ApJ...673..999P}, but this approach has several drawbacks. Galaxies with AGN cannot be used because the strong line methods for gas-phase metallicity estimate are no longer valid. As we discuss in Appendix \ref{sec:re}, a large fraction of the galaxies in our sample have AGNs. More specifically, 59\% of the star-forming SN~Ia hosts (27 of the 46), 22\% of SN~Ibc/IIb hosts (6 out of 27) and 32\% of SN~II hosts (15 out of the 47) have AGNs.
As we have already shown in Paper I, there is a higher ratio of SN~Ia host galaxies that have an AGN compared to the galaxies that hosted other SN types.
This ratio would increase to 67\% if the 11 elliptical galaxies, all of which host an AGN, are included.

\begin{figure}   
\centering
\includegraphics*[trim=0.22cm 0.12cm 0.08cm 0.04cm, clip=true,height=3.5cm]{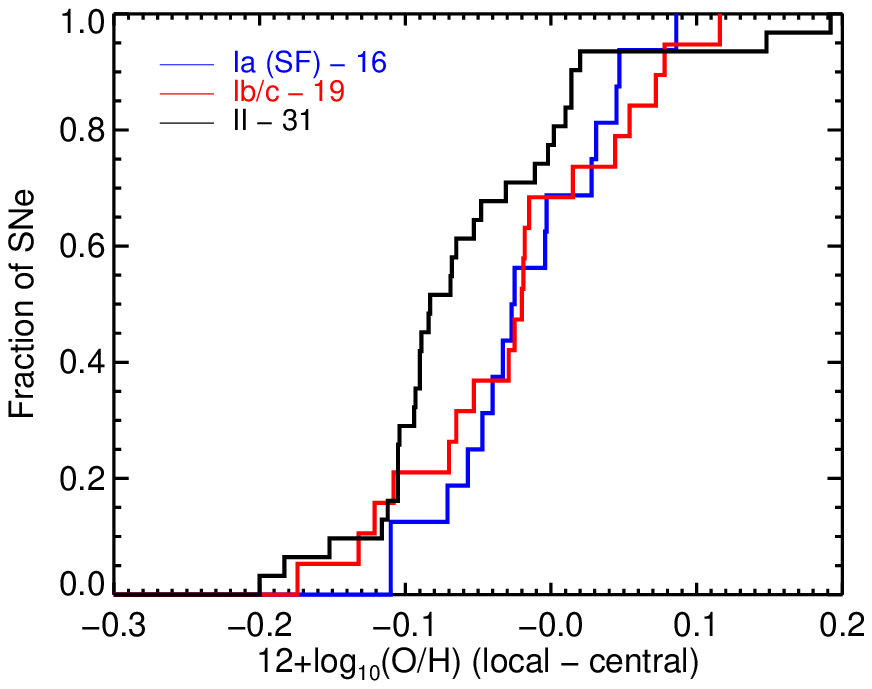} 
\includegraphics*[trim=0.52cm 0.12cm 0.08cm 0.04cm, clip=true,height=3.5cm]{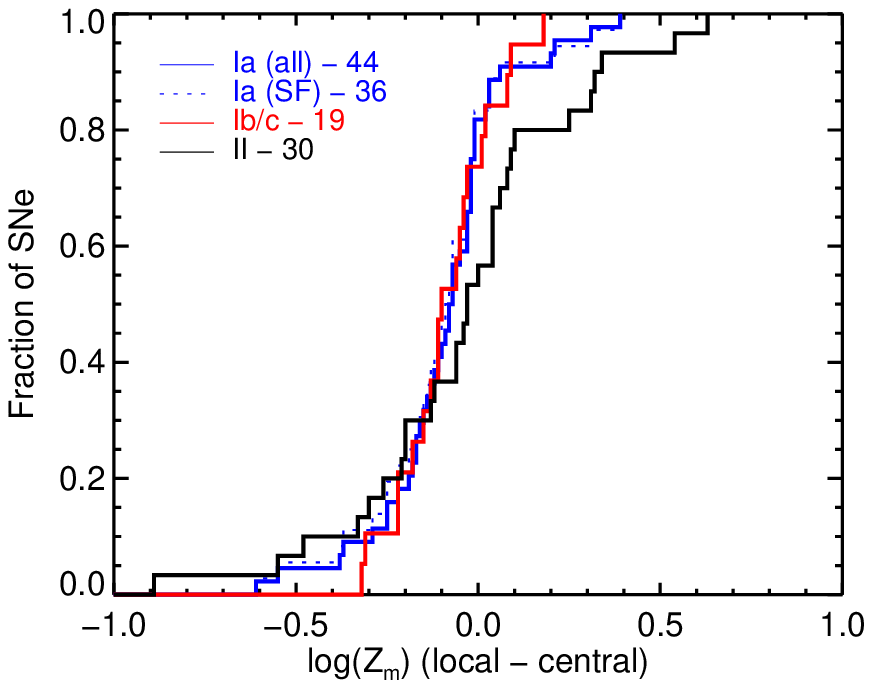} 
\caption{Distributions of the differences between local and central metallicity.}
\label{fig:loc-cen}
\end{figure}

Figure~\ref{fig:loc-cen} shows the distribution of the differences between the local metallicity and the metallicity measured at the galaxy core only for those galaxies with both measurements available. For the gas-phase metallicity all distributions have negative mean values and thus the central metallicity is an overestimation of the metallicity at the SN position. The offsets are not large, $-0.018$, $-0.025$ and $-0.055$~dex for SN~Ia, Ibc and II, respectively, and only the one for SN~II is statistically significant. 
The smallest offset for SNe~Ia could be explained by the  metallicity decrease  towards the center in some galaxies (see previous section). More than half of SN~Ia hosts in our sample have such decrease, which would center the distribution toward zero. The fraction of galaxies with the central decrease among the CC~SN hosts is lower and hence the offset is larger.
The difference of the stellar metallicities are $-0.088$, $-0.083$ and $-0.030$~dex  for SN~Ia, Ibc and II, respectively, with only the first two being statistically significant. The corresponding standard deviations are $\sigma_{\rm Ia}$=0.19, $\sigma_{\rm Ibc}$=0.13 and $\sigma_{\rm II}$=0.32~dex.
These results indicate the error if the central metallicity is used as a proxy for the local metallicity. However, because the standard deviation and the offset from zero are both larger than that for the total metallicity, its accuracy is lower.

Our findings agree with those of \cite{2011ApJ...731L...4M} who found that local metallicities of SN~Ibc in their sample were generally lower than the central ones. However their central estimation came from host galaxy luminosity and not from gas emission line ratios.
Our results also agree with \cite{2012ApJ...758..132S}, who claims that central and local metallicities are equal within 0.1~dex (in PP04 scale).

\subsubsection{Central metallicity plus characteristic metallicity gradient}

\begin{figure}   
\centering
\includegraphics*[width=9cm]{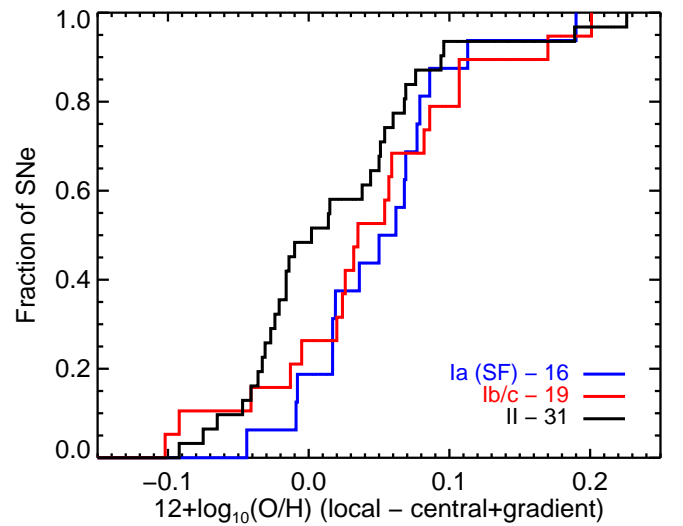}  
\caption{Distribution of differences between local metallicity and the proxy using central metallicity plus characteristic gradient. }
\label{fig:cen-grad}
\end{figure}

Another approach to estimate the local metallicity is to use the central metallicty plus the characteristic metallicity gradient (in normalized units). This can be very useful to utilize galaxy surveys performed with multi-fiber spectrographs like SDSS, which for low-redshift galaxies provide only the central metallicity.  
We here normalized the individual metallicity gradients to a common ruler, the disc effective radius $r_e$, the radius containing half of the total integrated flux coming from the disc component. The procedure followed to estimate $r_e$ is described in Appendix \ref{sec:re}.
We have tested this approach and Fig.~\ref{fig:cen-grad} shows CDFs of the differences. Of all proxies considered here, this one shows the largest mean differences and appears to be the least accurate one (Table~\ref{tab:proxies}). Only the difference for SN~II is not statistically significant. The larger offset for SN~Ia is likely due to the larger fraction of hosts with central metallicity decrease, which will bias this estimator.

\subsubsection{Metallicity gradient and radial distance}

\begin{figure}   
\centering
\includegraphics*[trim=0.22cm 0.12cm 0.08cm 0.04cm, clip=true,height=3.5cm]{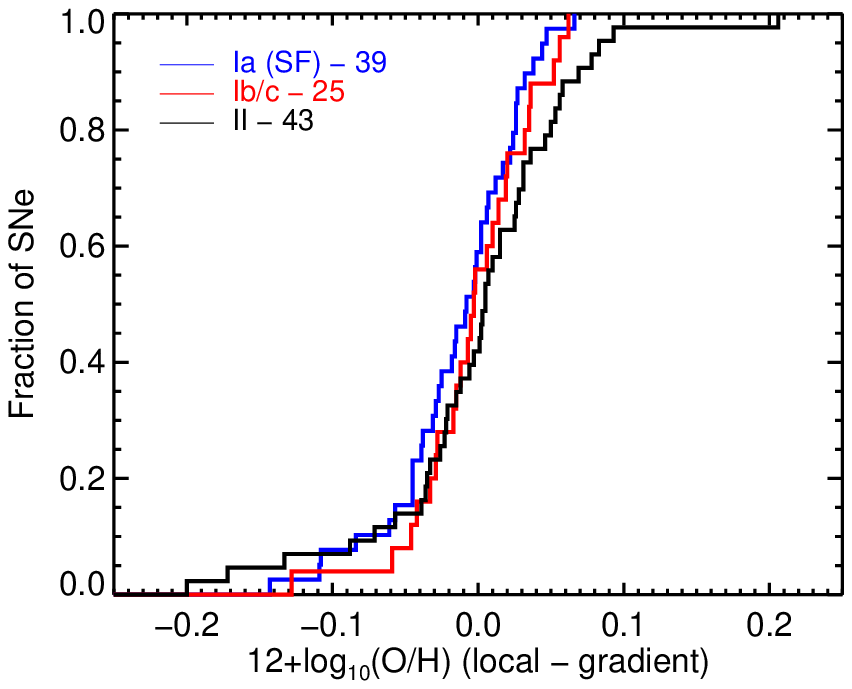}  
\includegraphics*[trim=0.52cm 0.12cm 0.08cm 0.04cm, clip=true,height=3.5cm]{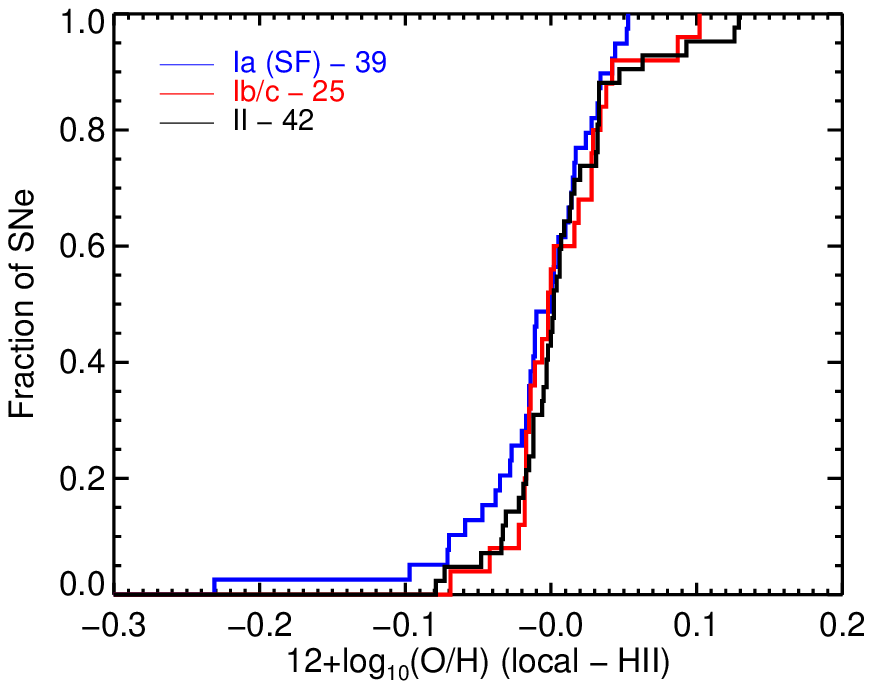}  
\caption{Distribution of differences between local metallicity and the proxy using gradients (left) and the metallicity of the closest HII region (right). All measurements performed with O3N2 index.}
\label{fig:ind-grad-hii}
\end{figure}

Metallicity gradients measured with long-slit spectroscopy have been used to estimate the metallicity at the location of SN explosions. The gradients are usually measured by placing the slit along the galaxy major axis and then the metallicity at the radial distance of the SN is estimated. We estimated the gas-phase metallicity at the SN location using the individual metallicity gradient for each galaxy obtained in Section \ref{sec:grad} and the deprojected GCD of the SN.
In the left panel of Figure \ref{fig:ind-grad-hii} we show the distributions of the differences between the metallicity at the SN position and that estimated using the gradients. 
We see that all SN types have narrow distributions with standard deviations of $\sigma_{\rm Ia}$=0.085, $\sigma_{\rm Ibc}$=0.053 and $\sigma_{\rm II}$=0.071~dex, and no significant differences in the average values can be seen for different SN types.
This is an indication that a (well-measured) metallicity gradient could be a good approximation to the value at SN position for SNe~Ibc and II. For SN~Ia the total metallicity appears to be a better approximation.
 
\subsubsection{Closest HII region}

As in Paper I, {\sc HIIexplorer} \citep{2012A&A...546A...2S} was used to extract HII regions from the extinction corrected H$\alpha$ emission maps. The caveats of this method are fully described in Paper~I, but the main ones are that (i) we are actually selecting clumps of HII regions (1 to 6,  \citealt{2014A&A...561A.129M}) instead of individual regions, especially in more distant galaxies; and (ii) this method tends to select regions with similar sizes, although this is not the case for real \ion{H}{ii} regions in galaxies.

For each SN, we determined the SN host \ion{H}{ii} region by selecting the clump that is closest to the SN location in terms of deprojected distance, 
and the spectra of the spaxels belonging to that \ion{H}{ii} region were co-added and analyzed. 
It is reasonable to consider that for CC~SNe the progenitor might be formed in fainter \ion{H}{ii} regions not detected by this method. For SNe~Ia it is difficult to associate the selected \ion{H}{ii} region as the place were the progenitor was formed as SNe~Ia may have old progenitors, which might have migrated away from their birthplace.

In the right panel of Fig.~\ref{fig:ind-grad-hii} we show the distributions of the differences among the metallicity at the SN position and that of the closest \ion{H}{ii} region.
All distributions are centered at zero within 0.01~dex and have small standard deviations $\sim0.04$~dex\footnote{The width of the distribution for SNe~Ia larger, but this is because of a single deviating point. Once this point is removed, the distribution becomes the narrowest of all with standard deviation 0.03~dex.}. The standard deviations are smaller than both the local metallicity estimates using gradients and the total metallicity, which means that the metallicity of the nearest \ion{H}{ii} region is a better approximation. This is expected because compared to the other proxies of the local metallicity, the nearest \ion{H}{ii} region is the most direct measurement.  

\begin{table}
\centering
\caption{Properties of the distributions of the differences between the metallicity estimated from local measurements and the metallicity estimated using different proxies. Shown are the mean, the standard deviation $\sigma$, the standard deviation of the mean $\sigma_m$, the median and the number of SNe.}
\label{tab:proxies} 
\begin{tabular}{lccccc}
\hline\hline
SN type & mean & $\sigma$ & $\sigma_m$ & median & $N$ \\
\hline
\multicolumn{6}{c}{12+log(O/H)$_\mathrm{local-total}$} \\
\hline
Ia\tablefootmark{a}   &    0.029 &  0.050 &  0.008 &    0.032 &   39  \\
Ib/c &    0.012 &  0.064 &  0.013 &    0.020 &   25  \\
II   &    0.006 &  0.078 &  0.012 &    0.008 &   44  \\
\hline
\multicolumn{6}{c}{12+log(O/H)$_\mathrm{local-H\,II}$} \\
\hline
Ia   & $-$0.004 &  0.035 &  0.006 &    0.001 &   39  \\
Ib/c &    0.007 &  0.037 &  0.007 & $-$0.001 &   25  \\
II   &    0.008 &  0.042 &  0.006 &    0.002 &   42  \\
\hline
\multicolumn{6}{c}{12+log(O/H)$_\mathrm{local-gradient}$} \\
\hline
Ia   & $-$0.014 &  0.045 &  0.007 & $-$0.007 &   39  \\
Ib/c & $-$0.003 &  0.041 &  0.008 & $-$0.002 &   25  \\
II   &    0.001 &  0.069 &  0.011 &    0.005 &   43  \\
\hline
\multicolumn{6}{c}{12+log(O/H)$_\mathrm{local-central}$} \\
\hline
Ia   & $-$0.018 &  0.056 &  0.014 & $-$0.026 &   16  \\
Ib/c & $-$0.024 &  0.077 &  0.018 & $-$0.019 &   19  \\
II\tablefootmark{a}    & $-$0.056 &  0.082 &  0.015 & $-$0.082 &   31  \\
\hline
\multicolumn{6}{c}{12+log(O/H)$_\mathrm{local-central+gradient}$} \\
\hline
Ia\tablefootmark{a}  & 0.052 &  0.056 &  0.014 & 0.057 &   16  \\
Ib/c\tablefootmark{a} & 0.043 &  0.077 &  0.018 & 0.036 &   19  \\
II    & 0.020 &  0.071 &  0.013 & 0.003 &   31  \\
\hline
\multicolumn{6}{c}{log($Z_m$)$_\mathrm{local-total}$} \\
\hline
Ia   & $-$0.004 &  0.150 &  0.023 & $-$0.011 &   44  \\
Ib/c &    0.004 &  0.150 &  0.034 &    0.047 &   19  \\
II   &    0.042 &  0.178 &  0.032 &    0.047 &   30  \\
\hline
\multicolumn{6}{c}{log($Z_m$)$_\mathrm{local-central}$} \\
\hline
Ia\tablefootmark{a}    & $-$0.086 &  0.185 &  0.028 & $-$0.070 &   44  \\
Ib/c\tablefootmark{a}  & $-$0.083 &  0.131 &  0.030 & $-$0.099 &   19  \\
II   & $-$0.029 &  0.311 &  0.057 & $-$0.024 &   30  \\
\hline
\end{tabular}\\
\tablefoot{
\tablefoottext{a}{statistically significant difference.}
}
\end{table}

\subsection{Aperture effects}

\begin{figure*}
\sidecaption
\includegraphics*[trim=0.0cm 0.0cm 0.3cm 0.4cm, clip=true,width=6.25cm]{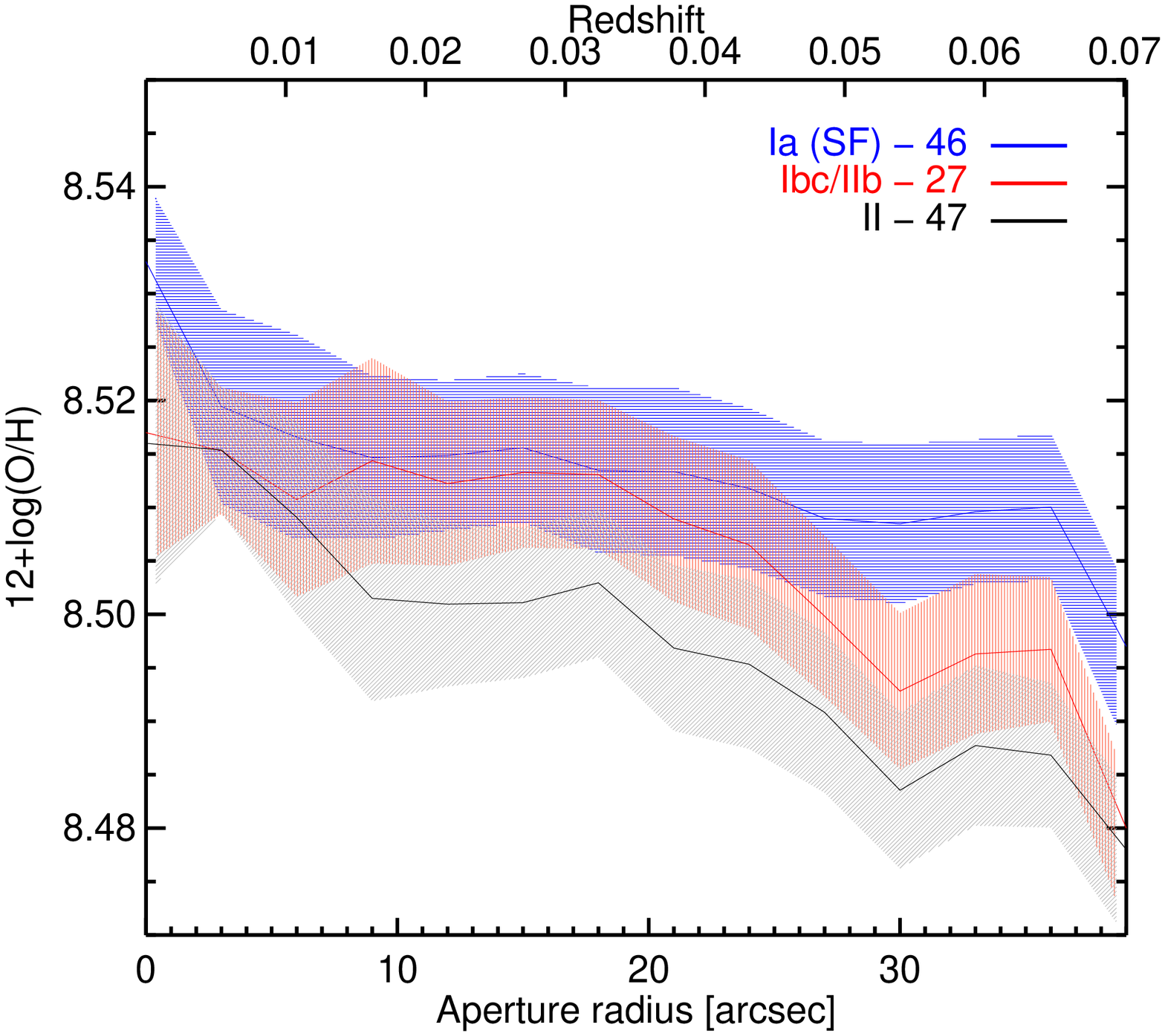}
\includegraphics*[trim=0.0cm 0.5cm 0.3cm 0cm, clip=true,width=6.55cm]{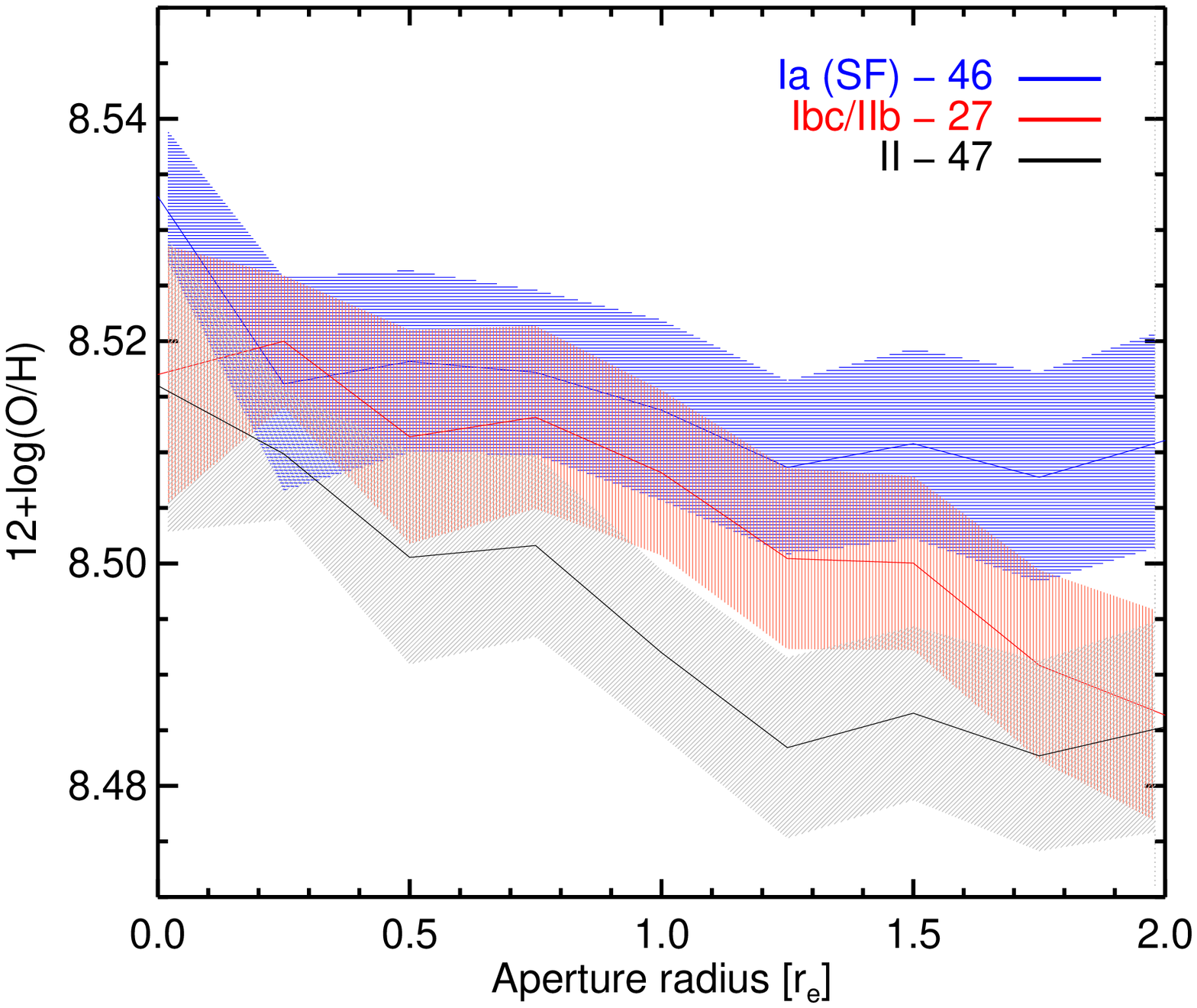}
\caption{Left: Mean values for oxygen abundance from the central spaxel (value at 0 arcsec) to the total spectra (value at 39 arcsec), increasing the aperture in steps of 3 arcsec. Error bands are errors of the mean values. 
The upper x-axis shows where the aperture with 3" diameter is projected for different redshifts.
Right: Same measurements but normalizing the unit of the aperture to the effective radius of the galaxies.}
\label{fig:metal-aper}
\end{figure*}

At low redshift fiber spectrographs usually measure only the light from the central region of galaxies. As the redshift increases, the fraction of the galaxy light integrated by the fibers also increases.
Comparing measurements obtained from different integration fractions of the galaxy introduces a bias in the results, and aperture effects must be accounted for.
IFS makes it possible to extract spectra integrating the datacubes in different apertures to study how the galactic parameters change with the aperture. In this way we are able to simulate observations obtained with different fixed-size fibers or, what would be equivalent, observations with a fixed-size fiber of galaxies at different redshifts.
This has been studied in detail in \cite{2013A&A...553L...7I} with H$\alpha$ emission, in Iglesias-P\'aramo et al. (sub. to A\&A) for oxygen abundances using the whole CALIFA galaxy sample, and in \cite{2015arXiv151101300G} with CALIFA early-type galaxies.

We calculated the spectra within 12 circular apertures centered at the galaxy cores and increasing the radii from 3 to 36\arcsec~in steps of 3\arcsec. 
The same  procedure used for the individual spaxels was applied to measure the metallicities from each of the increasing aperture spectra.
The left panel of Figure \ref{fig:metal-aper} shows the average values of the oxygen abundance for each SN type hosts as a function of an increasing aperture.
The measurements laying in the AGN region of the BPT diagram are not considered in the construction of the averages of this measurement.
On the right panel we show the same measurements but changing the radial scale to the disc effective radius of each galaxy.
The differences between the central and the total values shown in the previous subsection arise here with the intermediate behaviors. 
We see that the average gas abundances decrease with increasing the aperture, and the hosts of the three SN types remain ordered as it is shown in the total distributions (See Fig. \ref{fig:metal-loc}). 
The SN~Ia band is above the two CC~SN bands, and SN~II show lower values than SN~Ibc host galaxies.
This behavior is easily explained (and expected) by the presence of metallicity gradients.
As the aperture increases, new regions that are farther from the center are included in the integrated spectra. 
If these new regions are metal-poorer than their inner counterparts, the average of the aperture would be reduced, but not at the level of the metallicity of the new regions.  
Also remarkable is that in the two figures, SN~Ia galaxies band is shallower compared to the CC~SN host bands. Although passive galaxies are not included in the SN~Ia group we show in Appendix \ref{app:metgrad} that the metallicity gradients of SN~Ia hosts were also shallower.
The maximum difference for any SN host is around 0.04 dex, and this would be an estimation of the systematic error when using fiber spectra measurements of galaxies spanning a wide range of redshifts.

A similar approach was used by \cite{2015A&A...581A.103G} who instead of studying how the stellar metallicity varies at increasing circular apertures, studied the deprojected radial profiles and gradients, splitting 300 galaxies from the CALIFA sample by morphology and mass. Their results are in agreement with our findings. Elliptical and S0 galaxies have higher stellar metallicities and flatter gradients than the late-type galaxies. This corresponds to our SN~Ia group since these types of galaxies only host thermonuclear supernovae. 


\section{Discussion} \label{sec:disc}

\subsection{Total metallicity}
\label{sec:totmass}

We have seen that the gas-phase and the mass-weighted stellar metallicities derived from the integrated galaxy spectra show different behavior for the three SN types (Fig.~\ref{fig:metal-loc}). The average gas-phase metallicity for the three SN types is nearly equal with differences less than 0.01~dex. The distributions are also similar. However, both CC~SN types, Ibc and II, show more extended tail toward low metallicity, which is more prominent for SNe~II. The average mass-weighted stellar metallicity on the other hand shows differences of $\sim0.05$~dex between the three SN types. This behavior results from the distribution of the masses of the galaxies in our sample on one hand and  the shape of the mass-metallicity ($M-Z$) relation for the gas-phase and the stellar metallicities on the other. 

\begin{figure}   
\centering 
\includegraphics[trim=0cm 0cm 0cm 0cm, clip=true,width=\hsize]{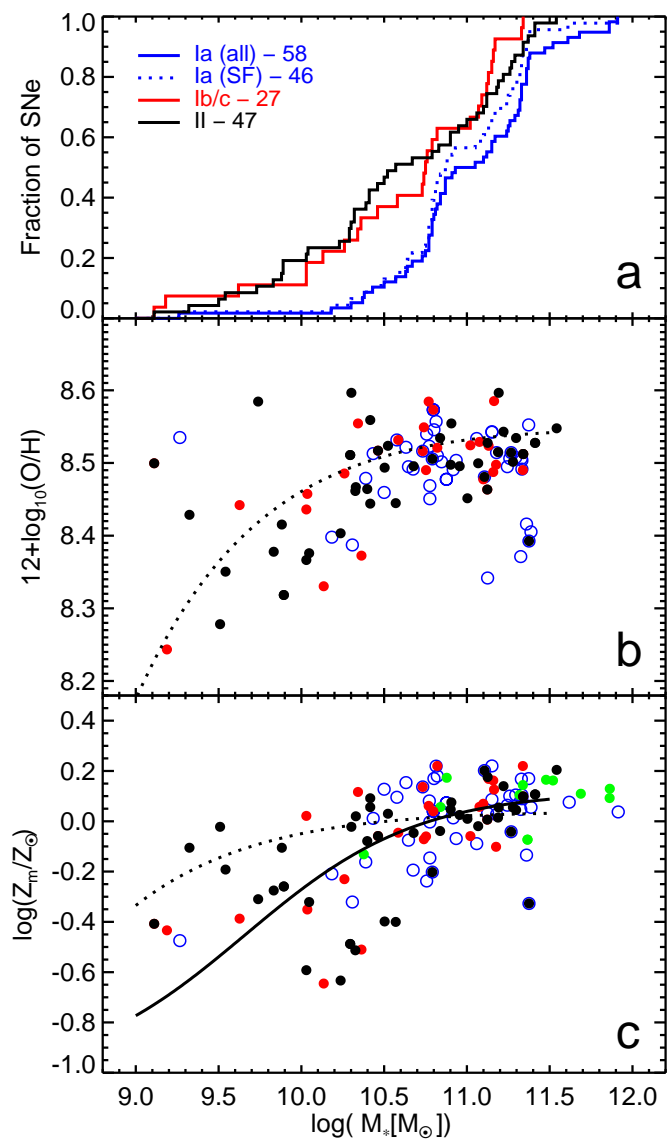}
\caption{$a$ - cumulative distributions of the galaxies stellar mass; $b$ - mass $vs.$ gas-phase metallicity. The dotted line is the $M-Z$ relation from \citet{2013A&A...554A..58S} converted from PP04 to M13 version of the O3N2 calibrator. $c$ -  mass $vs.$ mass-weighted stellar metallicity. The dotted line is the $M-Z$ relation from \citet{2013A&A...554A..58S} and the solid line is the \cite{2008MNRAS.391.1117P} relation. The green dots show the elliptical SN~Ia hosts.}
\label{fig:massmag}
\end{figure}

Figure~\ref{fig:massmag}$a$ shows the cumulative distribution of the current stellar mass of the galaxies in our sample by SN type. As already noted in Paper~I, the mass distribution shows $0.3-0.4$~dex difference between the mean masses of SN~Ia hosts and those of the two CC~SNe types. It can also be seen that about 90\% of the SN~Ia hosts and 60--70\% of the CC~SN hosts have masses above $\log(M/M_\sun)\simeq 10.3$ dex. 

\begin{figure}[!th]
\centering
\includegraphics*[width=8.8cm]{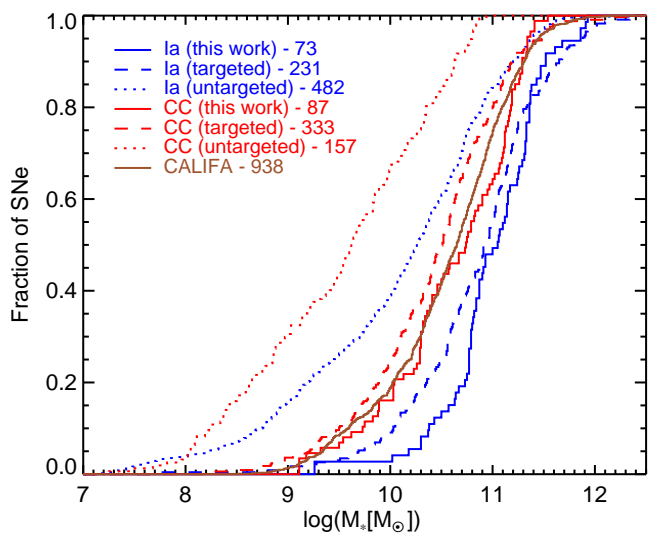}\\
\caption{CDFs of the host galaxy masses of various SN~Ia and CC~SNe samples (see the text for details). }
\label{fig:masscomp}
\end{figure}

The $M-Z$ relations for the gas-phase metallicity and mass-weighted stellar metallicities are shown in Fig.~\ref{fig:massmag}$b$ and $c$. The dotted lines in Figs.~\ref{fig:massmag}$b$ and $c$ show the $M-Z$ relation for the gas-phase metallicity derived \citet{2013A&A...554A..58S} from CALIFA observations. This relation describes the $M-Z$ relation for the gas-phase metallicity of the galaxies in our sample very well\footnote{Strictly speaking this relation is derived for oxygen abundance at the effective radius $r_\mathrm{e}$. However, as \citet{2013A&A...554A..58S} also noted, we found that the difference between the metallicity at $r_\mathrm{e}$ is very close to the total metallicity estimated from the integrated spectra. The average difference we found was $\sim0.02$~dex with 1$\sigma$ scatter around 0.03~dex. }, but from Fig.~\ref{fig:massmag}$c$ it is evident that the mass-weighted stellar metallicity $M-Z$ relation is much steeper, in accordance with the findings of \cite{2014ApJ...791L..16G}. The solid line in Fig.~\ref{fig:massmag}$c$ shows the \cite{2008MNRAS.391.1117P} $M-Z$ relation, which clearly describes the stellar metallicity $M-Z$ relation better. One can also see that both, the gas-phase and the stellar metallicity $M-Z$ relations present lower scatter above $\log(M/M_\sun)\simeq 10.4$ dex.

The fact that the gas-phase $M-Z$ relation is tighter and flatter than the stellar metallicity and that $\gtrsim$80\% galaxies in our sample have masses higher than $\log(M/M_\sun)\simeq 10.3$ dex is the main reason why there are only very small differences between the average gas-phase metallicities for the three SN types.  On the other hand, the stellar metallicity $M-Z$ relation is steeper and this is the most likely reason for the observed larger differences in the average stellar metallicities.
 
The small differences between the total gas-phase metallicity of different SN type host galaxies that we find is somehow at odds with the previously reported larger differences \citep[see, e.g.,][]{2008ApJ...673..999P,2014ApJ...791...57S}. This is however most likely due to the bias toward higher masses in our sample. 
To clarify the possible impact in our results of this bias we have extended the study using data from the literature.
Figure~\ref{fig:masscomp} shows the CDFs of the host masses of our SN~Ia and  CC~SN samples, including the galaxies with SNe outside of FoV, compared to the targeted and the untargeted samples from \cite{2009ApJ...707.1449N}, \cite{2010ApJ...715..743K}, \cite{2012ApJ...759..107K}, \cite{2013ApJ...770..107C}, \cite{2013ApJ...773...12S} and \cite{2014MNRAS.438.1391P}. 
Table~\ref{tab:masses} gives details of the properties of the corresponding distributions.
In these works different IFMs have been employed and we applied small corrections \citep[$\sim0.04-0.20$~dex; ][]{2003ApJS..149..289B,2010MNRAS.406..782S} to convert the  masses to \cite{1955ApJ...121..161S} IMF. However, additional systematic differences are almost certainly present because of the use of different recipes to construct the fitting galaxy templates,  different observational data used (photometry and spectroscopy) and different fitting approaches. Quantifying these differences is beyond the scope of this paper but they should be kept in mind when combining galaxy masses from different sources.

The data from the literature clearly shows that the targeted surveys suffer from a bias toward massive galaxies, which is $\sim0.77$~dex for SNe~Ia and $\sim0.9$~dex for the CC~SNe. It is also evident that our sample, which consists of SNe discovered by targeted searches, is biased even more toward higher masses; with respect to the targeted sample from the literature this bias for both SN~Ia and CC~SN is $\sim0.15$~dex. The mean masses of both SN~Ia and CC~SNe hosts found in targeted searches are in the range $\log(M/M_\sun)\simeq10.4-10.7$ and with only a small fraction of SNe ($\sim$10 and 20 \% for SN Ia and CC SN, respectively) present in galaxies with masses below $\log(M/M_\sun)\simeq10.3$ dex. Because of the lower scatter in the $M-Z$ relation above $\log(M/M_\sun)\simeq10.3$ dex, the metallicity differences between the SN types should be small in samples from targeted searches. The untargeted searches on the other hand include galaxies with lower masses in the range where the $M-Z$ relation is steeper and the metallicity differences, if significant, will be more pronounced.

\begin{table}
\caption{Statistics of the distributions of galaxy masses in different SN samples in units of $\log(M/M_\sun)$. For each SN sub-type are shown the mean, the standard deviation $\sigma$  and the median.}
\label{tab:masses}
\setlength{\tabcolsep}{4pt}
\begin{tabular}{@{}lrcccrcc@{}}
\hline\hline
         &   \multicolumn{3}{c}{SN~Ia} & & \multicolumn{3}{c}{CC~SN}\\
              \cline{2-4}\cline{6-8}
               & mean  &  $\sigma$ & median &   & mean  & RMS & median \\
\hline
This work    &  11.00 & 0.50  & 11.06 &  &  10.61 & 0.62  & 10.74 \\
Targeted:    &  10.85 & 0.66  & 10.94 &  &  10.42 & 0.68  & 10.51 \\
Untargeted  &  10.08 & 0.98  & 10.27 &  &   9.52 & 0.88  &  9.62  \\
\hline\hline
\end{tabular}
\end{table}

\subsection{Local metallicity}\label{sec:locmet}

We have estimated both the gas-phase and stellar metallicities at the SN locations.  As discussed in \cite{2012A&A...545A..58S} the stellar metallicities are estimated with much greater uncertainty due to various factors including astrophysical and numerical
degeneracies, noise in the observed spectra, uncertainties in the recipes to construct the SSPs, etc. Even without these uncertainties,  the S/N in the continuum of the binned spectra at the SN location may not be enough to estimate the metallicity accurately enough. For this reason we focus our discussion of the local metallicity to the gas-phase oxygen abundance.

We found only small differences between the average metallicity at the SN locations for the three SN types, 8.522, 8.500 and 8.484~dex for SN~Ia, Ibc and II, respectively. These differences are considerably smaller than the 1$\sigma$ scatter, which is 0.056 dex for SN~Ia and 0.112~dex for the CC~SNe. However, one question that should be addressed is whether what we measure is the local metallicity at the SN position. As the distance to the galaxies increases it becomes more difficult to measure local quantities because of the decrease of the spatial resolution when observing with fixed-size slits or fibers.
Recently, \cite{2015MNRAS.449.2706N} studied how the variation of the metallicity within a galaxy and the spatial resolution affect the local metallicity estimation. They found that when the resolution is lower than 500~pc, the local metallicity could be accurately estimated. Resolution up to 1~kpc also provide reasonable estimates, although some systematic effects are present.
When the resolution is larger than 1~kpc the constrain of the local metallicity becomes more uncertain. 
For the redshift range of our targets from 0.005 to 0.03 the resolution of the CALIFA maps (PSF$\sim$2.57 arcsec) corresponds to a spatial resolution between $\sim$250~pc and 1.5~kpc. However, only a small fraction of our targets are at the higher end of the redshift distribution and on average the resolution effects will be small in our sample.

\begin{figure}   
\centering 
\includegraphics[trim=0cm 0cm 0cm 0cm, clip=true,width=\hsize]{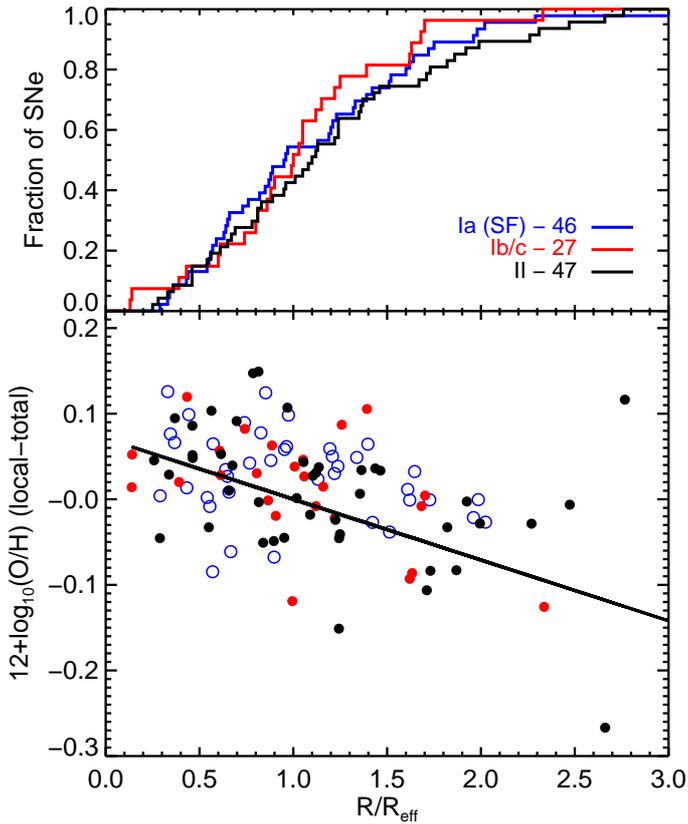}
\caption{Upper panel: CDFs of the normalized radial distribution of the different SN types. Lower panel: difference between the local and the total metallicity as function of the normalized GCD. The line is a guide-to-the-eye and has a slope of -0.074 -- the average metallicity gradient of the galaxies in our sample.}
\label{fig:dohdist}
\end{figure}

The fact that the differences between the local metallicities of the different SN types in our sample are small are not surprising because the total metallicities estimated from the integrated spectra are on average also the same and the galaxies seem to have a characteristic metallicity gradient in normalized radii units. Given this, large differences between the local metallicities can be observed if (i) the different SN types have significantly different radial distributions or (ii) the galaxies contain regions with metallicity, which are significantly different from the galaxy average. In the latter case, if certain SN type prefers to explode in low or high-metallicity environment, then the local metallicity of this SN type will tend to differ from the galaxy average. However, a careful inspection of the 2D metallicity maps shows that regions with significantly different metallicity from the average are rare or absent, because the dynamical range of abundances across the entire galaxy is small.

The analysis of the SN radial distances shows that SNe~Ibc tend to explode slightly closer (by less than 1~kpc) to the galaxy center than SNe~II, and SNe~Ia on average 1~kpc further away than SNe~II. However, once the distance are normalized to the effective radius $R_\mathrm{eff}$ all distributions have similar averages of 1.1, 1.0 and 1.2$R_\mathrm{eff}$ for SN~Ia, Ibc and II, respectively. This is in agreement with previous statistical studies of SNe radial distributions \citep{1992A&A...264..428B, 1997AJ....113..197V, 2004AstL...30..729T, 2005AJ....129.1369P, 2008ApJ...673..999P, 2009A&A...508.1259H, 2009MNRAS.399..559A}. 
The upper panel of Figure~\ref{fig:dohdist} shows the distribution of the de-projected distances measured in disk effective radius units. In the bottom panel of Fig.~\ref{fig:dohdist} the difference between the local and the total metallicities as function the normalized deprojected distances is shown. The difference decreases with the radial distance for all SN types at a rate consistent with the  average metallicity gradient (the line with slope of -0.074 guides the eye). From the figure one can see that SNe~Ia tend to be slightly above the line and hence they show larger average difference than the CC~SNe. However, the difference is rather small and could be caused by other effects, for example by slightly biased estimation of the total galaxy metallicity.

From Fig.~\ref{fig:dohdist}$b$ one can see that the metallicity at the SN location follows the galaxy mean metallicity and that it also changes with the radial distance following the metallicity gradients. However. the scatter of the difference between the local and the total metallicities, and the scatter around the line is considerable, $\sim0.05$~dex.
One interesting question is whether part of this scatter arises because some SNe preferably explode in regions with higher  metallicities. Because many galaxies have negative metallicity gradients, if that was the case, SNe may tend to explode closer to the galaxy center. However, the density of the stars in the galaxies and the SFR also increase toward the center and statistically one can expect to find more SNe toward the central regions. Thus, the two effects will be difficult to disentangle. 

\begin{figure}   
\centering 
\includegraphics[trim=0cm 0cm 0cm 0cm, clip=true,width=\hsize]{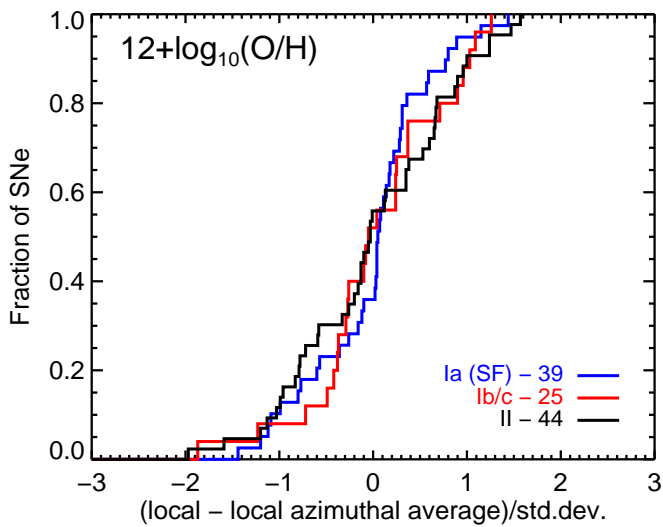}
\caption{Difference between the oxygen abundance at SN position compared to the azimuthal average at that distance, normalized to the standard deviation of the abundance at that distance.}
\label{fig:locazi}
\end{figure}

Galaxies typically have central symmetry. One way to test the hypothesis that some SNe prefer to explode in high-metallicity regions is to look at the difference between the metallicity at the SN position and the mean value of the galaxy metallicity at the radial distance of the SN. 
To estimate the latter, the mean metallicity and the standard deviation of the individual spaxels within 1~kpc of the deprojected GCD of the SNe were computed. In Fig.~\ref{fig:locazi} are shown the distributions of the differences divided by the standard deviation.
For all SN types the mean differences are essentially zero and the distributions are narrow with standard deviations of about one. SN~II show slightly wider distribution than the other SN types. This indicates that on average the metallicity at the SN location does not differ significantly from the galaxy average at the SN radial distance. There is only a small hint for an extended tail toward lower metallicities for SN~II.

\subsubsection{Proxies of local metallicity}

We have compared the directly estimated local metallicites obtained with IFS with several other approaches that have been used as proxies of local metallicity estimations. We conclude that the central metallicity, with and without a correction assuming characteristic metallicity gradient is the least accurate one. It has both larger bias and scatter. Not surprisingly, the metallicity of the closest \ion{H}{ii} region and the one estimated from each galaxy's individual metallicity gradient are the best proxies of the local metallicity. However, these two methods are applicable to nearby galaxies only and are likely of little practical use, e.g. for high-redshift SN~Ia studies. On the other hand, the total galaxy metallicity appears to be the best compromise. It has relatively small bias and scatter. In addition, it is a quantity that can be easily measured for galaxies at high redshifts.

The likely reason for the similarity between the local and total metallicities is that in low-redshift galaxies the metallicity gradients are fairly shallow. However, one should have in mind that this may not be the case for high-redshifts galaxies and hence the total metallicity may not be as good proxy as at low redshift.

\subsubsection{Differences among SN subtypes}

\begin{figure*}
\centering
\includegraphics*[width=0.99\hsize]{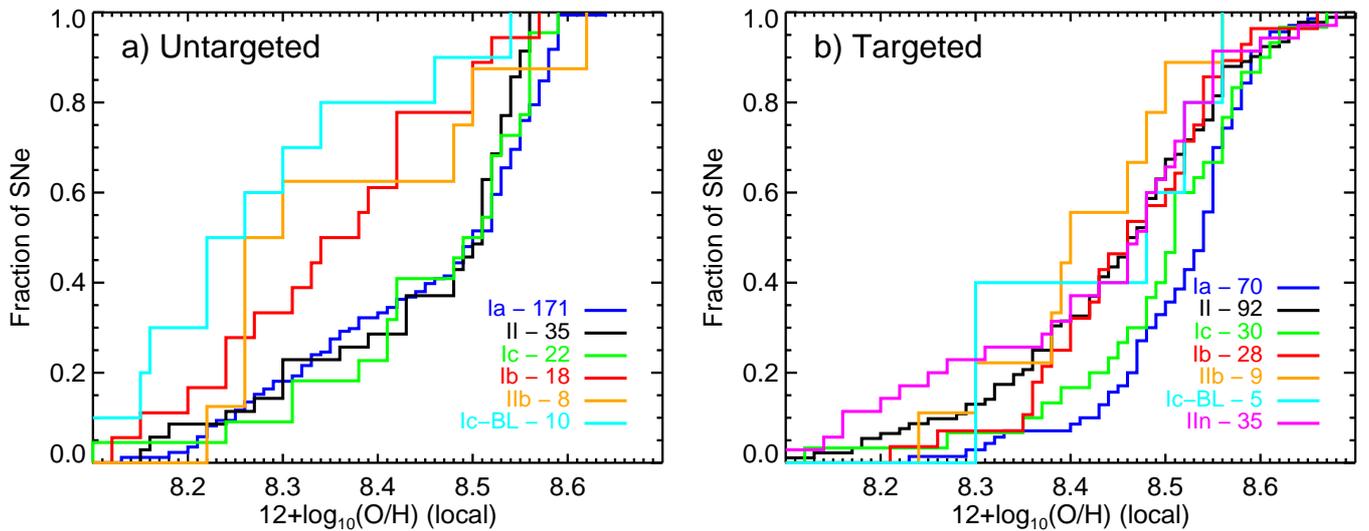}
\caption{Compilation of the oxygen abundances at the SN location measured in this work combined with previously published from the literature. We combined published measurements using the O3N2 calibrator with measurements using the N2 converted to O3N2 with \cite{2008ApJ...681.1183K} relation. $a)$ -- for SNe discovered in untargeted searches; $b)$ -- in targeted searches. The CALIFA sample presented in this work contributes essentially to the targeted sample.} 
\label{fig:others}
\end{figure*}

\begin{table*}
\caption{Statistics of the local metallicity distributions of SNe discovered in untargeted and targeted searches. For each SN sub-type are shown the mean, the standard deviation $\sigma$, the standard deviation of the mean $\sigma_m$, the median. and the number of SNe.}
\label{tab:all}
\begin{center}
\begin{tabular}{lrcccccrcccc}
\hline\hline
              &   \multicolumn{5}{c}{untargeted} & & \multicolumn{5}{c}{targeted}\\
              \cline{2-6}\cline{8-12}
              type & mean  & $\sigma$ & $\sigma_m$ & median & $N_\mathrm{SN}$ &   &  mean  & $\sigma$ & $\sigma_m$ & median & $N_\mathrm{SN}$ \\
\hline
Ia\tablefootmark{a}   &  8.455 &  0.129 &  0.010 &  8.503  &  171 & & 8.520 &  0.083 &  0.010 &  8.543 &  70   \\
Ic      &  8.458 &  0.124 &  0.026 &  8.515  &   22 & & 8.495 &  0.112 &  0.020 &  8.511 &  30   \\
II      &  8.447 &  0.126 &  0.021 &  8.511  &   35 & & 8.450 &  0.127 &  0.013 &  8.475 &  92   \\
IIb     &  8.367 &  0.149 &  0.053 &  8.302  &    8 & & 8.415 &  0.101 &  0.034 &  8.404 &   9   \\
Ib      &  8.355 &  0.130 &  0.031 &  8.388  &   18 & & 8.461 &  0.102 &  0.019 &  8.468 &  28   \\
Ic-bl   &  8.274 &  0.151 &  0.048 &  8.261  &   10 & & 8.436 &  0.123 &  0.055 &  8.481 &   5   \\
\hline
Ib/IIb  &  8.359 &  0.133 &  0.026 &  8.344  &   26 & & 8.450 &  0.100 &  0.016 &  8.448 &  39   \\
Ibc     &  8.418 &  0.133 &  0.020 &  8.428  &   44 & & 8.485 &  0.107 &  0.014 &  8.511 &  62   \\
IIn     &  \multicolumn{5}{c}{---}                  & & 8.427 &  0.150 &  0.025 &  8.475 &  35   \\
CC    &  8.421 &  0.137 &  0.015 &  8.488  &   88 & & 8.455 &  0.126 &  0.009 &  8.481 & 196   \\
\hline\hline
\end{tabular}\\
\tablefoot{
\tablefoottext{a}{The metallicities of the untargeted SN~Ia sample come from \cite{2014MNRAS.438.1391P} and \cite{2013ApJ...770..107C}. These measurements correspond to the total galaxy metallicity and have been corrected by +0.028~dex to account for the bias. See text for details.}
}
\end{center}
\end{table*} 

Our primary goal is to explore the connection between the SN environmental metallicity and the nature of the progenitors for each SN type. At present, the CALIFA sample used in our analysis is not sufficiently large to allow splitting of the two main CC~SN groups into further sub-classes with enough events. In addition, the vast majority of our sample comes from targeted searches and is biased toward high-mass, metal-rich galaxies. As a result the local metallicity we measure is very similar for all SNe.
For the above reasons we combined our values with measurements of the metallicity at SN locations from the literature. This enlarged sample contains SNe discovered by both targeted and untargeted surveys.
We considered only measurements performed at the SN location or close vicinity and whenever repeated we rely on our measurements.
The compiled sample includes all the measurements from: \cite{2009ApJ...698.1307T, 2010A&A...512A..70Y, 2011ApJ...731L...4M, 2011MNRAS.416..567A, 2012ApJ...745...70P, 2013AJ....146...30K,2013AJ....146...31K, 2013MNRAS.434.1636T}; and only those reported as local from: \cite{2010MNRAS.407.2660A, 2011A&A...530A..95L, 2012ApJ...758..132S,2012MNRAS.424.2841H}, and \cite{2016ApJ...818L..19M}.
In addition, we included metallicities of 11 SNe in galaxies observed with the new MUSE IFS \citep{2016MNRAS.455.4087G}. 
Following \cite{2012ApJ...758..132S}, we did not include \cite{2012ApJ...759..107K} since those are measurements from the center of the galaxies.

Since many different calibrators are used, we considered only those using PP04 O3N2 calibrator and converted them to the M13 scale. 
However, this sample can be further enlarged if we also consider those measurements computed with the PP04 N2 calibrator, and converted to O3N2 scale using the relation produced by \cite{2008ApJ...681.1183K}.
In this enlarged sample we used data from: \cite{2013ApJ...773...12S, 2009ApJ...697..676S, 2010ApJ...709L..26L, 2013ApJ...767...71M, 2013A&A...555A.142I, 2012AJ....143...19V, 2013A&A...558A.143T} and \cite{2015A&A...580A.131T}.
We also added the metallicities of the SN~Ia hosts from the untargeted searches reported in \cite{2014MNRAS.438.1391P} and \cite{2013ApJ...770..107C}. These metallicities were measured with long-slit spectroscopy and the manner the spectra were extracted from the 2D images means that they are close to the galaxies total metallicity. However, in our sample the local and the total metallicities of SN~Ia hosts differ on average by $0.029$~dex and we can use \cite{2014MNRAS.438.1391P} and \cite{2013ApJ...770..107C} measurements as proxy of the local metallicity .
  
We considered the following main SN classes: Ia, Ib, Ic, broad-line Ic (IcBL), II, IIn and IIb. 
Recently, the presence of two different populations of SN~II (IIP and IIL) has been put in question. While some works \citep{2012ApJ...756L..30A, 2014MNRAS.445..554F} present evidence for two separated populations, this has been questioned by a number of works \citep{2014ApJ...786...67A, 2015ApJ...799..208S,2016AJ....151...33G} which argued that this is really a continuous between slow- and fast-decliners, and the traditional division was due to the low number of SNe IIL. Following the latter works, we put SNe~IIP and IIL in one group II. 
We also consider separate classes of SNe~Ib and IIb, but because these are probably closely related (by a continuous sequence of the amount of Hydrogen in their outer layers) we also computed the mean metallicity of SNe~Ib and IIb together. Many SNe~Ib/c have uncertain classification. These were grouped with SNe~Ib/c with certain classification to compute the mean metallicity of the combined SN~Ib/c class. The mean metallicity of the whole CC~SN class was also computed excluding the broad-line SNe~Ic.

Further, the sample was divided into two groups -- SNe discovered by targeted or untargeted searches.
The resulting distributions of the oxygen abundance are shown in Figure \ref{fig:others}.
The properties of the distributions for all SN types are summarized in Table~\ref{tab:all} and the p-values of the KS tests are listed in Table \ref{tab:ks}. The CALIFA sample presented in this work comprises mostly of SNe discovered in targeted searches and thus it only makes significant contribution to the targeted SN sample.

$ $\\
\noindent{\it Metallicity at the SN~Ia locations}\\

The locations of SNe Ia are most likely not related at all to the birth place of the progenitor star due to their long delay times (100 Myrs to Gyrs), thus progenitors could have travelled across the galaxy a considerable distance before explosion as supernovae. 
Any association between SN environment observed properties and the SN progenitor population is much more problematic, however, the study of the properties of SN~Ia environments can give insights on the characteristics of the explosion that might be useful, for example, for cosmological analyses.

SNe~Ia have the highest mean metallicity at the SN location in the targeted sample (8.52~dex) and similar to SN~II and Ic in the untargeted (8.46~dex). However, most the untargeted SN~Ia come from measurements similar to the total metallicity, corrected for the small bias.
\cite{2014MNRAS.438.1391P} sample has higher mean metallicity 8.53~dex (corrected for the bias), but the mean of the whole sample is pushed toward low metallicity by the sample of \cite{2013ApJ...770..107C}. This comes from the fact that the latter sample also has significantly lower mean mass.

If we take the mean masses of SN~Ia hosts from Table~\ref{tab:masses} in the targeted and untargeted samples and compute the corresponding metallicities with the $M-Z$ relation of \citet{2013A&A...554A..58S} we get  8.53 and  8.47~dex for SNe~Ia in the combined targeted and untargeted samples, respectively. the same analysis for CC~SNe hosts gives metallicities of 8.50  and  8.36~dex. However, if we apply the $M-Z$ relation for our sample only, we obtain metallicitties of 8.53 and 8.50 dex, for SN~Ia and CC~SN hosts respectively. These numbers are very close to the actual estimated mean metallicities, except for CC~SNe in the untargeted sample, which deviate by 0.06~dex. This suggest that the difference in the mean metallicity on the different samples is to large extent due to the $M-Z$ relation.  
        
$ $\\
\noindent{\it Metallicity at the CC~SN locations}\\

From Table \ref{tab:all} we see that all SN types, except for SN~II, show bias toward higher metallicity in the targeted samples. The bias is different for the different types, and SN~Ib and Ic-BL are particularly affected ($>$0.1 dex).
In the targeted sample SN~Ic locations show higher $Z$ than both SN~Ib and SN~II locations, and the latter also show $Z$ similar to SN~IIn and SN~Ic-bl. SN~IIb is the CC~SN subtype with the lowest average $Z$.
The picture changes for the untargeted sample: while SN~II and SN~Ic now show similar $Z$, SN~Ib have significantly lower $Z$, but similar to SN~IIb. SN~Ic-BL are, as expected, in the lowest $Z$ end. 
So, from the untargeted SNe we have the following sequence:
\begin{equation}
\indent\indent\indent\indent Z_{Ic} \gtrsim Z_{II} > Z_{Ib} \gtrsim Z_{IIb} > Z_{Ic-BL}.
\end{equation}

\begin{figure*}[t]
\centering
\includegraphics[trim=1.0cm 0.2cm 0.3cm 0.7cm, clip=true,height=7.8cm]{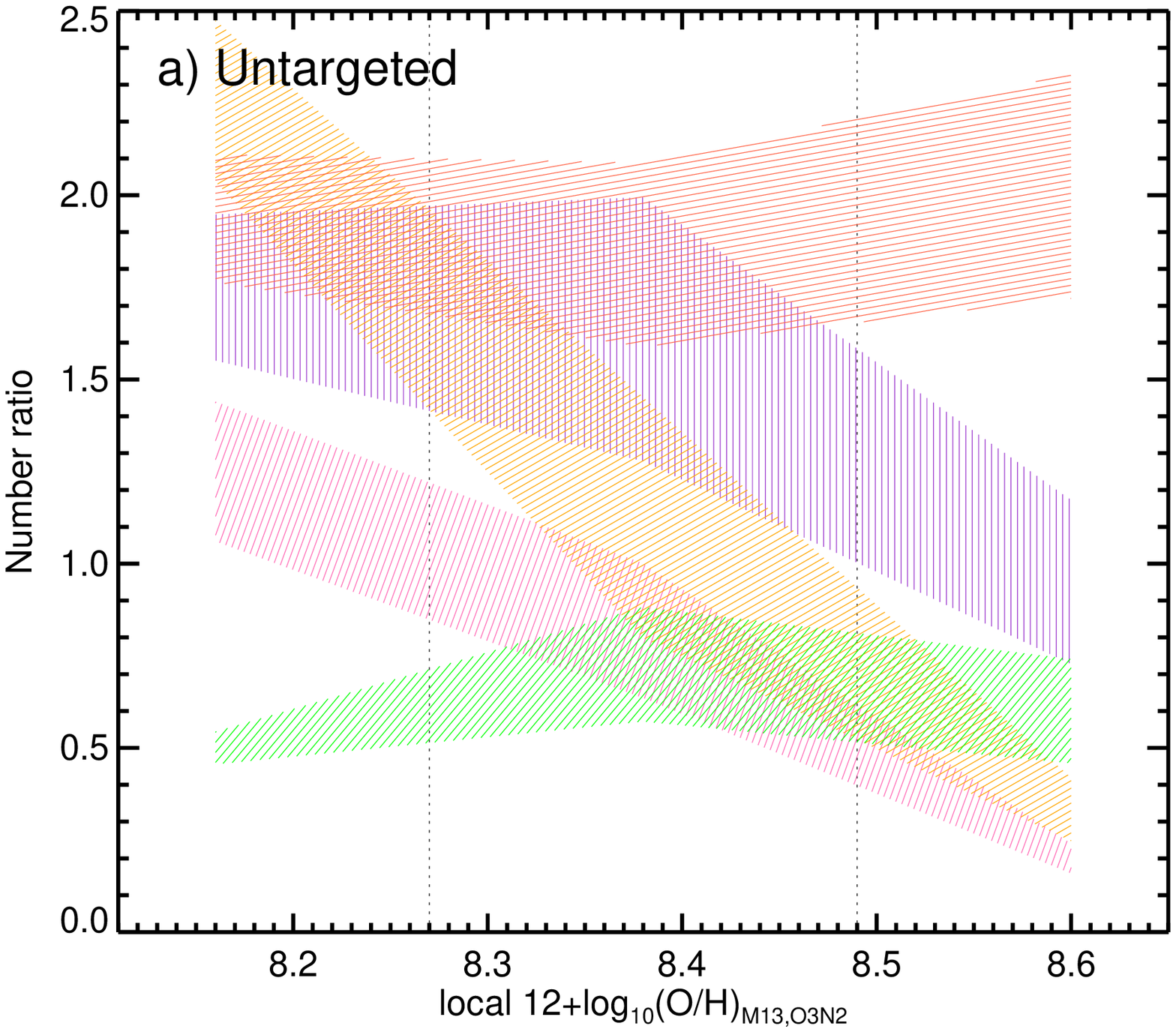}
\includegraphics[trim=1.0cm 0.2cm 0.3cm 0.7cm, clip=true,height=7.8cm]{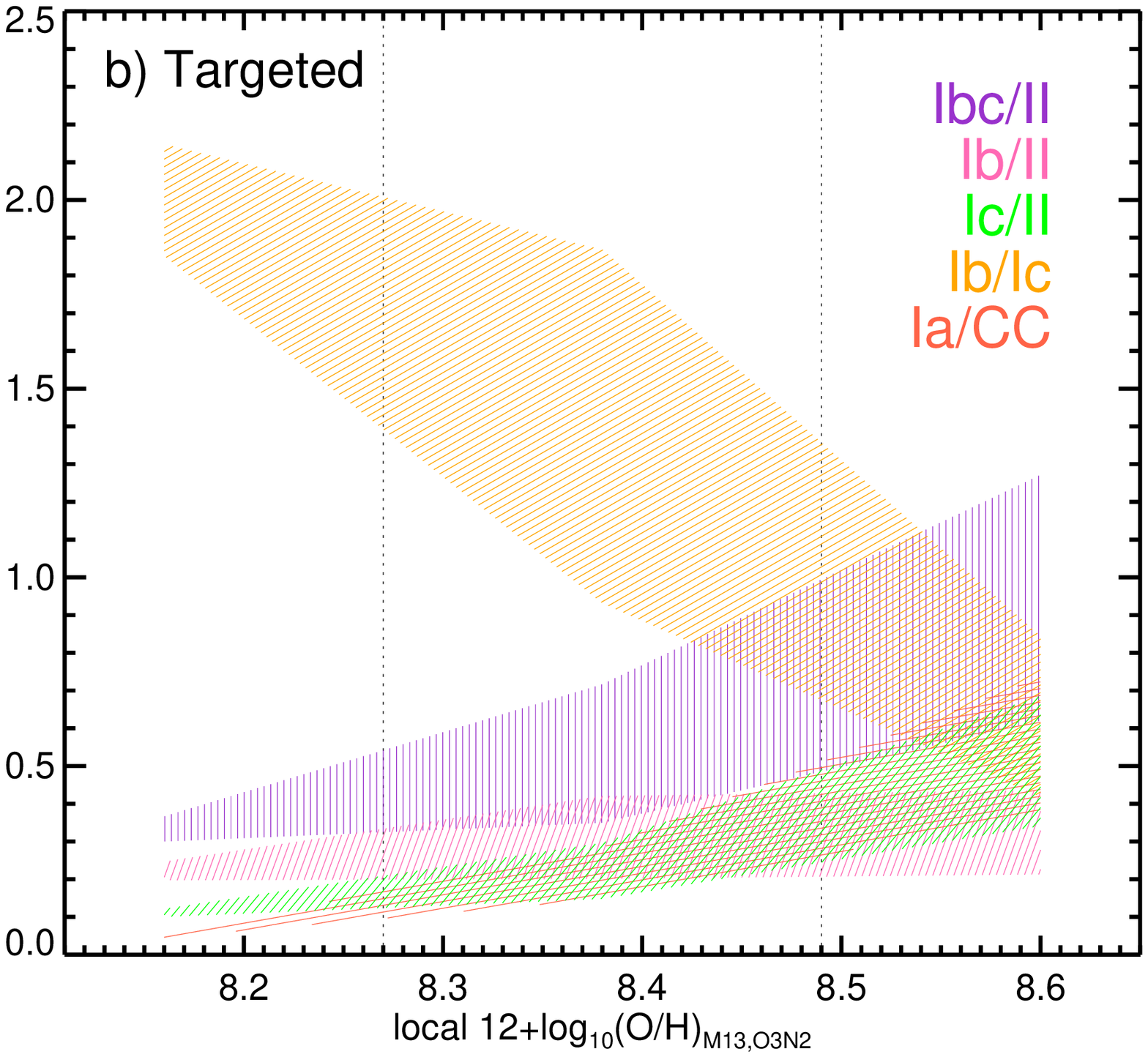}
\caption{
Ratio of SN subtypes as a function of local metallicity from the compilation of the oxygen abundances at the SN location measured in this work combined with previously published from the literature. $a)$ -- for SNe discovered in untargeted searches; $b)$ -- in targeted searches.} 
  \label{fig:ratio}
\end{figure*}

The differences in the mean local metallicites are generally small and in most cases probably unsignificant. KS test indicated that SNe Ic and II probably came from the same metallicity distribution, which is different from that of SN~Ib, IIb and Ic-BL. 

Furthermore, we studied how the number ratios between two subtypes evolve with the environmental metallicity increase.
In Figure \ref{fig:ratio} we present the results of several of these ratios measured in three metallicity bins centered at 8.16 (below solar), 8.38 (around solar), and 8.60~dex (above solar). The analysis is split for untargeted and targeted samples.

In general, the behavior of the ratios is different in the targeted and untargeted samples, with the exception of the Ib/Ic ratio that decreases with metallicity in both samples.
While in the untargeted sample the Ic/II ratio is practically flat with a soft increase to high metallicity, in the targeted sample it clearly increases. Indeed, as we have seen in Figure \ref{fig:others}, there is no significant difference in the untargeted SN~Ic and II distributions, while the difference is higher in the targeted sample.
We also found an increasing rate of targeted SNe Ibc over SNe II as previously found by \cite{2010ApJ...721..777A, 2003A&A...406..259P, 2008ApJ...673..999P} and \cite{2012ApJ...759..107K},
however, when we consider only the untargeted sample this ratio goes in the opposite direction.
By studying the relative rates between the different components we see that this evolution is driven by the decreasing number of SNe Ib at higher metallicities: while the ratio between Ib and II is practically flat for the three regimes in the targeted sample, it shows a steep decrease in the untargeted sample.
Therefore, any difference in the evolution of SN Ibc when compared to SN II is caused by the lower metallicity of SN Ib compared to Ic, in any targeted or untargeted samples.

Also, the proportion of SNe Ia over all CC~SN types increases with the environmental metallicity, as we already found in this and previous sections: although SNe Ia also occur in metal-poor environments, their ratio is significantly higher in the metal-rich regime compared to all CC~SN types.
The increase is more pronounced in the targeted sample, and flatter in the untargeted sample.

$ $\\
\noindent{\it The influence of metallicity in determining CC~SN subtype}\\

Differences in the evolution of SN type ratios with metallicity may be used to connect SN types to possible progenitor scenarios.
Both the wind strength and the binary mass transfer rate are the plausible scenarios through which SNe Ibc lose their external layers.
Another possibility may be that external H and He layers of the progenitor suffer severe depletion by enhanced mixing preventing their detection \citep{2013ApJ...773L...7F}, although this would be difficult to assess from environmental studies.

According to the models based on the evolution of single massive stars (e.g. \citealt{2003ApJ...591..288H,2009A&A...502..611G}) both high metallicities and high initial masses are required to produce SNe Ibc in single stars, while in binary systems the mass loss is dominated by mass transfer and the models predict less significant trend with metallicity.
If the primary SN~Ibc progenitor channel is a single WR star then the rate of SN~Ibc relative to SNe II would be enhanced at higher metallicities.
Although this is what we found when considering SN discovered on targeted searches, the picture gets more complicated in untargeted SN: while the ratio between SN Ibc and II decreases with increasing metallicity, it is driven by the lower local metallicities at SN Ib positions. 
The strongest result we found is that SNe Ic, which are stripped of their H and He external layers, happen on average in significantly higher metallicity environments than SN~Ib, which only lacks of the H envelope.
These SN progenitors are in general short-lived, so there is a higher probability of SN local environmental metallicity to be related to the metallicity of the progenitor.
The ratios presented in Figure \ref{fig:ratio} would imply that at least part of the progenitors of SNe Ic are single stars in strong metallicity driven wind environments, while the remaining SN~Ic and a significant part of SN~Ib and also SN IIb arise from binary progenitors \citep{2011A&A...528A.131C, 2012Sci...337..444S}.
This picture is supported by the results from \cite{2013AJ....146...30K} that inferred lower masses from the stellar popullation ages for SN Ib compared to SN Ic.
In addition, the flat ratio of untargeted SN Ic over SN II would imply that metallicity is not playing a role between these types, and that other factors (e.g. progenitor mass) could be more important.

On the other hand, the radial distributions of different SNe types (see section \ref{sec:locmet}) have been used as a proxy of environmental metallicity.
If central regions of galaxies have higher metallicity, from distance distributions one can argue that the progenitors of SNe Ibc, being more centered, should be metal-richer than those producing other SNe types. 
This argument is consistent with the models claiming that SNe Ibc progenitors are massive stars whose envelopes have been stripped by stellar winds driven by metallicity prior to explosion \citep{1996A&A...305..171P, 2000ARA&A..38..613K, 2007A&A...473..603M}.
Since stars in higher metallicity environments suffer from stronger driven winds, SNe Ibc progenitors lose more mass through these winds prior to explosion, compared to SNe II which are in average further and in lower metallicity environments. 
However, we showed in section \ref{sec:grad} that a significant fraction of SN host galaxies ($\sim$30\%) present a decreasing gradient towards the center which, once accounted for, would imply lower environmental metallicities for objects closer the the center in those galaxies.

These two findings together question the importance of metallicity as the relevant factor in determining the SN type and supports that at least a fraction of them could come from binary systems, both compatible with being at lower radial distances.
This is also in agreement with \cite{2012MNRAS.424.2841H} and \cite{2015PASA...32...19A} who found that SN~Ibc are more centralized compared to SNe II in interacting galaxies, where metallicity gradients are flatter or not present, pointing that other effects are playing a role besides metallicity in determining the SN type.

\begin{table}
\centering
\caption{P-values of the KS test for local distributions presented in Figure \ref{fig:others}.}
\label{tab:ks}                                                                                         
\begin{tabular}{lcc}
\hline\hline   
           &  Targeted        &   Untargeted   \\
\hline         
Ia--II     &  0.0002  &   0.0975   \\      
Ia--Ic     &  0.1434  &   0.6226   \\     
Ia--Ib     &  0.0108  &   0.0037   \\    
Ia--Ibc    &  0.0633  &   0.0420   \\    
Ia--IIb    &  0.0075  &   0.0690   \\   
Ia--Ic-BL  &  0.4416  &   0.0029   \\     
Ia--Ib/IIb &  0.0004  &   0.0008   \\  
Ia--CC     &  0.0001  &   0.0202   \\       
\hline                                         
II--Ic     &  0.0944  &   0.6855   \\    
II--Ib     &  0.7992  &   0.0038   \\     
II--Ibc    &  0.0886  &   0.1209   \\    
II--IIb    &  0.6427  &   0.0964   \\     
II--Ic-BL  &  0.8178  &   0.0068   \\     
II--Ib/IIb &  0.7849  &   0.0034   \\     
\hline                                         
Ic--Ib     &  0.3380  &   0.0622   \\     
Ic--IIb    &  0.0568  &   0.0439   \\    
Ic--Ic-BL  &  0.6309  &   0.0054   \\     
Ic--Ib/IIb &  0.0537  &   0.0400   \\    
\hline                                         
Ib--IIb    &  0.5754  &   0.6524   \\     
Ib--Ic-BL  &  0.6546  &   0.2813   \\     
\hline                                         
IIb--Ic-BL &  0.6342  &   0.1912   \\     
\hline
\end{tabular}
\end{table}

\section{Conclusions} \label{sec:conc}

We present an updated sample of 115 galaxies with mean redshift $z=0.015$ observed with IFS by the CALIFA survey. These galaxies hosted 132 SNe (47 type II, 27 type Ib/c+IIb, 58 Ia) that were within the FoV covered by the instrument.
The spatially resolved spectroscopy used in this work has several advantages with respect to the most common approaches, namely multi-wavelength photometry and fiber or slit spectroscopy. Indeed, IFS allows to obtain 2D maps of the most important parameters from both the stellar populations and the ionized gas emission. 
Following our previous work on the role of the local star-formation in determining the SN type, presented in Paper I, here we have focussed on the study and comparison of environmental metallicity for the various SN types. The relevance of this study is due to the influence of the SN progenitors metallicity on the supernova explosion and possible implications in high redshift supernova studies and determination of cosmological parameters with SN. Our main goals were: 1) unveil possible relations between the host metallicity at the SN location and the properties of the supernova or their progenitors;  2) identify systematic errors of the various proxies commonly used to infer the host metallicity at SN locations or the metallicity of the SN progenitors; and 3) estimate the accuracy in the determination of the local metallicity of the various proxies in high redshift supernova studies.
To achieve a better comparison between galaxies, the radial dependence of galaxy properties was studied in units of the galaxy disk effective radius.  We determined the galaxy disk effective radius by fitting a pure exponential function to their brightness profile and calculated the metallicity gradients for all galaxies. Different proxies to estimate the local metallicity were used, which allowed to search for differences among SN types. It also allowed to compare different proxies and to study their accuracy. 
Our main findings are summarized below.

\begin{itemize}
\item The gas-phase average metallicity at SN location show significant differences of $\sim$0.04~dex between SN Ia and SN II, while the mass-weighted stellar metallicity at the SN location shows no significant difference between the three main SN groups.

\item The total gas-phase metallicity distributions are undistinguishable among the SN groups (centered around 8.49~dex), while the total stellar metallicity distribution for SNe Ia is slightly shifted to higher values than SN II ($\avg{Z_{Ia}}-\avg{Z_{II}}$=0.015).

\item The difference between the local and the total metallicities are on average also small, and being significant only for SN Ia (0.03~dex). This is likely the result of the sample being biased toward galaxies with high masses; most galaxies have $\log(M/M_\sun)>10.3$ dex. 

\item Several proxies of the local metallicity were studied. The most accurate proxies are those using the galaxy metallicity gradient plus the SN galactocentric distance and the metallicity of the closest HII region from the SN. 

\item The total galaxy metallicity is also a good proxy of the local metallicity for CC SNe, although we find a significant shift of $\sim$0.03~dex for SNe Ia. We also find that weak AGNs that cannot be seen in the total spectrum may only weakly bias the metallicity estimate from the galaxy integrated spectrum by 0.03-0.04~dex. These results are encouraging for high-redshift studies as it allows the local metallicity to be estimated, although possible evolution of the metallicity gradient in galaxies with redshift may render this proxy less accurate. 

\item On the other hand we find that the central metallicity, with and without correction for the characteristic metallicity gradient, is the least accurate proxy, showing significant differences for both gas-phase and stellar metallicity. This can be due to the fact that about 30\% of the galaxies show metallicity decrease toward the center, with further 10\% showing just flattening. 

\item The local metallicity at the SN positions is not significantly different from the mean galaxy metallicity at the galactocentric distance of the SN. This argues against that certain SN type may prefer to explode in environments with specific metallicity.

\item When the galaxy masses of our sample are combined with those from the literature it becomes clear that the targeted SN searches are biased toward high-mass galaxies. The average bias is $\log(M/M_\sun)=0.8$.

\item By extending our SN sample with published measurements of the metallicity at the vicinity of the SN, we studied the metallicity distributions for all SN subtypes split into SN discovered in targeted and untargeted searches. We confirm the known fact that there is bias toward high metallicity in the targeted searches.

\item  For the SN discovered in targeted searches, for which our sample is representative, we find that SN~Ia and Ic have the highest metallicity than the other SN types. The difference are however small and in most cases statistically insignificant.

\item In the untargeted searches SN~Ia, Ic and II have higher metallicity comparing to the other SN types. Thus, it appears that SN~Ib/IIb have significantly different local metallicities relative to SN~Ic.

\item Studying the evolution of the ratios between pairs of subtypes as a function of the metallicity, we found that the ratio of Ibc/II decreases with metallicity, and this is driven by SN Ib being at significantly lower metallicities. This gives support to the picture of SN Ib resulting from binary progenitors and (at least some) SN Ic resulting from single stars being stripped of their outer H and He layers by strong metallicity driven winds.
\end{itemize}

The CALIFA Survey has proven the use of Integral Field Spectroscopy as a powerful tool for SN environmental studies, in addition to the main aims of the survey which were the study of galaxy formation and evolution.
The Sydney-AAO Multi-object Integral-field spectrograph (SAMI \citealt{2012MNRAS.421..872C}) and the Mapping Nearby Galaxies at APO (MaNGA, \citep{2015ApJ...798....7B}) are the current largest IFS surveys and will increase the statistics by a factor $\sim$5 to 20.
However, they both lack of high spatial resolution which would be needed to get rid of the systematics effects studied in this work.
\cite{2013AJ....146...30K, 2013AJ....146...31K, 2015PKAS...30..139K} performed a similar study than the one presented here and in Paper I, but studying the SN parent stellar clusters with resolutions around $\sim$60~pc. 
Although this high resolution allows to properly distinguish among stellar clusters, the small field of view does not allow comparisons between local and other estimations of galaxy parameters across the galaxy. 
The advent of the new generation of instruments, such as the Multi-Unit Spectroscopic Explorer (MUSE) Integral Field Unit recently
mounted to the 8m VLT UT4 \citep{2010SPIE.7735E..08B}, would provide the required high spatial resolution together with larger field of view, joining the two requirements (See \citealt{2015A&A...574A..47S,2016MNRAS.455.4087G,2015PASA...32...19A}).
This and other CALIFA studies will be improved with such a combination of these two factors.


\begin{acknowledgements}

We acknowledge the A\&A editor, Rubina Kotak, for her timely and pertinent comments on preliminary versions of this manuscript.
We acknowledge Joseph P. Anderson and Hanindyo Kuncarayakti for fruitful discussions on SN environments (among many other topics).
This work was partly funded by FCT with the research grant PTDC/CTE-AST/112582/2009. 
Support for LG is partially provided by FCT, by CONICYT through FONDECYT grant 3140566, and from the Ministry of Economy, Development, and Tourism's Millennium Science Initiative through grant IC12009, awarded to The Millennium Institute of Astrophysics (MAS). 
V.S. acknowledges financial support from Funda\c{c}\~{a}o para a Ci\^{e}ncia e a Tecnologia (FCT) under program Ci\^{e}ncia 2008. 
CJW acknowledges support through the Marie Curie Career Integration Grant 303912.
This study makes use of the data provided by the Calar Alto Legacy Integral Field Area (CALIFA) survey (\href{http://califa.caha.es}{http://www.caha.es/CALIFA/}). 
CALIFA is the first legacy survey being performed at Calar Alto.
The CALIFA collaboration would like to thank the IAA-CSIC and MPIA-MPG as major partners of the observatory, and CAHA itself, for the unique access to telescope time and support in manpower and infrastructures.  
The CALIFA collaboration thanks also the CAHA staff for the dedication to this project. 
Based on observations collected at the Centro Astron\'omico Hispano Alem\'an (CAHA) at Calar Alto, operated jointly by the Max-Planck Institut f\"ur Astronomie and the Instituto de Astrof\'isica de Andaluc\'ia (CSIC). 
The {\tt STARLIGHT} project is supported by the Brazilian agencies CNPq, CAPES and FAPESP and by the France-Brazil CAPES/Cofecub program. This research has made use of the Asiago Supernova Catalogue, the SIMBAD database, operated at CDS, Strasbourg, France, the NASA/IPAC Extragalactic Database (NED), which is operated by the Jet Propulsion Laboratory, California Institute of Technology, under contract with the National Aeronautics and Space Administration, IAU Circulars presented by the Central Bureau for Astronomical Telegrams, and data products from SDSS and SDSS-II surveys. 
We acknowledge the usage of the HyperLeda database (http://leda.univ-lyon1.fr).
 
\end{acknowledgements}
\bibliographystyle{aa}
\bibliography{IFU_P2}

\appendix

\section{Disc effective radius (r$_\mathrm{e}$)} \label{sec:re}

\begin{figure}
\centering
\includegraphics[trim=0.15cm 0.05cm 0.2cm 0.05cm, clip=true,width=\hsize]{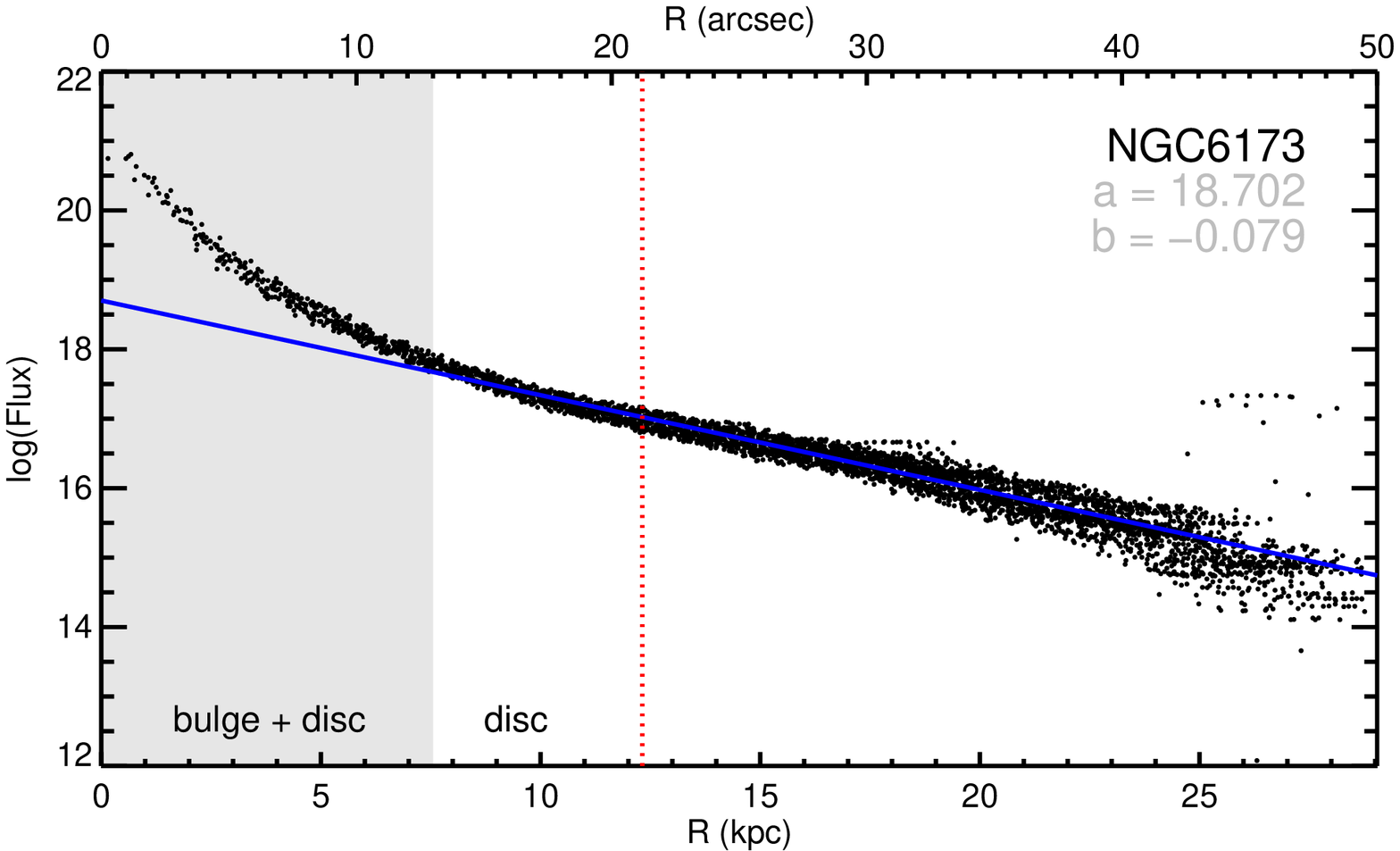}
\includegraphics[trim=0.15cm 0.05cm 0.2cm 0.05cm, clip=true,width=\hsize]{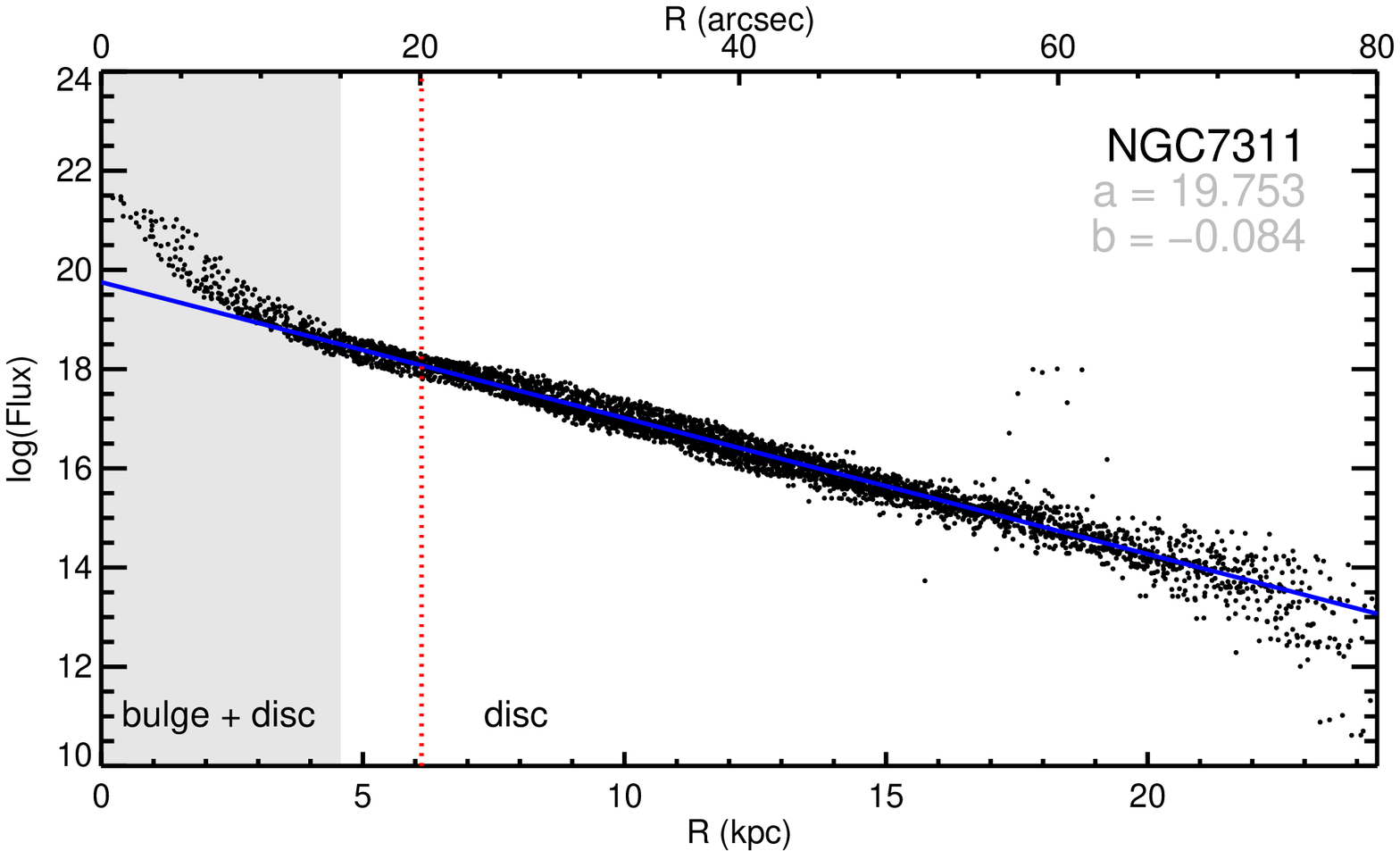}
\caption{Two examples showing the determination of the disk effective radius. The grey shadow defines the region where the bulge contributes to the total surface brightness and it is excluded from the fit which is performed to the disk profile (solid blue line). The red vertical dotted line correspond to the disk effective radius, $r_e$, obtained from the S\'ersic profile expression.}
\label{fig:reff}
\end{figure}

The disk effective radius, $r_\mathrm{e}$, is defined as the radius that contains half of the total integrated flux coming from the disk component.
Its measurement permits to have a common ruler to properly compare the size of the galaxies and the derived quantities at a certain distance. 
To determine $r_\mathrm{e}$ we have performed a morphological decomposition based on the brightness profile.
We integrated the flux under the $r$-band wavelength range in the spectrum of each spaxel, and then plotted the logarithm of these measurements against the deprojected GCD, as shown in Figure \ref{fig:reff} for NGC~6173 and NGC~7311.
One can clearly see the contribution of both the central bulge and the disk to the total brightness of the galaxy. 
This mixture produces a combined surface brightness profile which can be described by an expression that depends on $r^{1/n}$ \citep{1948AnAp...11..247D, 1968adga.book.....S, 2005AJ....130.1535G, 2012ApJ...755..125G}
\begin{eqnarray}
\log I(r)=\log I_0-1.6735~(r/r_\mathrm{e})^{1/n}.
\end{eqnarray}
Depending on the size of the bulge, a different $n$ is needed to properly describe the brightness profile. For example, elliptical galaxies that only have bulge are completely described by the de Vaucouleurs profile $r^{1/4}$ ($n=4$), while pure disk galaxies match with a pure exponential profile ($n=1$).
In order to measure the effective radius of the disk component, we visually determined where the bulge vanishes, and fitted a pure exponential profile to the disk contribution, excluding the central regions dominated by the bulge.
Finally, we obtained $r_\mathrm{e}$ as the distance at which the logarithm of the flux is 1.6735 less than the central value of the fit to the disk profile.
It is worth to note that fitting a line to these profiles is equivalent to obtain elliptical rings from the original image, since the inclination of the galaxies is taken into account in the deprojection. 

We compared the values for the disk effective radii calculated in this work to the values obtained by other works in the CALIFA collaboration. We used two sets of measurements: (i) those measured with a growth curve photometry analysis presented in \cite{2014A&A...569A...1W}, and (ii) those measured in \cite{2014A&A...563A..49S}, that used a similar procedure to the one presented here.
For the galaxies in common (basically those from the CALIFA mother sample), we obtained similar values (within 10\%) for almost all the galaxies. For six galaxies we obtained larger differences but, after repeating the analysis without those, the final results in terms of the common slope, do not change significantly when using  \cite{2014A&A...569A...1W} or  \cite{2014A&A...563A..49S} values.

\section{Characteristic metallicity gradient} \label{app:metgrad}

\begin{figure*}
\centering
\includegraphics[trim=1.1cm 0.4cm 0.5cm 0.8cm, clip=true,width=0.49\hsize]{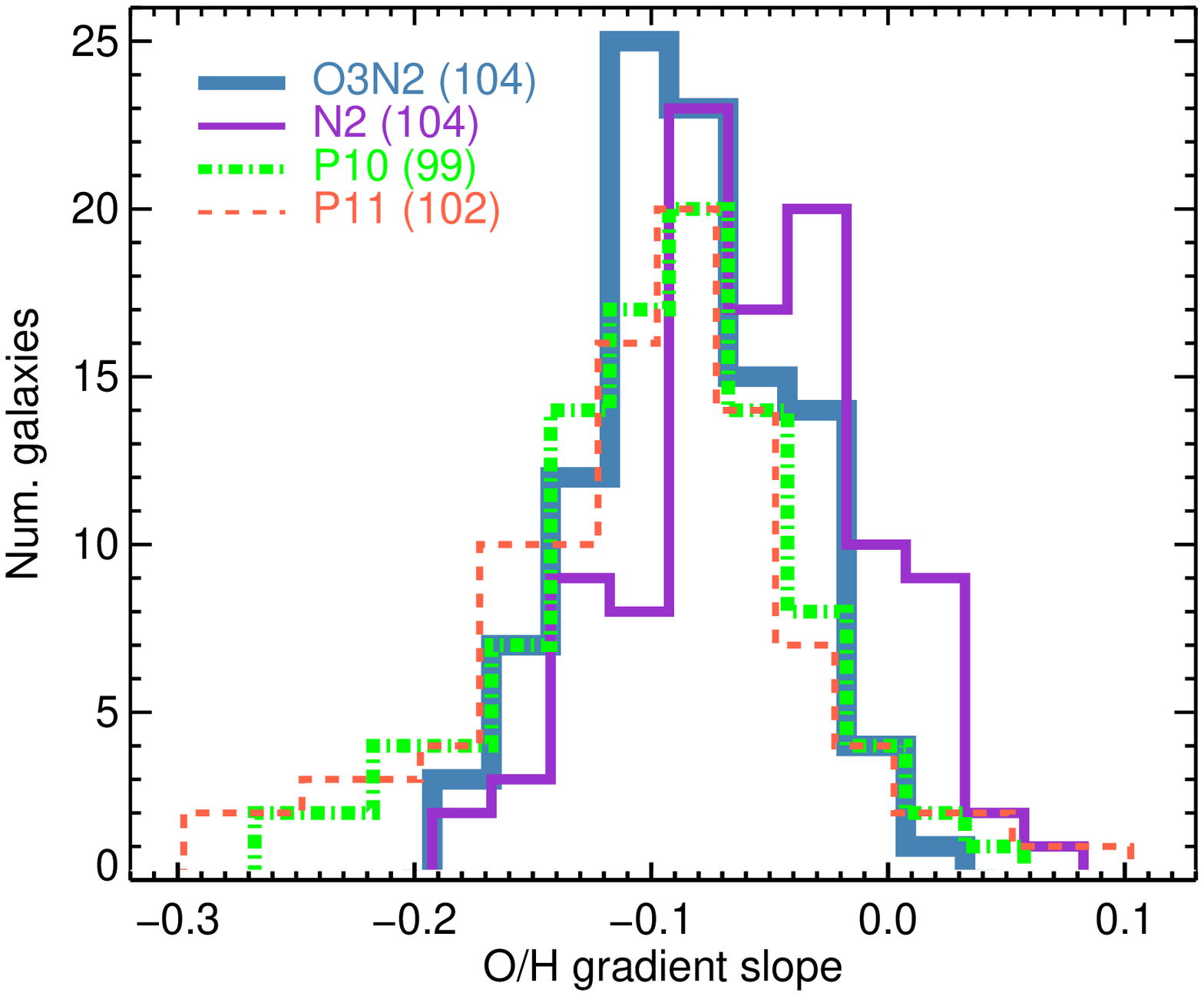}
\includegraphics[trim=1.1cm 0.4cm 0.5cm 0.8cm, clip=true,width=0.49\hsize]{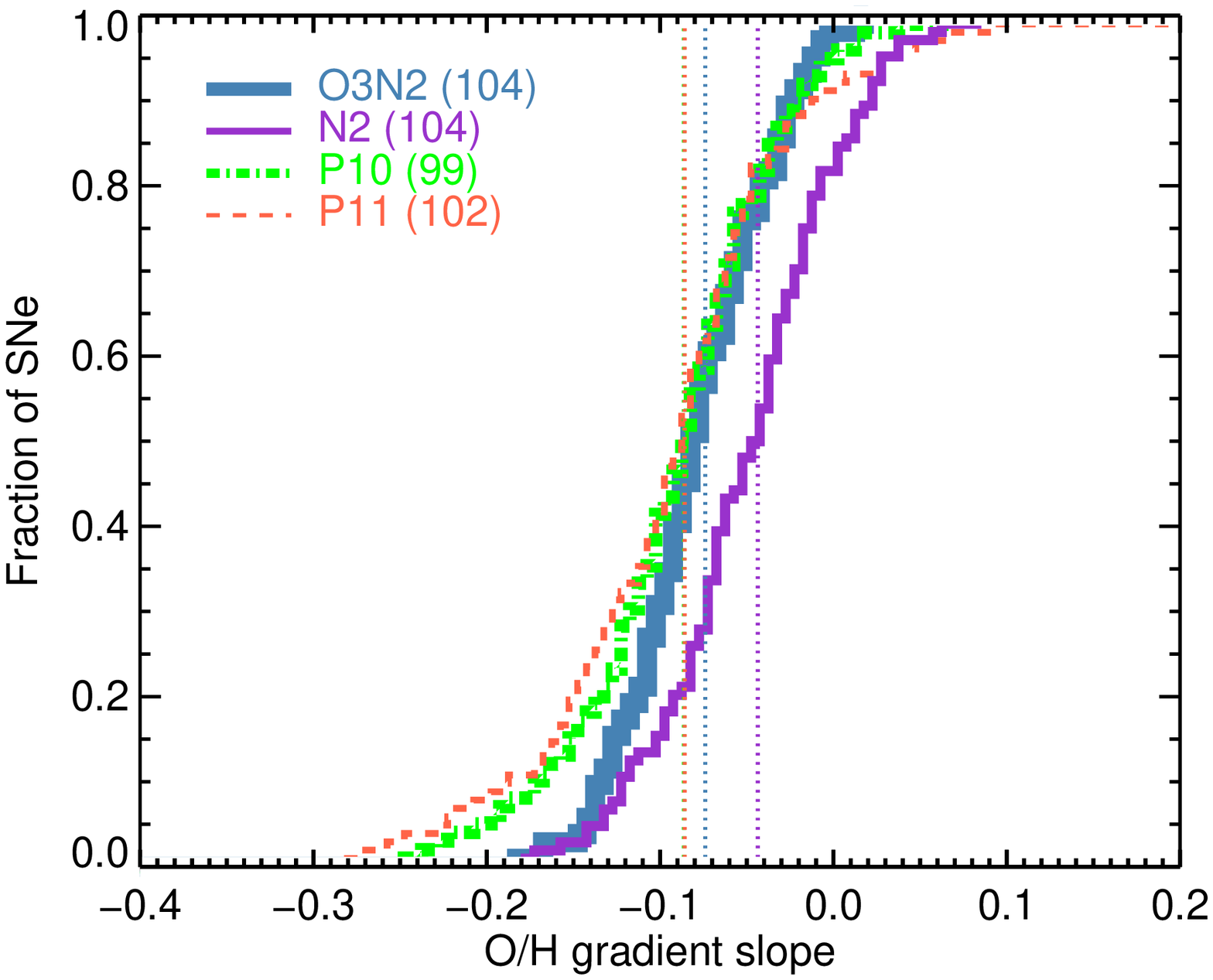}
\caption{Distribution of metallicity gradients using O3N2 (blue), N2 (purple), P10 (green), and P11 (red) methods. All fits were done considering the errors in each spaxel measurement.}
\label{fig:metal6}
\end{figure*}

In this appendix we present the distributions of slopes of the metallicity gradients for our SN host galaxy sample.
The central decrease was excluded in the metallicity gradient fits, resulting in different ranges for each galaxy up to 2 disc effective radius. The mean metallicity gradient of all 104 galaxies is $-0.016$\,($\pm0.013$)~dex\,kpc$^{-1}$ with the uncertainty being the standard deviation of the distribution.
When the distances are normalized to the disk effective radius $r_\mathrm{e}$, the average gradient becomes $-0.074$\,($\pm0.042$)~dex\,$r^{-1}_\mathrm{e}$ (using the O3N2 calibration from M13). The distribution of the normalized gradients is shown in Fig.~\ref{fig:metal6}. Note that the relative scatter of the normalized gradients is smaller than in physical units. In addition, the scatter is fairly low, suggesting a characteristic radial gradient for the oxygen abundance in normalized units.
This is in agreement with the results presented by \cite{2012A&A...546A...2S}, \cite{2014A&A...563A..49S} and \cite{LSM}, who reported an average gradient of $-$0.1~dex\,$r^{-1}_\mathrm{e}$ using the PP04 calibration (we find $-$0.105~dex\,$r^{-1}_\mathrm{e}$ with that calibration).

The statistics of the distributions using the main calibrator and the comparison methods are presented in Table~\ref{tab:slopes}. The mean values of the distributions of the metallicity gradients for P10 and P11 strong line methods are around $-0.08$~dex\,$r^{-1}_\mathrm{e}$.
Five galaxies (UGC~01087, UGC~04036, UGC~4107, NGC~6643, and NGC~3913 have no \OII~line measurement and no P10 abundance estimation can be computed, while IC~0307 and NGC~6166 have no \SII~lines measured and no P11 measurements can be performed.       
As expected, the distribution of metallicity gradients measured using N2 calibrator has lower values. This calibrator is not well defined in the high metallicity regime and the range of allowed values is narrower, thus flattening the measured gradients.
We note that although not significant, SNe Ia host galaxies show flatter gradients than both CC~SN hosts. 
Also noticeable is the narrower distribution of the O3N2 distribution. 

\begin{table*}
\centering
\caption{Average values of the metallicity gradient slopes and the standard deviation of the distributions.}
\label{tab:slopes}                                                                                        
\begin{tabular}{lcccccccc}
\hline\hline  
 & \multicolumn{2}{c}{O3N2} & \multicolumn{2}{c}{N2} & \multicolumn{2}{c}{P10} & \multicolumn{2}{c}{P11}\\
 & \# & avg.&\# & avg.&\# & avg.&\# & avg.\\
\hline    
All Galaxies & 104 & $-$0.074  (0.042) & 104 & $-$0.044  (0.052)  &99 &  $-$0.086   (0.059) & 102 & $-$0.086 (0.087) \\
\hline
SN~Ia hosts    & 46 & $-$0.064  (0.041) & 46 & $-$0 046  (0.050)  & 43 & $-$0.077   (0.064) & 46 & $-$0.077 (0.117) \\
SN~Ibc hosts   & 27 & $-$0.078  (0.040) & 27 & $-$0.040  (0.053)  & 25 & $-$0.087   (0.050) & 26 & $-$0.094 (0.049) \\
SN~II hosts   & 47 & $-$0.087  (0.037) & 47 & $-$0.037  (0.052)  & 46 & $-$0.091   (0.049) & 47 & $-$0.090 (0.050) \\
\hline
\end{tabular}
\end{table*}

\section{The effect of AGNs on total spectra}

The central spectra of about 29\% of CC~SN hosts, 59\% of the star-forming, and all passive SN~Ia hosts (see section \ref{sec:galnuc}) in our sample indicate the presence of AGN according to  the \cite{2001ApJ...556..121K} criterion in the BPT diagram\footnote{\cite{2012A&A...540A..11K}, \cite{2013A&A...558A..43S} and \cite{2013A&A...555L...1P} using CALIFA data have shown that in some galaxies narrow line emission that falls in the AGN section of the BPT diagram, is not due to AGN activity but from old post-AGB stars. } \citep{1981PASP...93....5B, 1987ApJS...63..295V}. The emission line measurements from the total spectra of most star-forming galaxies with AGN  fall in the star-forming region of the BPT diagram. This indicates that the AGNs are not strong enough to affect significantly the total spectrum. The passive galaxies are exception because most of the emission comes from the center and when AGN is present, its emission may dominate the total emission line spectrum. For this reason the passive galaxies were not considered in the analysis of the effect of AGNs.

Even though the AGNs in our galaxy sample are not particularly strong, their presence can still affect the metallicity estimates  and bias the results from the total spectra. To study this, for each cube two different extractions of the total spectrum were done. First, we just summed up the flux in all the spaxels with a S/N greater than 1.0. As an alternative, we did the same but excluding  those central spaxels that fall in the AGN region of the so-called BPT diagnostic diagram .
The first spectrum would correspond to the spectrum of a galaxy at high redshift or a small galaxy at lower redshift when the whole galaxy falls into the slit or the fiber of the spectrograph. 

\begin{figure}[!t]
\centering
\includegraphics*[trim=1.3cm 0.2cm 0.35cm 0.9cm, clip=true,width=\hsize]{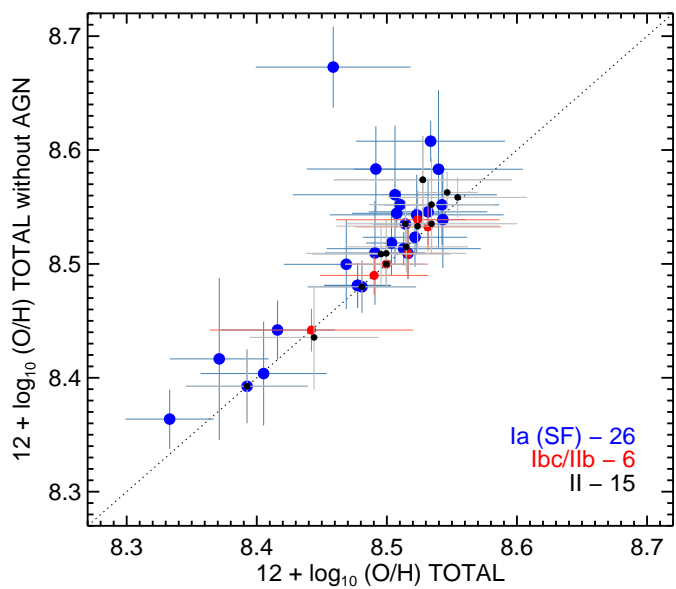}
\caption{Comparison between the average oxygen abundance with and without the AGN contribution.}
  \label{fig:total-agn}
\end{figure}

Figure~\ref{fig:total-agn} shows gas-phase metallicity estimated from the total spectra with and without including the central spaxels falling in the AGN region.
All values are close or above the diagonal, indicating that the metallicity measured from a spectrum including the AGN contribution is usually lower than the measurement excluding it. 
In general, there is a good agreement between the measurements from the two total spectra for the CC~SN hosts. However, for the SN~Ia host galaxies the abundances are on average higher by 0.03--0.04~dex when the AGN contribution is removed. 
This difference raises a potential problem in the metallicity estimation from spectra of high redshift SN~Ia hosts since the presence of a weak AGN contribution might be missed.  
Thus, the  metallicity measured from total galaxy spectra should be considered as a lower limit with a systematic error of at least 0.03 dex.
However, concerning the mean total gas-phase metallicity of the galaxies in our sample, the AGNs have little effect. When using the measurement with AGN contribution removed the mean metallicity of the SN~Ia hosts becomes larger by only 0.012~dex and no change is observed for CC~SN hosts.

\section{Measured quantities}
\begin{table*}\tiny
\caption{Parameters of the CALIFA galaxies that hosted SNe outside the instrument field of view.}
\label{tab:white}
\begin{center}
\begin{tabular}{@{}llccccccccccccc@{}}
\hline\hline                                                                                                     
Galaxy   & SN name&SN type& AGN &    z     &OH total&log (Z$_m$ / Z$_\sun$)&     Mass  \\
         &        &       &     &          &        &                      &[M$_\sun$] \\
\hline                                                                             
NGC3184  & 1921C  & I     & $-$ & 0.001975 & 8.82   & $-$0.02 &  10.30   \\  
NGC0477  & 2002jy & Ia    & $-$ & 0.019600 & 8.65   & $-$0.03 &  10.94   \\  
NGC2487  & 1975O  & Ia    & +   & 0.016148 & 8.71   & $-$0.05 &  11.14   \\  
NGC2513  & 2010ja & Ia    & +   & 0.015561 & $\dots$&    0.12 &  11.53   \\  
NGC3160  & 1997C  & Ia    & +   & 0.023083 & $\dots$& $-$0.06 &  11.07   \\  
NGC4185  & 1982C  & Ia    & $-$ & 0.013022 & 8.73   &    0.02 &  11.01   \\  
NGC5485  & 1982W  & Ia    & +   & 0.006428 & $\dots$&    0.10 &  10.90   \\  
NGC5532  & 2007ao & Ia    & +   & 0.024704 & $\dots$&    0.11 &  11.94   \\  
NGC5557  & 2013gn & Ia    & +   & 0.010737 & $\dots$&    0.16 &  11.48   \\  
IC1151   & 1991M  & Ia    & $-$ & 0.007235 & 8.53   & $-$0.38 &  10.02   \\  
NGC6027  & 2006ay & Ia    & $-$ & 0.014834 & 8.64   & $-$0.03 &  11.08   \\  
NGC6411  & 1999da & Ia    & +   & 0.012695 & $\dots$&    0.04 &  11.15   \\  
NGC7549  & 2009nq & Ia    & $-$ & 0.015799 & 8.74   & $-$0.22 &  10.78   \\  
NGC7782  & 2003gl & Ia    & +   & 0.017942 & 8.67   &    0.21 &  11.44   \\  
NGC3913  & 1979B  & Ia    & $-$ & 0.003182 & 8.73   & $-$0.47 &   9.27   \\  
\hline                                                                       %
NGC0171  & 2009hf & IIP   & +   & 0.013043 & 8.69   &    0.09 &  10.98   \\  
NGC0180  & 2001dj & IIpec & +   & 0.017616 & 8.81   &    0.02 &  11.12   \\  
NGC0309  & 2012dt & II    & c   & 0.018886 & 8.82   &    0.05 &  11.19   \\  
NGC0941  & 2005ad & IIP   & $-$ & 0.005364 & 8.58   & $-$0.34 &   9.76   \\  
UGC10710 & 2000bs & IIP   & $-$ & 0.027976 & 8.65   &    0.02 &  10.87   \\  
NGC1058  & 1969L  & IIP   & +   & 0.001728 & 8.68   & $-$0.41 &   9.11   \\  
NGC3184  & 1937F  & IIP   & $-$ & 0.001975 & 8.82   & $-$0.02 &  10.30   \\  
NGC3184  & 1921B  & II    & $-$ & 0.001975 & 8.82   & $-$0.02 &  10.30   \\  
\hline                                                                       %
NGC0214  & 2006ep & Ib    & +   & 0.015134 & 8.73   &    0.04 &  11.30   \\  
UGC04308 & 1962F  & IIb   & $-$ & 0.011895 & 8.69   & $-$0.17 &  10.68   \\  
NGC6060  & 1997dd & IIb   & $-$ & 0.014807 & 8.67   &    0.04 &  11.30   \\  
NGC0309  & 2008cx & IIb   & c   & 0.018886 & 8.82   &    0.05 &  11.19   \\  
NGC0628  & 2002ap & Ic-bl & c   & 0.002192 & 8.69   & $-$0.49 &  10.30   \\  
\hline\hline
\end{tabular}
\end{center}
\end{table*}
\begin{table*}\tiny
\caption{Type Ia SNe results.}
\label{tab:res1}
\begin{center}
\begin{tabular}{@{}llcccccccccccccc@{}}
\hline\hline                                                                                                     
       &           &         &      &     &  &  &   \multicolumn{3}{c}{12 + log (O / H)}  &  & \multicolumn{2}{c}{log (Z$_m$ / Z$_\sun$)} \\
\cline{8-10} \cline{12-13} 
\multicolumn{1}{c}{Galaxy}&\multicolumn{1}{c}{SN name}&\multicolumn{1}{c}{SN type}&\multicolumn{1}{c}{AGN}&\multicolumn{1}{c}{$r_\mathrm{e}$}&\multicolumn{1}{c}{dep-GCD}&\multicolumn{1}{c}{Mass}&  Gradient  & Local& Total  & & Local & Total \\
       &           &         &      & [kpc] & [kpc] & [M$_\sun$] &  [dex\,$r^{-1}_\mathrm{e}$] & & & & &\\
\hline
NGC 2565       & 1960M        & I      &   + &    4.8 &    9.4 &  11.11 & $-$0.082 & 8.46    & 8.48    & & 0.04    & $-$0.04 \\
NGC 7319       & 1971P        & I      &   + &   11.1 &   14.7 &  11.32 & $-$0.067 & $\dots$ & 8.37    & & $\dots$ & $-$0.04 \\
\hline
NGC 5668       & 1954B        & Ia     &   + &    3.4 &    2.2 &  11.38 & $-$0.108 & 8.42    & 8.39    & & $-$0.07 & $-$0.10 \\
NGC 3913       & 1963J        & Ia     & $-$ &    1.4 &    0.9 &   9.27 & $-$0.071 & 8.54    & 8.53    & & $\dots$ & $-$0.06 \\
NGC 3811       & 1969C        & Ia     & $-$ &    3.9 &    2.5 &  10.79 & $-$0.118 & 8.54    & 8.51    & & 0.09    & 0.18    \\
NGC 7619       & 1970J        & Ia     &   + &    5.3 &   10.5 &  11.52 & $-$0.056 & $\dots$ & $\dots$ & & $\dots$ & $-$0.40 \\
NGC 4874       & 1981G        & Ia     &   + &   11.9 &    9.2 &  11.69 &    0.013 & $\dots$ & $\dots$ & & 0.01    & 0.10    \\
NGC 1667       & 1986N        & Ia     &   + &    5.6 &    6.8 &  11.29 & $-$0.078 & 8.56    & 8.51    & & $\dots$ & $-$0.23 \\
NGC 3687       & 1989A        & Ia     &   + &    3.1 &    4.7 &  10.43 & $-$0.084 & 8.47    & 8.51    & & $\dots$ & 0.05    \\
NGC 5378       & 1991ak       & Ia     &   + &    6.6 &    8.1 &  10.78 & $-$0.037 & 8.50    & 8.47    & & $-$0.45 & $-$0.51 \\
NGC 0932       & 1992bf       & Ia     &   + &    7.1 &    4.1 &  11.26 & $-$0.018 & 8.42    & 8.51    & & 0.12    & $-$0.26 \\
UGC 04036      & 1995E        & Ia     &   + &    9.5 &    5.4 &  10.94 & $-$0.089 & 8.57    & 8.50    & & 0.15    & $-$0.07 \\
NGC 5157       & 1995L        & Ia     &   + &   10.7 &    9.4 &  11.33 & $-$0.003 & 8.54    & 8.49    & & 0.16    & $-$0.05 \\
NGC 5557       & 1996aa       & Ia     &   + &    4.8 &    1.4 &  11.48 & $-$0.114 & $\dots$ & $\dots$ & & $\dots$ & $-$0.07 \\
NGC 3982       & 1998aq       & Ia     &   + &    1.2 &    1.6 &  10.82 & $-$0.068 & 8.56    & 8.51    & & $-$0.08 & 0.22    \\
UGC 00139      & 1998dk       & Ia     & $-$ &    5.5 &    1.8 &  10.18 & $-$0.136 & 8.52    & 8.40    & & 0.18    & 0.10    \\
UGC 01087      & 1999dk       & Ia     &   c &    5.8 &    8.2 &  10.65 & $-$0.135 & 8.47    & 8.49    & & $\dots$ & $-$0.08 \\
NGC 0495       & 1999ej       & Ia     &   + &    4.1 &    7.5 &  10.84 & $-$0.019 & $\dots$ & $\dots$ & & 0.22    & $-$0.02 \\
NGC 2623       & 1999gd       & Ia     &   + &    4.2 &    6.9 &  10.92 &    0.018 & 8.52    & 8.49    & & 0.22    & 0.11    \\
UGC 04195      & 2000ce       & Ia     & $-$ &    6.4 &    8.0 &  10.85 & $-$0.091 & 8.54    & 8.50    & & $\dots$ & $-$0.06 \\
NGC 0523       & 2001en       & Ia     &   c &    7.4 &    8.3 &  10.78 & $-$0.090 & 8.47    & 8.45    & & $\dots$ & $-$0.13 \\
UGC 05129      & 2001fe       & Ia     &   + &    3.8 &    3.6 &  10.63 & $-$0.075 & 8.58    & 8.52    & & $\dots$ & $-$0.41 \\
2MFGC 13321    & 2002aw       & Ia     &   + &    4.8 &    3.2 &  10.76 & $-$0.053 & 8.48    & 8.54    & & $\dots$ & $-$0.06 \\
NGC 7253       & 2002jg       & Ia     & $-$ &   13.1 &    7.3 &  10.80 &    0.003 & 8.54    & 8.55    & & 0.18    & 0.05    \\
MCG -02-02-086 & 2003ic       & Ia     &   + &   18.4 &   10.9 &  11.91 & $-$0.130 & $\dots$ & $\dots$ & & $-$0.11 & $-$0.31 \\
UGC 00005      & 2003lq       & Ia     &   + &   11.0 &   19.2 &  11.27 & $-$0.101 & $\dots$ & 8.51    & & 0.13    & 0.07    \\
UGC 10097      & 2004di       & Ia     &   + &   11.4 &   22.0 &  11.32 & $-$0.035 & $\dots$ & 8.46    & & 0.11    & 0.06    \\
NGC 1060       & 2004fd       & Ia     &   + &    7.0 &    2.1 &  11.86 & $-$0.023 & $\dots$ & $\dots$ & & $-$0.46 & $-$0.26 \\
MCG -01-09-006 & 2005eq       & Ia     &   c &   10.7 &   21.7 &  11.24 & $-$0.101 & 8.47    & 8.49    & & $-$0.25 & $-$0.19 \\
NGC 7311       & 2005kc       & Ia     &   + &    6.1 &    5.9 &  10.77 & $-$0.089 & 8.58    & 8.52    & & $-$0.03 & 0.04    \\
UGC 04468      & 2006bb       & Ia     &   + &   20.2 &   14.8 &  11.34 &    0.030 & $\dots$ & $\dots$ & & $-$0.02 & 0.11    \\
NGC 5587       & 2006dy       & Ia     &   + &    3.7 &    6.0 &  10.58 & $-$0.024 & 8.53    & 8.53    & & 0.16    & $-$0.07 \\
CGCG 207-042   & 2006te       & Ia     & $-$ &    6.5 &    6.3 &  10.50 & $-$0.075 & 8.56    & 8.46    & & $\dots$ & $-$0.41 \\
NGC 0105       & 2007A        & Ia     &   + &    4.2 &    3.6 &  10.88 & $-$0.106 & 8.60    & 8.48    & & $\dots$ & $-$0.49 \\
UGC 04008      & 2007R        & Ia     &   c &   10.4 &    3.0 &  11.37 & $-$0.026 & 8.56    & 8.55    & & $\dots$ & $-$0.49 \\
NGC 2577       & 2007ax       & Ia     &   + &    3.3 &    1.5 &  10.88 & $-$0.063 & $\dots$ & $\dots$ & & $\dots$ & $-$0.51 \\
UGC 04455      & 2007bd       & Ia     &   + &    8.2 &    6.3 &  11.15 & $-$0.053 & 8.58    & 8.54    & & $-$0.53 & $-$0.65 \\
NGC 7469       & 2008ec       & Ia     &   + &    3.7 &    5.2 &  11.36 & $-$0.001 & 8.48    & 8.42    & & 0.21    & 0.18    \\
NGC 6166       & 2009eu       & Ia     &   + &   11.5 &   20.9 &  11.37 & $-$0.008 & $\dots$ & $\dots$ & & 0.15    & 0.08    \\
NGC 5525       & 2009gt       & Ia     & $-$ &   12.2 &   28.1 &  11.33 & $-$0.112 & $\dots$ & 8.50    & & $\dots$ & $-$0.33 \\
NGC 6146       & 2009fl       & Ia     &   + &    8.8 &    7.7 &  11.62 & $-$0.049 & $\dots$ & $\dots$ & & $-$0.62 & $-$0.43 \\
NGC 6173       & 2009fv       & Ia     &   + &   12.3 &    5.8 &  11.86 & $-$0.011 & $\dots$ & $\dots$ & & $-$0.01 & 0.14    \\
NGC 7364       & 2009fk       & Ia     &   c &    4.9 &    2.6 &  10.80 & $-$0.018 & 8.58    & 8.57    & & 0.02    & 0.04    \\
UGC 11975      & 2011fs       & Ia     & $-$ &    5.2 &   16.8 &  10.81 & $-$0.045 & $\dots$ & 8.56    & & $\dots$ & 0.05    \\
NGC 7364       & 2011im       & Ia     &   c &    4.9 &    9.7 &  10.80 & $-$0.018 & 8.57    & 8.57    & & $\dots$ & 0.16    \\
NGC 5421       & 2012T        & Ia     &   c &    6.8 &    3.0 &  11.33 & $-$0.049 & 8.52    & 8.51    & & $-$0.09 & 0.08    \\
NGC 5611       & 2012ei       & Ia     &   + &    2.2 &    2.1 &  10.38 & $-$0.076 & $\dots$ & $\dots$ & & 0.09    & 0.09    \\
UGC 08250      & 2013T        & Ia     & $-$ &    7.3 &   12.6 &  10.31 & $-$0.064 & 8.39    & 8.39    & & $\dots$ & $-$0.16 \\
NGC 7321       & 2013di       & Ia     &   + &    7.6 &   12.2 &  10.73 & $-$0.103 & 8.53    & 8.52    & & 0.12    & 0.04    \\
NGC 2554       & 2013gq       & Ia     &   + &    6.0 &    2.6 &  11.39 & $-$0.009 & 8.50    & 8.41    & & $\dots$ & 0.03    \\
NGC 6166       & PS15aot      & Ia     &   + &   11.5 &    5.4 &  11.37 & $-$0.008 & $\dots$ & $\dots$ & & $\dots$ & 0.14    \\
NGC 0938       & 2015ab       & Ia     &   + &    5.1 &    3.8 &  11.12 & $-$0.013 & 8.43    & 8.34    & & $\dots$ & $-$0.33 \\
\hline
UGC 03151      & 1995bd       & Iapec  &   + &    6.0 &    7.1 &  11.06 & $-$0.011 & 8.59    & 8.53    & & 0.16    & 0.07    \\
NGC 0105       & 1997cw       & Iapec  &   + &    4.2 &    3.5 &  10.88 & $-$0.106 & 8.56    & 8.48    & & 0.03    & 0.15    \\
NGC 2595       & 1999aa       & Iapec  & $-$ &    9.1 &    8.2 &  11.18 & $-$0.038 & 8.45    & 8.51    & & 0.19    & 0.17    \\
NGC 6063       & 1999ac       & Iapec  &   c &    5.1 &    7.7 &  10.39 & $-$0.120 & $\dots$ & 8.48    & & $-$0.11 & 0.13    \\
NGC 0976       & 1999dq       & Iapec  &   + &    7.5 &    2.6 &  11.15 & $-$0.030 & 8.62    & 8.54    & & 0.02    & 0.07    \\
NGC 2691       & 2011hr       & Iapec  &   c &    5.7 &    2.1 &  10.68 & $-$0.055 & 8.56    & 8.49    & & 0.16    & 0.04    \\
\hline\hline
\end{tabular}
\end{center}
\end{table*}

\begin{table*}\tiny
\caption{Type II SNe results.}
\label{tab:res2}
\begin{center}
\begin{tabular}{@{}llcccccccccccccc@{}}
\hline\hline                                                                                                     
\multicolumn{1}{c}{Galaxy}&\multicolumn{1}{c}{SN}&\multicolumn{1}{c}{SN type}&\multicolumn{1}{c}{AGN}&\multicolumn{1}{c}{r$_{eff}$}&\multicolumn{1}{c}{dep-GCD}&\multicolumn{1}{c}{Mass}&\multicolumn{3}{c}{12 + log (O/H)}  &  & \multicolumn{2}{c}{Z$_m$} \\
\cline{8-10} \cline{12-13} 
       &           &         &      &     &  &  &      grad  & local& total  & & local & total \\
\hline
NGC 3811       & 1971K        & II     & $-$ &    3.9 &    7.6 &  10.79 & $-$0.118 & 8.50    & 8.51    & & $-$0.16 & $-$0.40 \\
NGC 2565       & 1992I        & II     &   + &    4.8 &   13.3 &  11.11 & $-$0.082 & 8.60    & 8.48    & & $-$0.01 & 0.10    \\
NGC 3057       & 1997cx       & II     & $-$ &    3.6 &    1.7 &   9.51 & $-$0.095 & 8.33    & 8.28    & & 0.09    & $-$0.06 \\
NGC 2916       & 1998ar       & II     &   + &    7.8 &   10.8 &  11.07 & $-$0.135 & $\dots$ & 8.50    & & $\dots$ & $-$0.39 \\
UGC 03555      & 1999ed       & II     & $-$ &    4.7 &    6.4 &  11.12 & $-$0.134 & 8.50    & 8.46    & & 0.09    & 0.10    \\
NGC 0309       & 1999ge       & II     &   c &   12.0 &    6.8 &  11.19 & $-$0.087 & 8.70    & 8.60    & & $-$0.03 & 0.04    \\
UGC 05520      & 2000L        & II     & $-$ &    5.3 &    6.6 &  10.03 & $-$0.180 & 8.33    & 8.37    & & 0.15    & 0.06    \\
UGC 00005      & 2000da       & II     &   + &   11.0 &    7.7 &  11.27 & $-$0.101 & 8.61    & 8.51    & & 0.05    & $-$0.21 \\
MCG -01-10-019 & 2001H        & II     & $-$ &    8.8 &    3.2 &  10.40 & $-$0.123 & 8.56    & 8.46    & & 0.04    & 0.06    \\
NGC 2604       & 2002ce       & II     & $-$ &    3.4 &    3.0 &  10.05 & $-$0.092 & 8.33    & 8.38    & & 0.07    & 0.22    \\
NGC 7771       & 2003hg       & II     & $-$ &    3.7 &    3.6 &  11.28 & $-$0.055 & 8.61    & 8.50    & & $-$0.40 & 0.01    \\
UGC 00148      & 2003ld       & II     & $-$ &    7.5 &    2.5 &  10.24 & $-$0.040 & 8.43    & 8.40    & & $-$0.67 & $-$0.63 \\
NGC 5668       & 2004G        & II     &   + &    3.4 &    5.6 &  11.38 & $-$0.108 & $\dots$ & 8.39    & & $-$0.04 & 0.10    \\
NGC 5980       & 2004ci       & II     &   c &    5.5 &    6.1 &  11.12 & $-$0.120 & 8.56    & 8.53    & & $-$0.47 & $-$0.09 \\
NGC 6786       & 2004ed       & II     & $-$ &    5.2 &    5.3 &  11.54 & $-$0.089 & 8.55    & 8.55    & & 0.10    & 0.10    \\
NGC 5056       & 2005au       & II     &   c &    6.6 &    8.1 &  11.00 & $-$0.122 & 8.43    & 8.45    & & 0.14    & 0.13    \\
NGC 5682       & 2005ci       & II     & $-$ &    3.4 &    1.9 &   9.54 & $-$0.083 & 8.32    & 8.35    & & $\dots$ & 0.06    \\
NGC 5630       & 2005dp       & II     & $-$ &    3.3 &    4.8 &   9.89 & $-$0.057 & 8.35    & 8.32    & & $\dots$ & 0.03    \\
UGC 04132      & 2005en       & II     & $-$ &    8.1 &    6.8 &  11.34 & $-$0.104 & 8.46    & 8.51    & & 0.07    & 0.13    \\
NGC 0774       & 2006ee       & II     &   + &    7.8 &    4.8 &  10.42 & $-$0.161 & 8.50    & 8.44    & & 0.04    & 0.05    \\
NGC 5888       & 2007Q        & II     &   + &   11.2 &   12.6 &  11.41 & $-$0.053 & 8.56    & 8.53    & & $\dots$ & 0.22    \\
NGC 6643       & 2008ij       & II     & $-$ &    4.5 &    3.0 &  10.46 & $-$0.101 & 8.53    & 8.52    & & 0.20    & 0.10    \\
NGC 5888       & 2010fv       & II     &   + &   11.2 &    9.1 &  11.41 & $-$0.053 & 8.68    & 8.53    & & 0.13    & $-$0.08 \\
NGC 1056       & 2011aq       & II     &   c &    2.6 &    0.7 &  10.57 & $-$0.075 & 8.49    & 8.44    & & $-$0.38 & $-$0.15 \\
NGC 5425       & 2011ck       & II     & $-$ &    2.7 &    2.8 &   9.83 & $-$0.113 & 8.42    & 8.38    & & 0.07    & 0.17    \\
NGC 5732       & ASASSN-14fj  & II     & $-$ &    4.8 &    8.3 &  10.33 & $-$0.162 & 8.36    & 8.47    & & $-$0.24 & 0.04    \\
NGC 5406       & PSN J14002117 +3854517& II     &   + &    7.1 &    2.0 &  11.22 & $-$0.072 & 8.50    & 8.54    & & $\dots$ & 0.20    \\
\hline
IC 0758        & 1999bg       & IIP    & $-$ &    1.8 &    3.4 &   9.32 & $-$0.049 & 8.35    & 8.43    & & 0.21    & $-$0.02 \\
NGC 1637       & 1999em       & IIP    &   c &    2.5 &    1.1 &   9.74 & $-$0.020 & 8.64    & 8.58    & & 0.22    & 0.08    \\
NGC 3184       & 1999gi       & IIP    & $-$ &    2.2 &    2.5 &  10.30 & $-$0.057 & 8.63    & 8.60    & & $\dots$ & $-$0.02 \\
NGC 2347       & 2001ee       & IIP    &   + &    4.6 &    6.2 &  11.19 & $-$0.093 & 8.52    & 8.51    & & 0.05    & 0.17    \\
NGC 5772       & 2002ee       & IIP    &   + &    7.5 &   12.9 &  10.52 & $-$0.069 & 8.44    & 8.52    & & 0.17    & 0.03    \\
NGC 0628       & 2003gd       & IIP    &   c &    3.2 &    7.3 &  10.29 & $-$0.132 & 8.48    & 8.51    & & $\dots$ & 0.14    \\
NGC 1093       & 2009ie       & IIP    &   + &    6.5 &   16.2 &  10.96 & $-$0.074 & 8.49    & 8.50    & & 0.05    & 0.20    \\
UGC 09356      & 2011cj       & IIP    & $-$ &    2.4 &    2.6 &   9.88 & $-$0.081 & 8.40    & 8.42    & & $-$0.10 & 0.01    \\
NGC 0628       & 2013ej       & IIP    &   c &    3.2 &    7.4 &  10.29 & $-$0.132 & $\dots$ & 8.51    & & 0.20    & 0.01    \\
NGC 7691       & 2014az       & IIP    & $-$ &    8.2 &    7.8 &  10.50 & $-$0.025 & 8.45    & 8.49    & & $-$0.16 & $-$0.32 \\
UGC 10123      & 2014cv       & IIP    &   c &    4.0 &    2.7 &  10.68 & $-$0.044 & 8.53    & 8.50    & & $-$0.09 & 0.16    \\
\hline
UGC 01635      & 2003G        & IIn    &   c &    6.1 &    2.8 &  10.42 & $-$0.025 & 8.64    & 8.56    & & 0.19    & 0.12    \\
NGC 0214       & 2005db       & IIn    &   + &    7.1 &    5.6 &  11.30 & $-$0.033 & 8.68    & 8.53    & & $-$0.08 & 0.06    \\
NGC 2906       & 2005ip       & IIn    &   + &    3.2 &    4.0 &  10.91 & $-$0.081 & 8.40    & 8.55    & & 0.02    & 0.01    \\
NGC 5630       & 2006am       & IIn    & $-$ &    3.3 &    2.7 &   9.89 & $-$0.057 & 8.31    & 8.32    & & $\dots$ & $-$0.00 \\
NGC 4644       & 2007cm       & IIn    &   + &    4.2 &    8.4 &  10.84 & $-$0.052 & 8.51    & 8.53    & & 0.05    & 0.09    \\
NGC 5829       & 2008B        & IIn    & $-$ &    6.3 &    9.0 &  10.33 & $-$0.098 & 8.50    & 8.46    & & 0.19    & 0.17    \\
UGC 09842      & 2012as       & IIn    &   c &   11.0 &   20.1 &  10.90 & $-$0.087 & 8.46    & 8.50    & & $-$0.40 & $-$0.28 \\
UGC 04132      & 2014ee       & IIn    & $-$ &    8.1 &   10.1 &  11.34 & $-$0.104 & 8.47    & 8.51    & & $\dots$ & $-$0.32 \\
\hline
NGC 1058       & 1961V        & IIpec  &   + &    1.0 &    2.7 &   9.11 & $-$0.072 & 8.23    & 8.50    & & 0.15    & 0.16    \\
\hline\hline
\end{tabular}
\end{center}
\end{table*}

\begin{table*}\tiny
\caption{Type Ibc/IIb SNe results.}
\label{tab:res3}
\begin{center}
\begin{tabular}{@{}llcccccccccccccc@{}}
\hline\hline                                                                                                     
\multicolumn{1}{c}{Galaxy}&\multicolumn{1}{c}{SN}&\multicolumn{1}{c}{SN type}&\multicolumn{1}{c}{AGN}&\multicolumn{1}{c}{r$_{eff}$}&\multicolumn{1}{c}{dep-GCD}&\multicolumn{1}{c}{Mass}&\multicolumn{3}{c}{12 + log (O/H)}  &  & \multicolumn{2}{c}{log (Z$_m$/Z$_\sun$)} \\
\cline{8-10} \cline{12-13} 
       &           &         &      &     &  &  &      grad  & local& total  & & local & total \\
\hline
NGC 3655       & 2002ji       & Ibc    &   c &    1.8 &    3.0 &  10.74 & $-$0.064 & 8.54    & 8.55    & & $-$0.49 & $-$0.47 \\
NGC 5714       & 2003dr       & Ibc    & $-$ &    4.8 &    7.8 &  10.26 & $-$0.079 & 8.39    & 8.49    & & $-$0.03 & $-$0.35 \\
UGC 06517      & 2006lv       & Ibc    & $-$ &    3.0 &    2.6 &  10.04 & $-$0.130 & 8.52    & 8.46    & & $\dots$ & $-$0.11 \\
UGC 02134      & 2011jf       & Ibc    &   c &    6.6 &    0.9 &  11.02 & $-$0.028 & 8.58    & 8.52    & & $\dots$ & $-$0.07 \\
\hline
UGC 04107      & 1997ef       & Ibc-pec & $-$ &    5.9 &    5.4 &  10.82 & $-$0.096 & 8.50    & 8.52    & & $-$0.25 & $-$0.24 \\
NGC 5559       & 2001co       & Ibc-pec &   c &    6.2 &    7.8 &  11.10 & $-$0.088 & 8.56    & 8.48    & & 0.16    & $-$0.11 \\
\hline
NGC 0991       & 1984L        & Ib     &   + &    3.8 &    4.3 &   9.63 & $-$0.142 & 8.43    & 8.44    & & 0.22    & 0.09    \\
NGC 5480       & 1988L        & Ib     & $-$ &    3.1 &    1.9 &  10.34 & $-$0.045 & 8.58    & 8.55    & & $-$0.26 & 0.01    \\
NGC 0776       & 1999di       & Ib     &   c &    8.0 &    5.9 &  11.16 & $-$0.076 & 8.67    & 8.59    & & 0.19    & 0.06    \\
NGC 2596       & 2003bp       & Ib     & $-$ &    8.7 &    9.2 &  11.16 & $-$0.123 & 8.51    & 8.49    & & 0.11    & 0.01    \\
NGC 7364       & 2006lc       & Ib     &   c &    4.9 &    4.2 &  10.80 & $-$0.018 & 8.57    & 8.57    & & 0.09    & $-$0.19 \\
NGC 6186       & 2011gd       & Ib     & $-$ &    4.2 &    0.6 &  10.77 & $-$0.039 & 8.60    & 8.58    & & 0.04    & 0.09    \\
\hline
UGC 05100      & 2002au       & IIb    & $-$ &    4.6 &    7.9 &  11.18 & $-$0.058 & 8.50    & 8.50    & & 0.19    & 0.11    \\
IC 0307        & 2005em       & IIb    &   c &   11.7 &   12.1 &  11.34 & $-$0.077 & $\dots$ & 8.49    & & $\dots$ & $-$0.20 \\
NGC 5735       & 2006qp       & IIb    &   + &    5.7 &    9.3 &  10.75 & $-$0.139 & 8.40    & 8.49    & & 0.07    & $-$0.05 \\
NGC 6643       & 2008bo       & IIb    & $-$ &    4.5 &    3.7 &  10.46 & $-$0.101 & $\dots$ & 8.52    & & $-$0.03 & 0.02    \\
NGC 1070       & 2008ie       & IIb    &   + &    6.2 &    8.7 &  11.13 &    0.020 & 8.63    & 8.52    & & 0.05    & $-$0.20 \\
UGC 10331      & 2011jg       & IIb    & $-$ &    5.8 &    6.7 &  10.36 & $-$0.072 & 8.39    & 8.37    & & $\dots$ & $-$0.04 \\
\hline
NGC 3310       & 1991N        & Ic     &   c &    1.5 &    0.6 &  10.13 & $-$0.027 & 8.35    & 8.33    & & $-$0.07 & 0.03    \\
MCG -01-54-016 & 2001ch       & Ic     & $-$ &    4.4 &    4.4 &   9.19 & $-$0.065 & 8.12    & 8.24    & & $\dots$ & 0.10    \\
NGC 4210       & 2002ho       & Ic     &   + &    4.3 &    3.5 &  10.59 & $-$0.097 & 8.56    & 8.53    & & $\dots$ & 0.22    \\
NGC 5000       & 2003el       & Ic     &   c &    6.2 &    6.6 &  11.08 & $-$0.048 & 8.58    & 8.53    & & $-$0.02 & $-$0.13 \\
UGC 03555      & 2004ge       & Ic     & $-$ &    4.7 &    2.0 &  11.12 & $-$0.134 & 8.58    & 8.46    & & 0.22    & 0.14    \\
NGC 4961       & 2005az       & Ic     & $-$ &    3.0 &    1.8 &  10.03 & $-$0.104 & 8.49    & 8.44    & & 0.16    & 0.17    \\
UGC 04132      & 2005eo       & Ic     & $-$ &    8.1 &    9.9 &  11.34 & $-$0.104 & 8.49    & 8.51    & & 0.20    & 0.20    \\
NGC 1058       & 2007gr       & Ic     &   + &    1.0 &    1.0 &   9.11 & $-$0.072 & 8.54    & 8.50    & & $\dots$ & $-$0.59 \\
NGC 7321       & 2008gj       & Ic     &   + &    7.6 &   17.8 &  10.73 & $-$0.103 & 8.39    & 8.52    & & $\dots$ & 0.02    \\
\hline\hline
\end{tabular}
\end{center}
\end{table*}

\end{document}